\documentclass[a4paper,11pt]{article}
\pdfoutput=1 

\usepackage[T1]{fontenc}
\usepackage{amssymb}
\usepackage{enumerate}
\usepackage{amsmath}
\usepackage{graphicx}
\usepackage{subfig}
\usepackage{jheppub}
\usepackage{dsfont}
\usepackage{framed}
\usepackage{slashed}
\usepackage{natbib}

\newcommand{\la}[1]{\label{#1}}

\newcommand{\1}{\mathds{1}}

\def\[{\left[}
\def\]{\right]}
\def\({\left(}
\def\){\right)}

\newcommand{\beq}{\begin{equation}}
\newcommand{\eeq}{\end{equation}}
\newcommand\beqa{\begin{eqnarray}}
\newcommand\eeqa{\end{eqnarray}}

\renewcommand{\i}{\mathrm{i}}

\newcommand{\Q}{{\bf Q}}
\newcommand{\bQ}{\mathbf{Q}}
\newcommand{\bP}{\mathbf{P}}
\newcommand{\e}{\text{e}}

\title{Regge trajectories and bridges between them in integrable AdS/CFT}
\author[a]{Nicol\`o Brizio,}
\author[a]{Andrea Cavaglià,}
\author[a]{Roberto Tateo,}
\author[b]{Valerio Tripodi}
\affiliation[a]{Dipartimento di Fisica, Università di Torino and INFN, Sezione di Torino, Via P. Giuria 1, 10125, Torino, Italy}
\affiliation[b]{International School of Advanced Studies (SISSA), Via Bonomea 265, 34136, Trieste, Italy}
\emailAdd{nicolo.brizio@unito.it}
\emailAdd{andrea.cavaglia@unito.it}
\emailAdd{roberto.tateo@unito.it}
\emailAdd{vtripodi@sissa.it}

\abstract{We study the analytic continuation in the spin of the planar spectrum of ABJM theory using the integrability-based Quantum Spectral Curve (QSC) method. 
Under some minimal assumptions, we classify the analytic properties of the Q-functions appearing in the QSC  compatible with the spin being non-integer. In this way we find not one - but \emph{two} distinct possibilities. While one is related to standard Regge trajectories, we show that the second choice can be used to build bridges which connect leading and subleading Regge trajectories, thus giving a shortcut to reach infinitely many sheets of the spin Riemann surface without going  explicitly around the branch points in the complex spin plane.
Moreover, the bridges are exact spin reflections of standard Regge trajectories. Together, these results reveal the existence of a hidden symmetry which we call ``twist/co-twist symmetry'': every non-BPS local operator has an exact image -- living below unitarity on a different Regge trajectory -- with the same $\Delta$  and the spin flipped by a Weyl reflection.
We discuss how an analogous phenomenon, based on the same mechanism at the level of the QSC, also occurs at non-perturbative level in $\mathcal{N}$$=$$4$ SYM. This provides a framework to understand recent independent observations of the symmetry in this model at weak and strong coupling. 
We present numerical results for Regge trajectories in planar ABJM theory, in particular we compute exactly the coupling dependence of the position of the leading Regge pole in the correlator of four stress tensors. The shape of this leading trajectory shows a behaviour at weak coupling that strongly resembles the BFKL limit in $\mathcal{N}$$=$$4$ SYM.
}

\begin{document}
\maketitle 
\section{Introduction}
Analytic continuation of the mass spectrum in spin controls high-energy scattering processes in relativistic QFT \cite{Regge:1959mz}. In a beautiful parallel, in conformal field theories the spectrum of scaling dimensions is analytic in spin \cite{Caron-Huot:2017vep}, and arranges itself in Regge trajectories. This structure is important to understand Lorentzian physics, such as the Regge limit~\cite{Costa:2012cb} and chaos \cite{Mezei:2019dfv}, and underlies powerful developments in the analytic conformal bootstrap \cite{Hartman:2022zik}. 

Regge trajectories are usefully represented in a Chew-Frautschi plot, plotting $S$ vs $\Delta$. In the past few years, it has been understood that generic points on Regge trajectories, in between local operators, can be interpreted as non-local ``light-ray'' operators which represent detector observables at null infinity~\cite{Kravchuk:2018htv,Chang:2020qpj,Kologlu:2019mfz,Henriksson:2023cnh,Caron-Huot:2022eqs}. 

The $\mathcal{N}=4$ Super Yang-Mills (SYM) theory and ABJM theory are prominent examples of holographic CFTs, where we are fortunate enough that integrability arises in the planar limit (as it does also in some examples of AdS$_3$/CFT$_2$), see e.g. \cite{Beisert:2010jr,10.1093/oso/9780198828150.001.0001,Seibold:2024qkh} for reviews. 
While our understanding of correlation functions is still not complete, we have an exquisite control on the planar spectrum in these two theories: integrability encodes it through a simple mathematical structure known as the Quantum Spectral Curve (QSC), first found in \cite{Gromov:2013pga}. 
Since the QSC can also be used to describe non-local light-ray operators, these integrable models give us great opportunities to test properties of light-ray operators, make intuitions concrete and try to find new universal features of spin physics in CFT. Conversely, computing these observables gives us precious information on these models and their real time dynamics. 

 For the case of $\mathcal{N}$$=$$4$ SYM, the study of continuous spin with the QSC is well developed~\cite{Alfimov:2014bwa,Gromov:2014bva,Gromov:2015wca,Gromov:2015vua,Alfimov:2018cms,Klabbers:2023zdz,Ekhammar:2024neh}. The QSC gives an unequivocal interpolation of the physical spectrum, which aligns with the hypothesis of ``light-ray operators universality'' \cite{Kravchuk:2018htv,Homrich:2022mmd}, namely that the shape of Regge trajectories is universal, and it does \emph{not} depend on the choice of any particular 4-point function to define them via the Lorentzian inversion formula. 
 
Studies of Regge trajectories in $\mathcal{N}$$=$$4$ SYM via the QSC have connected the integrability description of local operators to a different (and older) kind of integrability in gauge theory, the one of BFKL-type physics~\cite{Kuraev:1977fs, Balitsky:1978ic}, which is also one of the points where $\mathcal{N}$$=$$4$ SYM and planar QCD have close affinities. Recently, these studies have also been enriched with a study of the Odderon and by a general understanding of BFKL-type behaviour on general trajectories~\cite{Klabbers:2023zdz,Ekhammar:2024neh}. These findings also showed, as expected from the general theory of light-ray operators, that leading and subleading trajectories are represented as sheets of the same Riemann surface, connected in accordance with global symmetries, linked via complex branch points in spin $S$, see e.g. the beautiful plots of this surface in  \cite{Gromov:2015wca, Klabbers:2023zdz}. 
 
 A strong motivation for this paper was starting to develop similar tools for ABJM theory, where the previous QSC formulation was only limited to the case of local operators. 

The QSC was first developed in \cite{Gromov:2013pga,Gromov:2014caa} for $\mathcal{N}$$=$$4$ SYM, and in \cite{Cavaglia:2014exa, Bombardelli:2017vhk} for ABJM theory, where it was used to study various observables in \cite{Gromov:2014eha,Anselmetti:2015mda,Bombardelli:2018bqz,Lee:2018jvn,Lee:2019oml,Ekhammar:2023cuj}. Historically, there have been many important steps for arriving at this formulation, starting from the understanding of the exact two-dimensional S-matrices describing the worldsheet excitations~\cite{Beisert:2005tm,Beisert:2005fw} and passing then through the Thermodynamic Bethe Ansatz (TBA) method~\cite{Arutyunov:2009ur,Gromov:2009tv, Bombardelli:2009ns, Bombardelli:2009xz, Gromov:2009at}, which leads to the QSC through a complex chain of simplifications~\cite{Cavaglia:2010nm, Gromov:2011cx}. 
A simpler derivation is, at the moment, not understood. However, for AdS$_3$/CFT$_2$ with Ramond-Ramond flux and target space $AdS_3\times S^4\times T^4$, the QSC equations were recently conjectured\footnote{More work is needed to establish the full range of applicability of these equations, and to solve them in generic regimes. The AdS$_3$ case is more complicated than the other two known examples of QSC.}, simultaneously, in \cite{Ekhammar:2021pys,Cavaglia:2021eqr}, based on a streamlined classification approach (dubbed ``monodromy bootstrap'' in \cite{Ekhammar:2021pys}) that tries to condense the fundamental properties of the QSC in these integrable AdS/CFT cases. 

 Some previous results for particular limits of Regge trajectories in ABJM were obtained in \cite{Gromov:2014eha}, giving the slope of the trajectory around the stress tensor, and in \cite{Lee:2018jvn,Lee:2019oml},  giving some trajectories at weak coupling, but a general   non-perturbative method was not developed until now. 
 
To adapt the QSC for ABJM theory to describe continuous spin, we need to tweak one element of the construction: the so-called \emph{gluing matrix} $\mathcal{G}(u)$, which enters the Riemann-Hilbert problem describing the Q-functions $Q(u)$ of the model. 
 The logic emerging from previous studies is very simple: this matrix is constant for local operators, and non-constant (but highly constrained) for light-ray operators. In this paper, we determine its form for light-ray operators in ABJM theory at finite coupling, by using consistency conditions that are similar in spirit to the ``monodromy bootstrap'' of \cite{Ekhammar:2021pys}  and the similar approach of \cite{Alfimov:2018cms,Gromov:2017blm}. With this result we start a non-perturbative  exploration of Regge trajectories in ABJM theory.
 
 Moreover, something unexpected emerges from our analysis: a new symmetry that predicts that all the non-BPS, local operator points on the Chew-Frautschi plot have exact spin-flipped images on other trajectories, and which we expect to be there also in $\mathcal{N}$$=$$4$ SYM theory! 
 
Our main results and the precise statement of the symmetry are discussed in detail in the rest of this introduction.

\begin{figure}[t!]
\centering
\includegraphics[width=8cm,height=9cm]{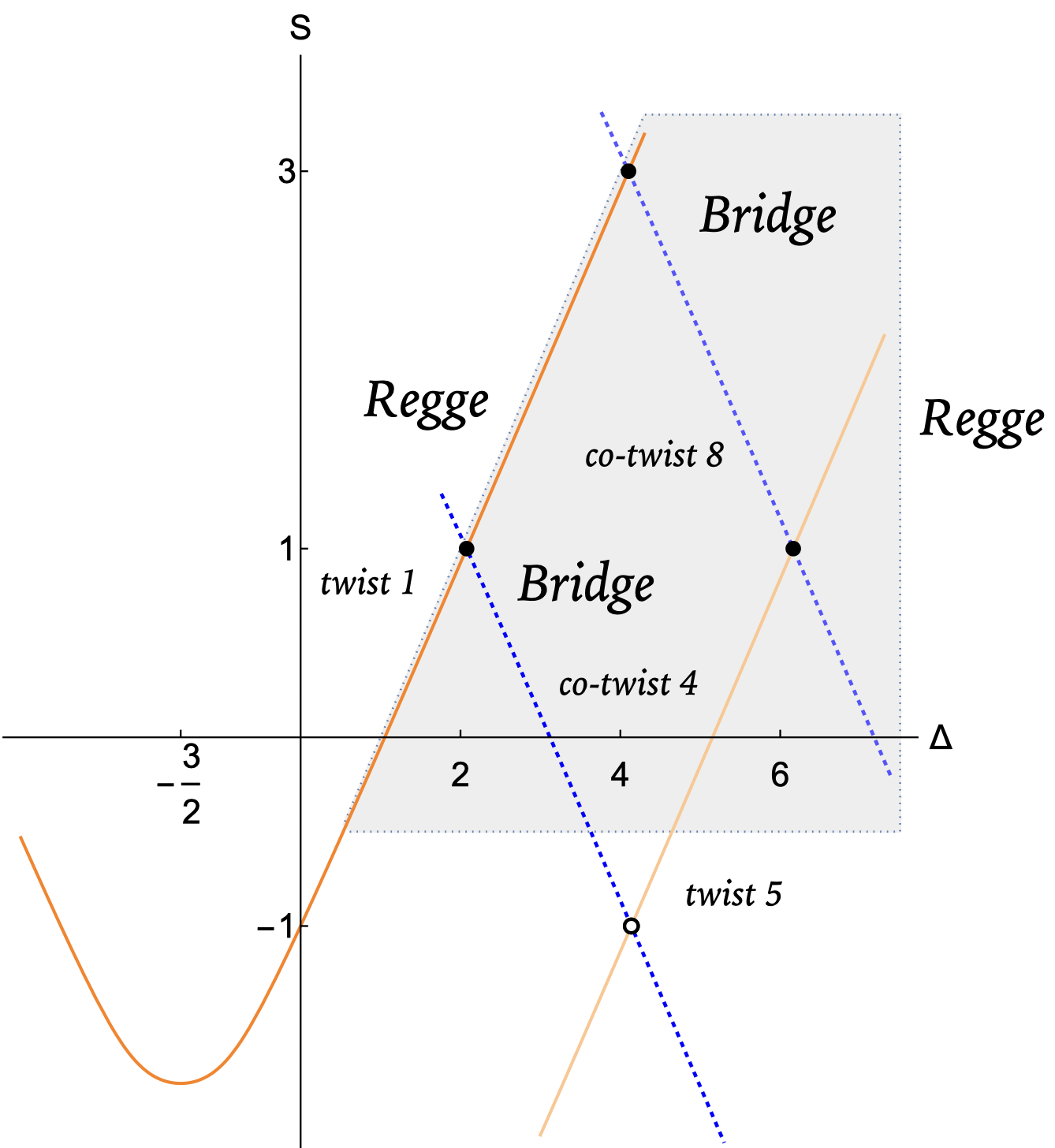}
\caption{The plot shows two Regge trajectories of ABJM theory of twists $1$ and $5$, and how they are connected by Bridging trajectories (here two are shown). The Bridges are spin shadows of two other Regge trajectories of twists $4$ and $8$ in this case. This structure underlies the twist/co-twist symmetry: each operator has a shadow sitting on a separate Regge trajectory and related by the exact non-perturbative map $\tau \leftrightarrow \tilde{\tau}$. This is on top of the usual symmetry $\Delta \leftrightarrow d - \Delta$ of the whole Chew-Frautschi plot~\cite{Kravchuk:2018htv}. 
The picture depicts some trajectories and bridges at finite coupling $h = 0.1$. The shaded area is the unitarity region, and marked points correspond to physical local operators (bullets) or spin-shadows of local operators (circles). 
The plot is given in the ``top'' grading,  which means all the shown trajectories interpolate between operators at the top of their supermultiplets.
}
\label{fig:bridgingintro}
\end{figure}

\subsection{Main results and Twist/Co-twist symmetry }
We examined how to modify the gluing conditions of ABJM theory in order to make the spin into a continuous parameter. This leads us to an unexpected discovery: there are in fact not one, but \emph{two} consistent forms of the gluing matrix for non-integer spin $S$. The two types of gluing matrix, which we dub $\mathcal{G}_{\text{Regge}}$ and $\mathcal{G}_{\text{Bridge}}$, are different and schematically related by\footnote{
The transposition refers to the formal structure of the matrices, not on the values of the coefficients.} 
\beq
\mathcal{G}_{\text{Regge}}(u)\cong \mathcal{G}_{\text{Bridge}}^T(u).
\eeq
Their precise form is given in the main text in Section \ref{sec:classify}. For local operators, the gluing matrix is simpler, as it becomes independent of $u$: in this case, its form is a subcase of both $\mathcal{G}_{\text{Regge}}$ and $\mathcal{G}_{\text{Bridge}}$. Starting from a QSC solution corresponding to a local operator and relaxing the gluing matrix to $\mathcal{G}_{\text{Regge}}(u)$, we can then trace the Regge trajectory passing through the operator. 

With this method we obtain several new results, in particular we compute the intercepts of the leading trajectories going through even and odd spins as functions of the coupling. These results are new, since the Pomeron eigenvalue in this theory was so far unknown. We also show that it exhibits a BFKL-type behaviour at weak coupling that invites further studies. 

Then, a natural question arises. What happens if we instead take the gluing condition to the \emph{other} form $\mathcal{G}_{\text{Bridge}}$? 
In this case we start tracing an upside-down trajectory like the ones shown in blue/dashed in Figure \ref{fig:bridgingintro}. We call such trajectories Bridging trajectories. \emph{What is their meaning? }Combining the consideration of the QSC and numerical experiments, we found  the properties listed below:\footnote{Preliminary results were announced in \cite{BrizioPoster}.}
\begin{enumerate}[1)]

\item We find that \emph{every} non-BPS local operator is crossed by both a regular Regge trajectory and a Bridging trajectory, obtained by using the two gluing matrices. 

\item The Bridging trajectories can be considered -- as the name suggests -- as Bridges, that allow to move between different Regge trajectories! See, for example, the figure \ref{fig:bridgingintro}. Using them, we can reach hidden sheets of the spectral surface without needing to go around branch points in the spin plane (which was the approach in previous studies \cite{Gromov:2015wca, Klabbers:2023zdz}).  

\item At the same time, from the structure of the QSC, we prove that the ``upside down'' trajectories computed with $\mathcal{G}_{\text{Bridge}}(u)$ are nothing else but spin-shadows of regular Regge trajectories. 
 The spin shadow map is one of the Weyl symmetries of the superconformal group. It is defined, for superconformal primaries at the top of their supermultiplets, by\footnote{In this work, we always treat supermultiplets as a whole. The shadow map for superdescendants is obtained by preserving the supermultiplet structure. }
\beq\label{eq:spinflip0}
S_{\text{top}}\rightarrow -d+2-S_{\text{top}}.
\eeq
\end{enumerate}
We have a precise proof of the statement 3) above, but property 1) can be considered a conjecture based on numerical experiments. Property 2) follows from 3) and 1) together.  
While we have performed a much less intensive numerical study, moreover, we found that all these three properties are valid not just in ABJM, but also in $\mathcal{N}$$=$$4$ SYM! Also in this case, at the level of the QSC they descend from the existence of \emph{two} gluing matrices, which was not noticed before. 
Therefore, the rest of this discussion will cover at the same time both ABJM theory and $\mathcal{N}$$=$$4$ SYM. 

Combining the properties 1) and 3) above, we arrive at the prediction that the Chew-Frautschi plot of the theory presents a certain symmetry between pairs of special points: we can call it \textbf{twist/co-twist symmetry}.\footnote{This terminology is inspired by the terminology of conformal spin and co-spin, introduced in \cite{JohanH} by Johan Henriksson, who noticed what appears to be the same symmetry in 1-loop perturbative results for  $\mathcal{N}$$=$$4$ SYM. We thank him for exchanges and sharing \cite{JohanH} with us. 
} Let us call
\beq\label{eq:twist}
\tau =\Delta_{\text{top}} - S_{\text{top}} 
\eeq the \textbf{twist} (as usual) and 
\beq\label{eq:defcotwist}
\tilde{\tau} =\Delta_{\text{top}} + d - 2 + S_{\text{top}}
\eeq
the \textbf{co-twist} of the supermultiplet.
 It has been noticed a number of times that $\tau \leftrightarrow \tilde{\tau} $ is a symmetry of the superconformal blocks.\footnote{We thank Johan Henriksson and Tobias Hansen for discussions.}
 Here we are  proposing that there is a deeper twist/co-twist symmetry, realized at the level of the Chew-Frautschi plot as follows:
\begin{framed}
\textbf{Twist/co-twist symmetry. }\emph{ For every non-BPS local operator sitting on a Regge trajectory of twist $\tau$, and having co-twist $\tilde{\tau}$, there will be a spin-flipped point belonging to a subleading Regge trajectory with twist $\tilde{\tau}$ and co-twist $\tau$.}
\end{framed}
This statement is valid at finite 't Hooft coupling. 
Notice that, since unitarity implies $\tau < \tilde{\tau}$, this creates a map between physical operators and points which sit below the unitarity bound on different, subleading trajectories. 

We are not aware of arguments pointing towards this symmetry in generic CFTs. For integrable AdS/CFT models, to the best of our knowlwedge this statement is new at the non-perturbative level. Perturbatively, what appears to be the same symmetry was noticed by Johan Henriksson at weak coupling in 1-loop studies of $\mathcal{N}$$=$$4$ SYM \cite{JohanH}, by inspecting the results of \cite{Henriksson:2017eej} 
for CFT data and noticing that they are symmetric with respect to $\tau \leftrightarrow \tilde{\tau}$. {At strong coupling, Julius Julius and Nika Sokolova recently discovered independently a manifestation of the symmetry: they point out that analytic continuation of subleading Regge trajectories is in some cases equivalent to flipping the points with precisely the map (\ref{eq:spinflip0}) (see \cite{Julius:2024ewf}, in particular Fig. 2). In particular, they found the image of the Konishi operator on a subleading trajectory, consistent with the picture described above.} 
 
Notice that our statement of the symmetry leaves something not clarified. Namely, we are saying that a given physical (non-BPS) local operator characterized by $\tau$, $\tilde{\tau}$ with $\tau < \tilde{\tau}$ will have an image on \emph{some} subleading trajectory with twist $\tilde{\tau}$. 
However, the number of Regge trajectories increases with respect to the tree-level value of the twist. Therefore, in general there will be multiple trajectories with degenerate tree-level twist approximately $\tilde{\tau}$, and we are not specifying on \emph{which one} we will find the image of our local operator. 
Relatedly, we might ask: since it is not possible that \emph{all} points with integer spin (in steps of 2) present below the unitarity bound on Regge trajectories correspond to spin-flips of local operators according to (\ref{eq:spinflip0}), what distinguishes the points involved in the symmetry from the ones that are not? Further studies are needed to answer this question and understand the precise pattern with which trajectories are linked by the symmetry, given their tree-level quantum numbers.   
In this paper, we will concentrate on presenting evidence for the general statement of the symmetry written above.\footnote{Regarding the more fine-grained question \emph{Which trajectory is linked with which by the bridges?}, we expect that the answer is hidden in the tree-level form of the Q-functions, which should encode the quantum numbers identifying different trajectories. The transformation rule for Q-functions under the symmetry is explained in this paper. {Inspecting the strong coupling data could also be very instructive for decoding the answer to this question, e.g. \cite{Julius:2024ewf} identifies precisely on which subleading trajectory lives the twist/co-twist image of the Konishi operator.}}

While we do not know of the possible physical origin of this symmetry,  we point out that it is not merely a direct effect of supersymmetry. Notice that supersymmetry implies that Regge trajectories come in groups, shifted by half-integer vectors in the $(\Delta, S)$, plane, replicating the supermultiplet structure, see e.g. \cite{Lemos:2021azv}. Here,  this structure is implicit since the QSC solutions are associated with the whole supermultiplets (disentangling the structure is a matter of susy representation theory). 
Instead, we are discussing something different: a symmetry that relates points on Regge trajectories which interpolate between \emph{different supermultiplets}.

To make a concrete example we present evidence (see Section \ref{sec:numericsSYM}) that a spin-shadow of the twist-2 Konishi operator $\mathcal{O}_{\text{Konishi, top}} = \text{tr}(\Phi_i \Phi^i)$ sits at any coupling on a trajectory of twist-4 in $\mathcal{N}$$=$$4$ SYM, once it is continued to the unphysical value of spin $S_{\text{top}}\rightarrow -2$. {This is consistent with the observation made at strong coupling by  Julius and Sokolova in \cite{Julius:2024ewf}.}
Explicit further checks of the symmetry using the explicit weak coupling results for ABJM theory of \cite{Papathanasiou:2009en,Papathanasiou:2009zm} are presented in Appendix \ref{app:weakc}.

It would be very interesting, of course, to find an explanation for the symmetry and understand whether it is valid also beyond the planar limit. More generally, is it a special feature of integrable holographic theories or perhaps a version of this symmetry might be more general? 

The remainder of this paper is organised as follows.

In Section \ref{sec:review}, we review the QSC formalism for ABJM theory in a concise way. In particular, we point out the role of the gluing matrix and its relation to the distinction between local and non-local light-ray operators. We also discuss how the Weyl reflections of the superconformal group are realized in the QSC.

In Section \ref{sec:classify}, we classify the consistent forms of the gluing matrix for both local and non-local light-ray operators in a sector of ABJM theory with LR symmetry and $u \leftrightarrow -u$ symmetry, finding  2 distinct results. We point out that the same situation arises in $\mathcal{N}$$=$$4$ SYM. This finding will be important in determining the presence of the symmetry. 
Section \ref{sec:numerics} contains the bulk of our results. After a brief review of the numerical method, we discuss new numerical results for the intercepts of the leading even and odd spins trajectories. Then we discuss the numerical data that corroborate the existence of the twist/co-twist symmetry, both in ABJM theory and $\mathcal{N}$$=$$4$ SYM. 

Finally, Section \ref{sec:conclusions}  presents some conclusions and discussions. The paper also contains some appendices with weak coupling checks and technical details. 
\section{Quantum spectral curve for integer and non-integer spin}\label{sec:review}
In this section, we review how the QSC encodes the spectrum and describe how to extend the setup to allow for analytic continuation in the spin. The readers familiar with the QSC may prefer to skip ahead to Subsections \ref{sec:Weyl} and \ref{local:vs}, where we introduce some new elements and summarize the distinction of the case of local and non-local light-ray operators at the level of the QSC. 
\subsection{The main ingredients of the QSC}
The QSC formalism encodes the spectral data of all non-BPS supermultiplets into the solutions of a particular nonlinear Riemann-Hilbert-type problem. 
This mathematical problem requires finding some functions of an auxiliary \emph{spectral parameter} $u$, called Q-functions, based on their analytic properties and the functional relations they satisfy, the so-called QQ-relations
\cite{Reshetikhin:1983vw, Beisert:2005fw, Dorey:2006an, Ferrando:2020vzk, Ekhammar:2021myw}. In short:
\beq
\texttt{QSC}=\texttt{QQ-relations}+ \texttt{Analyticity}.
\eeq
The QQ-relations are expected to be common to all integrable systems with the same global symmetry. For ABJM theory, these functional relations reflect the superconformal algebra $OSp(4|6)$. 
On the other hand, the analytic properties are generally model-dependent and will differ, in particular, for integer or non-integer spin.

Let us start by giving a brief description of the Q-system, focusing on a specific subset of functional relations sufficient to numerically solve the QSC.
\subsubsection{Quick summary of QQ-relations in ABJM theory}
The Q-functions are characterized by specific indices representing particular representations of the bosonic subgroups of $OSp(4|6)$, namely  $SO(6)\times Sp(4)$. 

We will restrict our attention to three fundamental types of Q-functions:
\begin{itemize}
\item 6 functions organized in the antisymmetric matrix $\bP_{ab}(u)$, with $a,b \in \left\{1,2,3,4\right\}$, $\bP_{ab}(u) = -\bP_{ba}(u)$.
\item 6 functions organized in the antisymmetric matrix $\Q_{ij}(u)$, with $i,j \in \left\{1,2,3,4\right\}$, $\bQ_{ij}(u) = -\bQ_{ji}(u)$.
\item 16 functions organized as a $4\times 4$ matrix $Q_{a|j}(u)$, with $a, i\in \left\{1,2,3,4\right\}$.
\end{itemize}
These functions are not all independent, as the matrices above satisfy the following constraints:\footnote{Where, for an antisymmetric matrix $A$, $\text{Pf}(A) \equiv A_{12} A_{34} + A_{14} A_{23} - A_{13} A_{24}$. } 
\beq\label{eq:quadraticP}
1 = -\text{Pf}( \bQ_{ij} )= -\text{det}( Q_{a|i} ) = \text{Pf}( \bP_{ab} ). 
\eeq
As we will see, the functions $\bP_{ab}$ and $\bQ_{ij}$ have the simplest analytic properties and play a central role in the construction. The essential QQ-relations will tell us how to construct $\bQ_{ij}$ from $\bP_{ab}$. This involves the following two steps:
\begin{enumerate}[1)]
\item Assuming knowledge of $\bP_{ab}$, we can construct $Q_{a|i}$ by solving the following system of difference equations
\beq\label{eq:shiftQ} 
Q_{a|i}\(u+\frac{\i}{2}\)=\bP_{ab}(u) \, Q^{b|k}\(u-\frac{\i}{2}\) \, \kappa_{ki}, 
\eeq
where repeated indices are summed over, the range of all indices is $\left\{1,\dots,4\right\}$, 
$Q^{a|i}$ denotes the inverse matrix, 
i.e. $Q^{a|i} Q_{a|j} = \delta^i_j$, and $\kappa$ is a constant antisymmetric matrix: 
\beq\label{eq:defkappa}
\kappa_{ij} = \left(
\begin{array}{cccc}
 0 & 0 & 0 & 1 \\
 0 & 0 & -1 & 0 \\
 0 & 1 &  0 & 0 \\
 -1 & 0 & 0 & 0
\end{array}\right).
\eeq
\item Once we have obtained $Q_{a|i}$, we can simply construct $\bQ_{ij}$ as
\beq\label{eq:defQij}
\bQ_{ij}(u) = - Q_{a|i}\(u-\frac{\i}{2}\) \bP^{ab}(u) Q_{b|j}\(u-\frac{\i}{2}\),
\eeq
where again raised indices denote inversion: $\bP_{ab} \bP^{bc} = \delta_a^c$. 
\end{enumerate}
Below, we will review how to numerically solve these equations, which, of course, depends critically on the analytic properties we want to impose on the Q-functions. This is the next crucial ingredient, which we will address next. 
\subsubsection{Branch cuts and gluing matrix}\label{analytic}
\begin{figure}[t!]
\centering
\includegraphics[width=6cm,height=6cm]{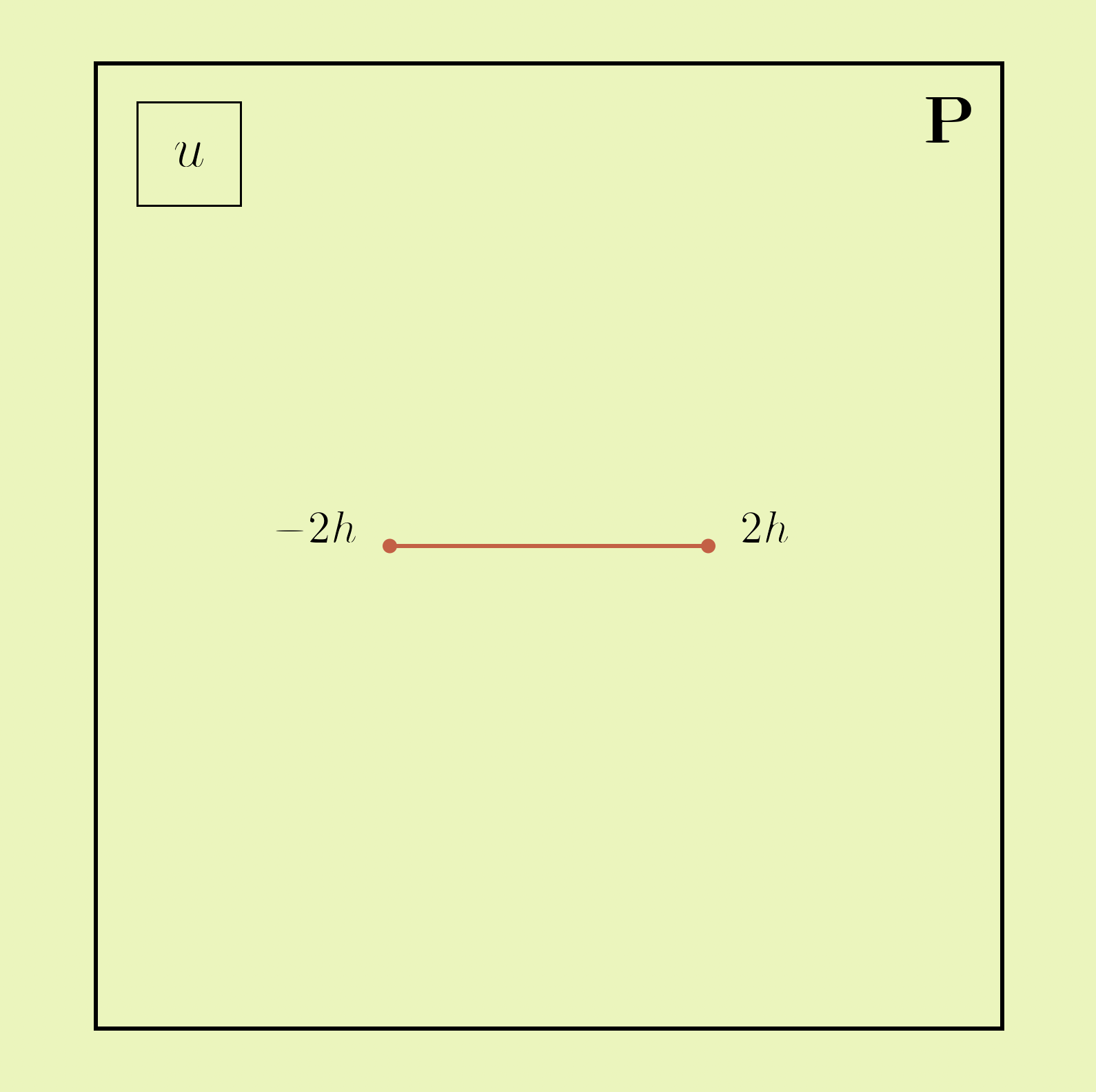}\quad\includegraphics[width=6cm,height=6cm]{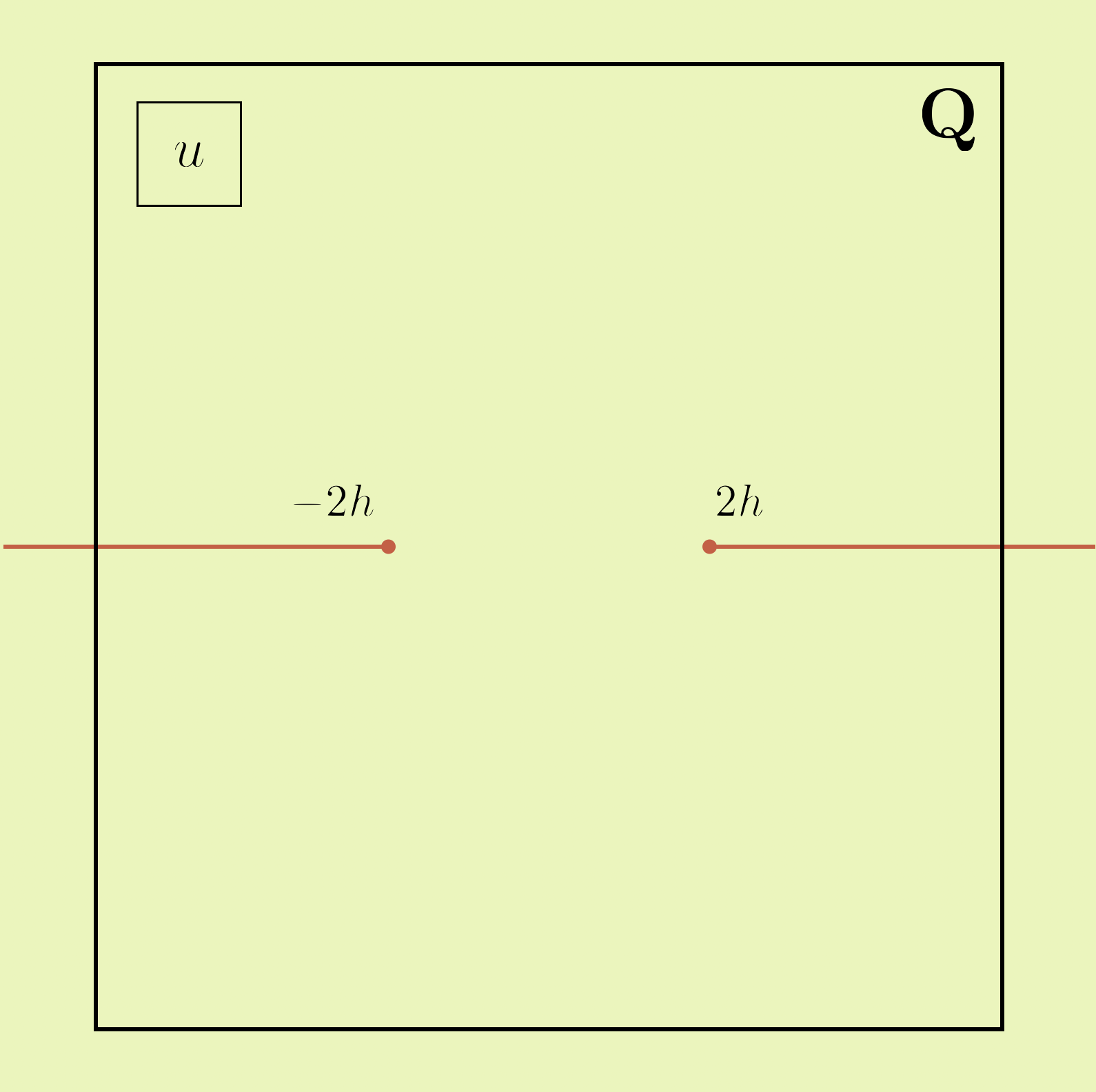}
\caption{The cut structure of $\bP$ and $\bQ$-type Q-functions on their defining Riemann sheet. The position of the branch points depends on the effective coupling $h \equiv h(\lambda)$. The QSC closure axioms describe how to identify the $u$ spectral parameter in the two pictures in such a way that the QQ-relations hold.}
\label{fig:cut1}
\end{figure}

\paragraph{Branch cut structure: minimal description.}
It turns out that, with respect to the spectral parameter, Q-functions live on a complicated Riemann surface with infinitely many branch points. 
Despite being intricate, this structure is highly constrained,  
and we can provide a minimal description of it 
by specifying the position of branch points for  \emph{some} of the functions on specific Riemann sheets. 
These particularly simple Q functions are the ones denoted as $\bP_{ab}$ and $\bQ_{ij}$, and the relevant properties are (see figure \ref{fig:cut1}):
\begin{itemize}
\item There is a special Riemann sheet in $u$ where the functions $\bP_{ab}(u)$ have a single \emph{short} branch cut for $u\in [-2 h, 2 h ]$, where $u = \pm 2 h$ are quadratic branch points. 

\item In analogy, the functions $\bQ_{ij}(u)$ admit a Riemann sheet where they have a single \emph{long} cut, running on the semi-infinite lines  $u\in (-\infty, -2 h] \cup[+2 h , +\infty)$. 

\item The upper half-planes of these two Riemann sheets are identified. As will become clear later, choosing the upper over the lower half plane is a convention. 

\item 
Apart from these branch points, Q-functions have no other singularities. In particular, they do not have any poles anywhere and are finite as the branch points are approached.
\end{itemize}

Finally, on top of the previous ones, there is an additional property, which is valid in AdS$_5$ and AdS$_4$ (but should be dropped in AdS$_3$ as found in \cite{Ekhammar:2021pys,Cavaglia:2021eqr}):
\begin{itemize}
\item All these branch points are of square-root type.  
\end{itemize}

\paragraph{Position of the branch points and 't Hooft coupling constant}
The position of the branch points is related to the 't Hooft coupling via the \emph{interpolating function} $h(\lambda)$, for which a very solid conjecture was made in \cite{Gromov:2014eha} (later also extended to the planar ABJ case~\cite{Cavaglia:2016ide}, which -- according to the proposal of that paper -- is therefore also implicitly covered by our analysis). 
For the ABJM case this reads~\cite{Gromov:2014eha}:
\beq
\lambda = \frac{\sinh(2 \pi h)}{2 \pi}\,_3 F_2\left( \frac{1}{2},\frac{1}{2}, \frac{1}{2}; 1, \frac{3}{2} ; -\sinh^2(2 \pi h)\right).
\eeq
\paragraph{Power-like asymptotics. }
For large $u$ the  Q functions $\bP_{ab}$ and $\bQ_{ij}$  have a power-like behaviour on the respective sheets, in particular, 
\beq
\bP_{ab}(u)\simeq\# \,u^{\mathcal{N}_{ab}},\quad u \rightarrow + \infty ,
\eeq
while\footnote{By convention, for functions with a long cut we require these asymptotics to hold as we approach infinity staying slightly above the cut. 
There is an alternative but equivalent formulation, involving a \emph{different} basis of Q-functions, where power-law asymptotic behavior is imposed at large values of $\text{Re}(u)$ just below the long cut.
}
\beq\label{eq:powQij}
\bQ_{ij}(u)\simeq\# \,u^{\hat{\mathcal{N}}_{ij}},\quad u \rightarrow + \infty + \i 0^+,
\eeq
where $\mathcal{N}_{ab} = \mathcal{N}_{ba}$ and $\hat{\mathcal{N}}_{ij} = \hat{\mathcal{N}}_{ji} $ are parameters which will encode, respectively, the R-symmetry and the conformal charges. Consistently with these equations, we will have
\beq
\label{eq:powQai}
Q_{a|j}(u)\simeq\# \,u^{\mathcal{N}_a + \hat{\mathcal{N}}_i },\quad u \rightarrow + \infty .
\eeq
For the moment we postpone the detailed discussion of the relations between these parameters and the charges of supermultiplets to Section \ref{sec:multipletscharges}. 

\paragraph{Q-system vs cut structure. }
With the proposed cut structure, the construction starts being mathematically tight due to tension between the QQ-relations and the assumed analytic properties. 
Let us review how this works~\cite{Gromov:2015wca,Gromov:2015vua,Bombardelli:2017vhk,Alfimov:2018cms}. 

The point is that if we view the QQ-relations (\ref{eq:shiftQ}) and (\ref{eq:defQij}) as rules to construct $\bQ_{ij}$ taking the single-cut functions $\bP_{ab}$ as an input, then we will obtain a solution that is apparently at odds with the cut structure required for $\bQ_{ij}$. This is due to the fact that finite difference equations such as $(\ref{eq:shiftQ})$ will tend to replicate the cut structure of $\bP_{ab}$, generating a solution for $\bQ_{ij}$ with an infinite ladder of short cuts. 

\begin{figure}[t!]
\centering
\includegraphics[width=6cm,height=6cm]{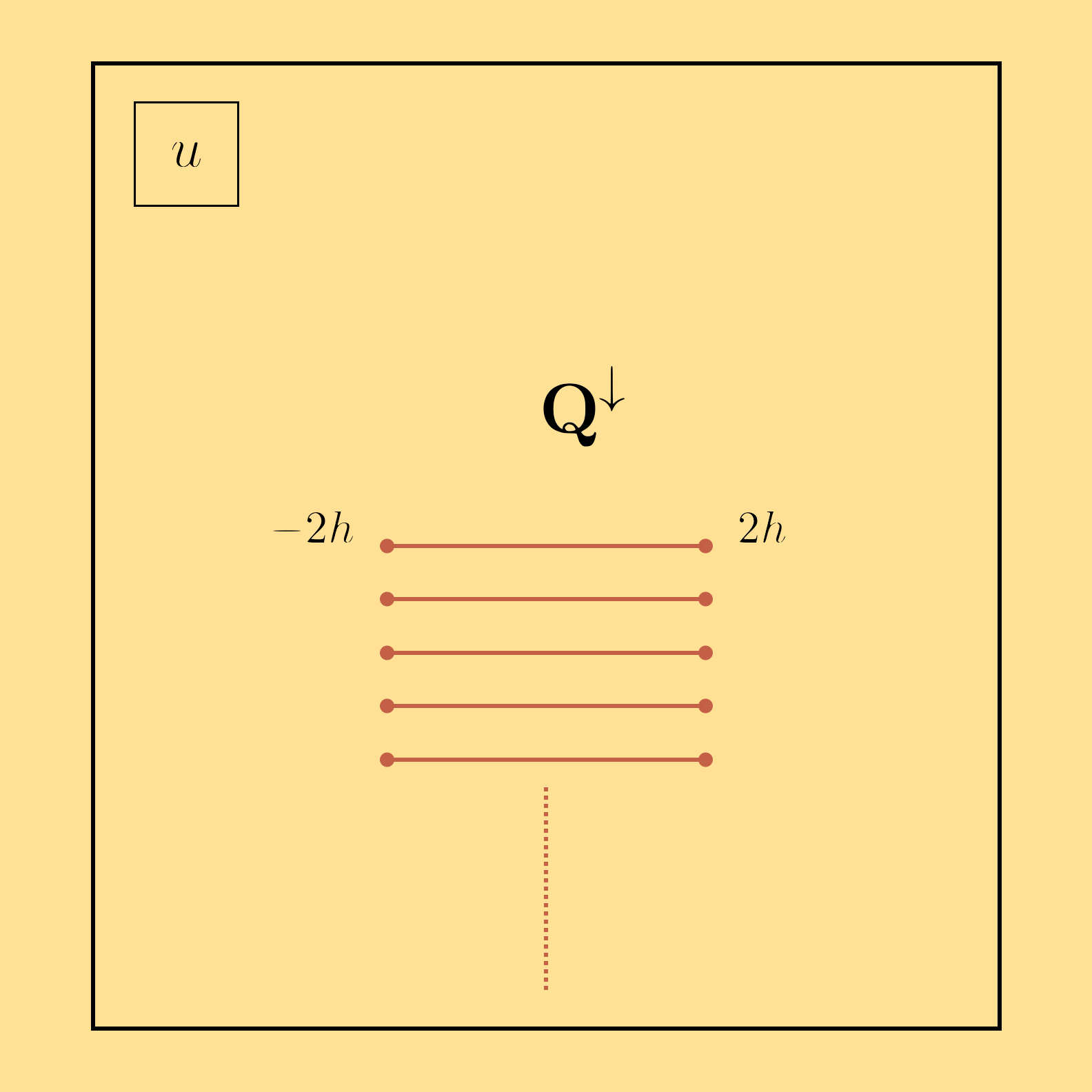}\quad\includegraphics[width=6cm,height=6cm]{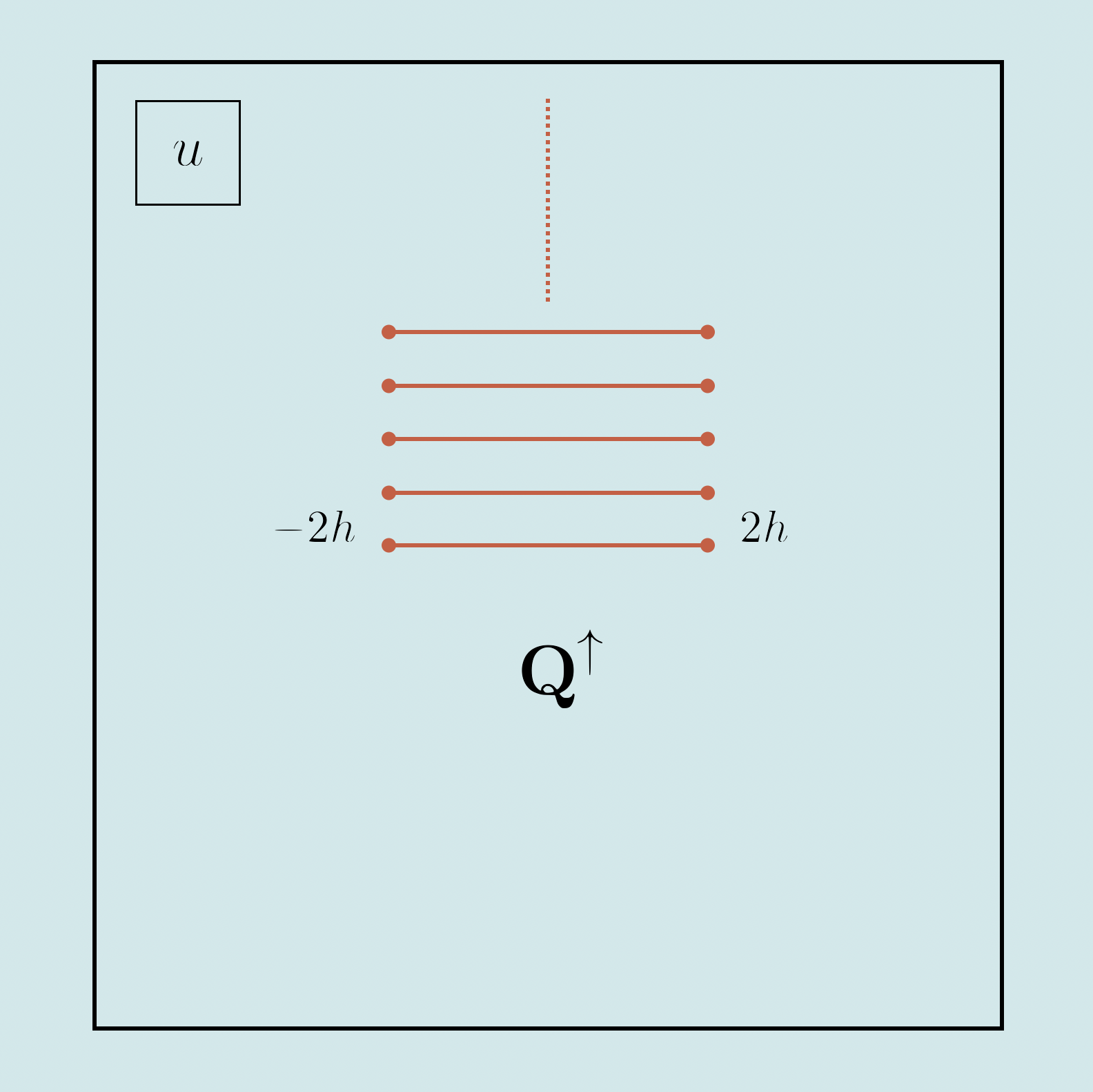}
\caption{Using the QQ-relations, with the $\bP_{ab}$ functions as input, we can construct two alternative bases for $\bQ_{ij}$. 
Each of these bases is analytic in half of the complex plane and has an infinite series of \emph{short} cuts in the other half.}
\label{updown}
\end{figure}

Postponing more details of the construction, we notice that two alternative bases of solutions for $\bQ_{ij}$ can be constructed by requiring that they are analytic in the upper or lower half-planes. These two alternative solutions are denoted as $\bQ^{\downarrow}_{ij}$ and $\bQ^{\uparrow}_{ij}$, respectively. Each is analytic in a half of the complex plane, but has an infinite ladder of branch cuts for $u \in [-2h, 2 h] \pm \i \mathbb{N}$ in the other half; see figure \ref{updown}. 
How do we square this finding with the prescription above, i.e. that $\bQ_{ij}$ should instead have a single long cut and no other cuts in the complex plane? The answer is that these should all be sections of the same function, glued by a symmetry of the Q-system. 

\begin{figure}[t!]
\centering
\includegraphics[width=6cm,height=6cm]{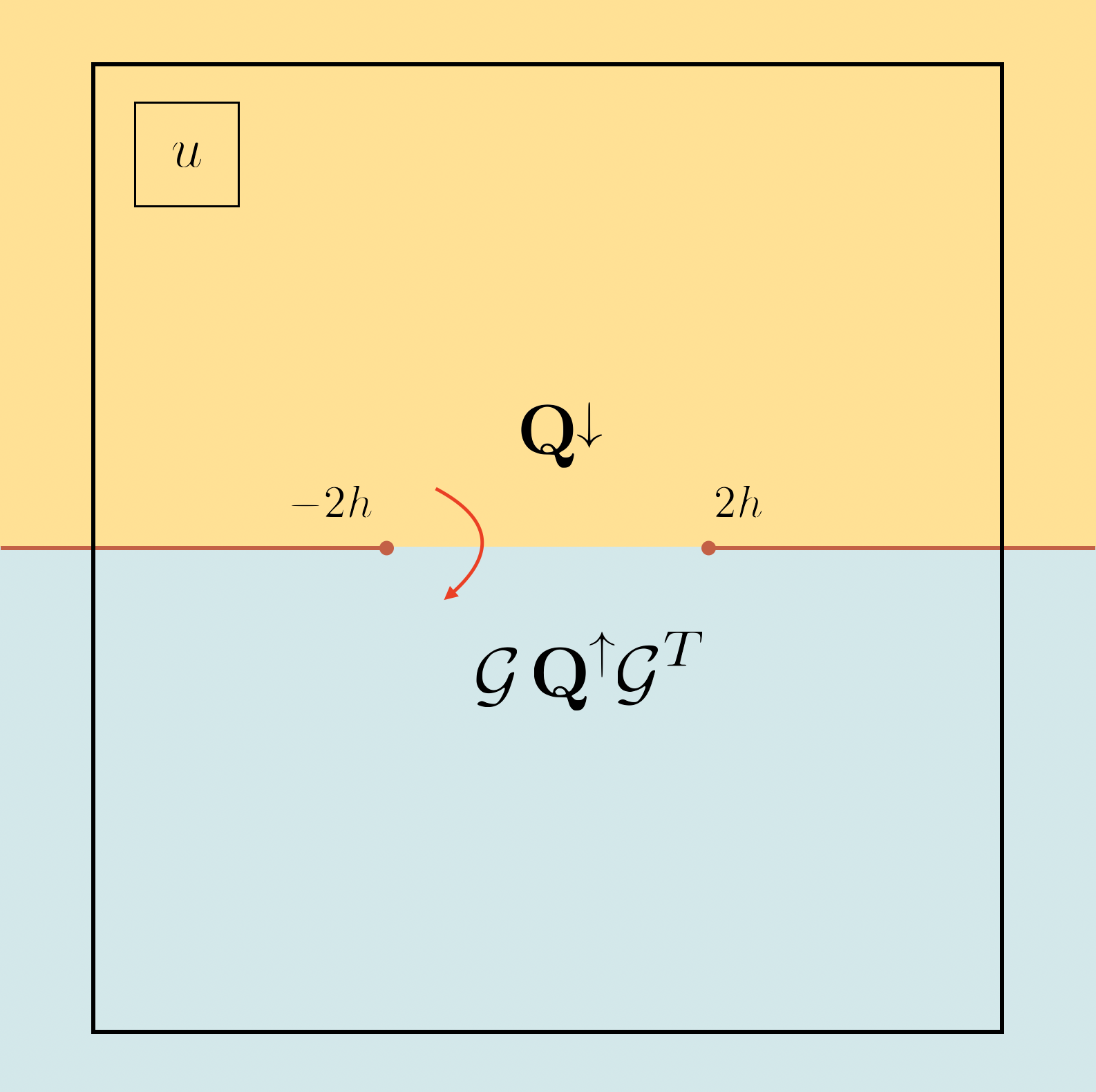}
\caption{The compatibility between the postulated cut structure of Figure \ref{fig:cut1} and the consequences of the QQ-relations in Figure \ref{updown} is ensured by \emph{gluing} the sheets of $\bQ^{\downarrow}$ and $\bQ^{\uparrow}$. The \emph{gluing} matrix $\mathcal{G}$ should be a symmetry of the Q-system.}
\label{gluing}
\end{figure}

\paragraph{The gluing matrix.}\label{det}
In particular, by convention, we shall demand that $\bQ_{ij}$ with a single long cut agrees with the function $\bQ^{\uparrow}$ constructed above in the upper half plane, where they are both holomorphic:
\beq\label{eq:glue1}
\left. \bQ_{ij}(u) \right|_{\text{long cut}} \equiv \bQ^{\downarrow}_{ij}(u), \quad\text{ for } \text{Im}(u) > 0 ,
\eeq
and that its analytic continuation to the lower half plane can be identified with the function $\bQ^{\downarrow}$, see figure \ref{gluing},
\beq\label{eq:glue2}
\left. \bQ_{ij}(u) \right|_{\text{long cut}} \equiv   \mathcal{G}_{i}^{m}(u) \, {\bQ}^{\uparrow}_{mn}(u) \, {\mathcal{G} }_{j}^{n}(u), \quad\text{ for } \text{Im}(u) < 0 ,
\eeq
where for generality, we included the possibility to allow for a recombination of the Q-functions, parametrised by a $4 \times 4$ \emph{gluing matrix} $\mathcal{G}_i^j(u)$, which will be a crucial character in our discussion.  
To preserve the integrity of the cut structure, the matrix $\mathcal{G}(u)$ should be free of any cuts or singularities. 
Additionally, we require $\bQ_{ij}$ to satisfy the QQ-relations 
in the lower half plane as well. 
Consequently, $\mathcal{G}$ must keep the Q-system  unchanged, and it is easy to verify that this leads to the following properties: 
\begin{itemize}
\item It should satisfy 
\beq\label{eq:constrG}
\mathcal{G}_{i}^{m}(u) \kappa_{mn} \mathcal{G}_{j}^{n}(u+\i) = \kappa_{ij},
\eeq
which in particular implies that it is a periodic matrix under shifts of $u\rightarrow u+2 \i$.
\item It should have determinant $1$. 
\end{itemize}
Moreover, as we already anticipated, consistency with the analytic properties of Q-functions ensures that
\begin{itemize}
\item $\mathcal{G}(u)$ should be holomorphic in the whole complex plane,
\end{itemize}
as otherwise it would introduce new singularities. 
The conditions (\ref{eq:glue1}),(\ref{eq:glue2}) indicate that, aside from the Q-system symmetry that is parametrized by the gluing matrix, 
$\bQ^{\downarrow}$ and $\bQ^{\uparrow}$ should be viewed as two sections of the same function, i.e. they are connected via analytic continuation through the cut at $u \in [-2h, 2h]$. This can be expressed as:
\beq\label{eq:basicgluing}
\tilde{ \bQ}^{\downarrow}_{ij}(u) = \mathcal{G}_{i}^{m}(u) \, {\bQ}^{\uparrow}_{mn}(u) \, {\mathcal{G} }_{j}^{n}(u) ,
\eeq
where we denote by \emph{tilde} the analytic continuation around the branch point $-2 h$ or $+2 h$. 

Finally, it is now easy to anticipate the main conjecture: solutions of the QSC with a gluing matrix that does not depend on the spectral parameter are in one-to-one correspondence with super conformal multiplets of local operators in the theory (or their Weyl reflections shadows). 
To describe non-local light-ray operators, one needs to relax this condition and allow $\mathcal{G}(u)$ to be a nontrivial function of $u$.

\subsubsection{In-depth discussion of the basis of Q-functions }\label{section}
In the present section, we specify some further details of the construction, which are needed to uniquely fix the basis of Q-functions, and thus to define 
unambiguously the gluing matrix. Readers who are less interested in these technical details may also safely skip to later sections. 
\paragraph{Specifying the basis of Q-functions. }
Let us start by spelling out how we can define unambiguously the basis of Q-functions. 
Given $\bP_{ab}(u)$, we define $Q_{a|i}^{\downarrow}(u)$ to be the solution to (\ref{eq:shiftQ}) analytic in the upper half plane with an asymptotic expansion 
\beq\label{eq:pure1}
Q_{a|i}^{\downarrow}(u) \simeq u^{\mathcal{N}_a + \hat{\mathcal{N}}_i} \left( B_{a|i}^{\downarrow}  + \sum_{n\ge1} \frac{b_{(a|i), n}^{\downarrow}}{u^n} \right),\quad u \rightarrow + \infty ,
\eeq
which can be analytically extended to the whole upper half plane. 
Once we fix the leading coefficients\footnote{These coefficients should satisfy some constraints which we report in Appendix \ref{ancil}. 
As an example of an obvious constraint, the coefficients $B_{a|i}^\downarrow$ should be compatible with the fact that $\text{det} \left(Q_{a|i}^\downarrow\right) = -1$. } $B_{a|i}^\downarrow$, this solution is fixed uniquely if we assume that the powers $\hat{\mathcal{N}}$ are all distinct and non-integer:
$$\label{Nhat}
\hat{\mathcal{N}}_1 \neq  \hat{\mathcal{N}}_2 \neq   \hat{\mathcal{N}}_3 \neq  \hat{\mathcal{N}}_4 \notin \mathbb{Z}.  
$$
For a generic point on a Regge trajectory, this will be the case, but for local operators, $\hat{\mathcal{N}}_2 - \hat{\mathcal{N}}_3$ will be integer since this quantity is identified with (twice) the spin, as we discuss later. 
When this is the case, we add an extra condition to make the definition of $Q_{a|i}^\downarrow(u)$ unique up to rescalings: if $\hat{\mathcal{N}}_2 = \hat{\mathcal{N}}_3 + \ell $ with $ \ell \in \mathbb{N}$, we demand that\footnote{In this paper, we study only states with $u \leftrightarrow -u$ symmetry, for which this condition is pleonastic, since this coefficient is then automatically zero by the symmetry. In \cite{Bombardelli:2018bqz} more general states were studied. }
$b_{(a|2), \ell}^\downarrow = 0$ to eliminate the resulting ambiguity in the definition of $Q_{a|2}^\downarrow(u)$, see \cite{Bombardelli:2018bqz} for full details.   
Similarly, and with the same subtleties, we define the basis of solutions analytic in the lower half plane as those which have the same asymptotics but for $u \rightarrow -\i \infty$:
\beq\label{eq:pure2}
Q_{a|i}^{\uparrow}(u) \simeq u^{\mathcal{N}_a + \hat{\mathcal{N}}_i} \left(B_{a|i}^\uparrow  + \sum_{n\ge1} \frac{b_{(a|i), n}^{\uparrow}}{u^n} \right).
\eeq
This asymptotic expansion can be analytically extended from the real axis to be valid in the lower half plane. 

The expansions in (\ref{eq:pure1}) and (\ref{eq:pure2}) are dubbed \emph{pure asymptotics} in the QSC literature. Notice that, while for generality we kept the coefficients $B^{\downarrow}$, $b^{\downarrow}$ vs $B^{\uparrow}$, $b^{\uparrow}$ distinct, one can always choose them to be the same: in fact, this is automatic if we specify the leading order normalization such that  $B_{a|i}^{\downarrow} = B_{a|i}^{\uparrow}$, which is always possible. However, it is convenient sometimes to normalize the two expansions differently.\footnote{In particular, in the symmetric sector we focus on in this paper it is convenient to define $Q_{a|i}^{\uparrow}(u)$ via  $Q_{a|i}^{\downarrow}(-u)$, as explained later. In this case the coefficients of the expansion have a precise relative normalization depending on this definition. } 

With these precise conventions in place, 
we then also have a unique definition of $\bQ^{\downarrow}$ and $\bQ^{\uparrow}$, given by equation (\ref{eq:defQij}) with $Q_{a|i}$ replaced by $Q_{a|i}^{\downarrow}(u)$ or $Q_{a|i}^{\uparrow}(u)$, respectively.
\paragraph{Additional useful objects and their properties. }

Apart from the Q-system compatibility requirements,
the gluing matrices must also satisfy some less obvious constraints, as we know from the analysis of \cite{Bombardelli:2017vhk} where these properties were initially deduced from the TBA analysis. Since we will use these properties in our classification of gluing matrices, let us present them. It is useful to introduce the following object
\beq\label{eq:defOmega}
\Omega_i^j(u) \equiv Q_{a|i}^{\uparrow }\(u-\frac{\i}{2}\)\, Q^{a|j \downarrow }\(u-\frac{\i}{2}\),
\eeq
which by definition is a matrix that rotates the two bases of $Q_{a|i}$ functions:
\beq\label{omega:def}
Q_{a|i}^{\uparrow}(u) = Q_{a|j}^{\downarrow}(u) \Omega_i^{\, j}\left(u+\frac{\i}{2}\right).
\eeq
By its very construction, it is immediate to prove that this matrix  is also a symmetry of the Q-system, i.e. it satisfies
\beq\label{eq:constrOm}
\Omega_{i}^{m}(u) \kappa_{mn} \Omega_{j}^{n}(u+\i) = \kappa_{ij} ,\quad \text{det}(\Omega) = 1,
\eeq
and in particular it is $2 i$-periodic. Notice that this matrix has an infinite ladder of short cuts at $u\in[-2h, 2h]+ i \mathbb{Z}$, and the shifts in the equations (\ref{eq:constrOm}) should be taken without crossing these cuts. 
A very useful property of $\Omega$ is that its asymptotic limit at $u \rightarrow \pm \infty$ is fixed by its definition. This will be exploited later to constrain the form of $\mathcal{G}(u)$. 

From this matrix, as well as the gluing matrix $\mathcal{G}(u)$, we can construct another natural object,
\beq\label{eq:deff}
f_i^j(u) \equiv \mathcal{G}_i^k(u) \, \Omega_k^j(u) ,
\eeq
which now allows connecting the analytic continuation of a Q-function to itself:
\beq\label{eq:fgluing}
\tilde{ \bQ}^{\downarrow}_{ij}(u) = f_{i}^{m}(u) \, {\bQ}^{\downarrow}_{mn}(u) \, f_{j}^{n}(u) .
\eeq
By construction, we see that also $f_i^{\, j}(u)$  is a Q-system symmetry, i.e. it satisfies
\beq\label{eq:constrf}
f_{i}^{m}(u) \kappa_{mn} f_{j}^{n}(u+\i) = \kappa_{ij} ,\quad \text{det}(f) = 1.
\eeq
However, $f_i^j$, like $\Omega_i^j$ is not an entire function, but rather is an $2\i$-periodic function with an infinite ladder of branch cuts. It is known from the TBA derivation in \cite{Bombardelli:2017vhk} that this matrix should have a very peculiar structure:
\begin{itemize}
\item It satisfies 
\beq\label{eq:ffinvconstr}
f_i^j(u) + \(f^{-1}(u) \)_i^j = 2 \delta_i^j.
\eeq
\item Moreover, the matrix $A_i^{j}(u) \equiv f_i^j(u)  - \delta_i^j$ has rank one, and in particular we have the following parametrisation in terms of $4+4$ functions $\tau_i(u)$, $\tau^i(u)$:
\beq\label{f:form}
f_i^j(u) = \delta_i^j - \tau_i(u) \tau^j(u) ,
\eeq
where
\beq\label{eq:constrtau}
\tau_i(u) \tau^i(u) = 0.
\eeq
Notice that the periodicity condition (\ref{eq:constrf}) then implies that $\tau_i(u)$, $\tau^i(u)$ should be $2 \i$-periodic, and more precisely  satisfy: 
\beq\label{tau:up}
\tau^i(u) \equiv e^{-i \mathcal{P}} \kappa^{ij} \, \tau_j(u+\i) ,
\eeq
where $\mathcal{P}$ is a constant.\footnote{This constant may depend on the particular solution of the QSC, and on all parameters except $u$.} 
The 4 functions $\tau_i(u)$ 
 should be free of singularities at finite $u$, apart from the branch points. 
\end{itemize}
These properties were found in the derivation of the QSC from the TBA~\cite{Bombardelli:2017vhk}. 
While we lack a complete argument,
we suspect that the conditions above stem from the consistency of the gluing equations with the fact that the branch points are quadratic, and that Q-functions should remain finite at the branch points.

All in all, the consequence is that the matrix $f_i^j(u)$ does not have 16 independent elements, but we can reduce its parametrisation to a constant $\mathcal{P}$ and four functionally independent degrees of freedom, $\tau_i(u)$, subject to (\ref{tau:up}) and (\ref{eq:constrtau}).  We will use the constraints above in our classification of gluing matrices.

\subsubsection{Asymptotics and quantum numbers}\label{sec:multipletscharges}

\paragraph{Constraints between powers. }
Up to this point, we have only specified that on the main Riemann sheet, the $\bP$ and $\bQ$ functions should have distinct power-like asymptotics. These powers satisfy certain constraints and can be expressed in terms of five quantities $M_A$ ($A\in\left\{1,2,5\right\}$) and $\hat{M}_I$ ($I\in\left\{1,2\right\}$). The powers of the $\bP$ functions can be represented as
\beq\label{P:asy}
\bP_{ab}\simeq\# \,u^{\mathcal{N}_a+\mathcal{N}_b}\quad\text{as}\quad u\gg1,
\eeq
where
\beq 
\mathcal{N}_a=\frac{1}{2}(-M_1-M_2-M_5,-M_1+M_2+M_5,M_1-M_2+M_5,M_1+M_2-M_5).
\eeq
Due to the single cut of $\bP$ functions, these powers should be integers, i.e. $M_A \in \mathbb{Z}$ for $A = 1,2,5$. The asymptotics of $\bQ$ functions are parametrised as 
\beq\label{Q:asy} 
{\bf Q}_{ij}\simeq\#\, u^{\hat{\mathcal{N}}_i+\hat{\mathcal{N}}_j-1}\quad\text{as}\quad u\gg1,
\eeq
with
\beq\label{eq:NhM}
\hat{\mathcal{N}}_i=\frac{1}{2}\(\hat{M}_1+\hat{M}_2,\hat{M}_1-\hat{M}_2,\hat{M}_2-\hat{M}_1,-\hat{M}_1-\hat{M}_2\).
\eeq
At finite coupling, these charges are generically all different. By consistency, the multiplicative prefactors in front of the asymptotics of $\bP$ and $\bQ$ are also constrained in terms of the charges, see \cite{Bombardelli:2017vhk} and appendix \ref{ancil} for details. 

The description we have given so far has some redundancies, since the symmetry of the Q-systems allows us to rotate the bases of $\bP$ functions and $\bQ$ functions. To partially remove this ambiguity we can choose a conventional ordering of the magnitude of the asymptotics, i.e. we demand 
\beq\label{eq:magnitudes}
\hat{M}_1>\hat{M}_2\geq M_2>M_1> M_5 \geq 0.
\eeq
Of these requirements, some are just conventional, i.e. the relative orderings of the powers among different $\bP$'s, or the ones among different $\bQ$'s. On top of this, the demand that the powers of $\bQ$'s should be (in absolute value) greater than the powers of $\bP$'s is a physical requirement that describes the unitarity region of the spectrum~\cite{Bombardelli:2017vhk}, as we recall below. 

\paragraph{Map to quantum numbers of supermultiplets. }\label{multi}
Let us now review the physical identification between the quantum numbers and the powers in the asymptotics of Q-functions; see \cite{Bombardelli:2017vhk} for more details and checks with the Asymptotic Bethe Ansatz limit. 

Every solution of the QSC represents a whole non-BPS superconformal multiplet of the planar theory. We can identify the multiplet by giving the R-symmetry and conformal charges of the superconformal primary operator(s), i.e. those with the lowest value of scaling dimension $\Delta$ inside the multiplet. Those will also be called the \emph{top} operators in the multiplet. The representation of the R-symmetry is specified by the $SU(4)$ Dynkin labels $[p_1, q, p_2]_{SU(4)}$, while the conformal charges are the spin $\Delta_{\text{top}}$ and $S_{\text{top}}$, where the subscript reminds us that we are specifying the charges for the superconformal primary at the top of the multiplet. The charges $\left[p_1, q, p_2; \Delta_{\text{top}}, S_{\text{top}}\right]$ are known as the charges of the multiplet in the ``distinguished grading''. 

The dictionary with the powers entering in the QSC formulation is
\beq
\begin{aligned}
&M_1 = 1+r_2,\quad M_2 =  2+r_1 , \quad M_5 = r_3,\\
&\hat{M}_1 = \Delta_{\text{top}} + S_{\text{top}} + 2,\quad\hat{M}_2 = \Delta_{\text{top}} - S_{\text{top}}+1,
\end{aligned}
\eeq
where  $r_1 \equiv 1/2\,(p_1 + p_2) + q$, $r_2 \equiv 1/2\,(p_1 + p_2)$, $r_3 \equiv 1/2\,(p_2 - p_1)$ are the corresponding $SO(6)$ weights, and in fact the condition (\ref{eq:magnitudes}) translates into the unitarity bound satisfied by physical multiplets
\beq
2 S_{\text{top}}+1 >0,\quad\Delta_{\text{top}} - S_{\text{top}} \geq 1 + r_1, \quad r_1+1>r_2, \quad1+r_2>r_3 .
\eeq
The second inequality is saturated only strictly at zero coupling for particular operators, including the $sl(2)$-like sector.

The quantum numbers of other conformal primary operators in the same multiplet can all in principle be reconstructed using supersymmetric representation theory.  In particular,  the  
dimensions of conformal primary operators in the supermultiplet range in steps of $\frac{1}{2}$ from $\Delta_{\text{top}}$ to $\Delta_{\text{bottom}} = \Delta_{\text{top}} + 3$, since we have $6$ fermionic supercharges which act nontrivially on the top operator and can be used to build superdescendants. Every superdescendant has dimension, spin and R-symmetry Dynkin labels shifted by half-integer steps compared to the top component. 

In particular, inside the multiplet there are operators with quantum numbers corresponding to the so-called $sl(2)$-like grading. Their quantum numbers $\left[p_{1},q, p_{2} ; \Delta, S \right]_{sl(2)}$ are shifted with respect to the ones of the top operators by
\beq\label{eq:sl2shifts}
\(\Delta_{sl(2)}, S_{sl(2)}\) = (\Delta_{\text{top}}, S_{\text{top}}  ) + (1,1),
\eeq
and
\beq\label{eq:sl2shifts2}
\left[p_{1},q,p_{2} \right]_{sl(2)} = \left[p_1 , q, p_2 \right]_{\text{top}} +\left[1,0,1\right].
\eeq
For the operators at the bottom of the supermultiplet (i.e., the conformal primaries with the highest scaling dimension in the supermultiplet), we have instead
\beq\label{eq:bottomshifts}
(\Delta_{\text{bottom}}, S_{\text{bottom}}) = (\Delta_{\text{top}}, S_{\text{top}}  ) + (3,3),
\eeq
and 
\beq\label{eq:bottomshifts2}
\left[p_{1},q,p_{2} \right]_{\text{bottom}} = \left[p_1 , q, p_2 \right]_{\text{top}} .
\eeq
We notice that it is the bottom component that gives the leading Regge trajectory in the sector we will consider. In the following, we will always plot only one representative for each multiplet, specifying which.

\subsection{Weyl symmetries in the QSC }\label{sec:Weyl}
The symmetry of Regge trajectories we present in this paper is intimately related to the way the QSC reflects the Weyl symmetries of the superconformal group. 

They appear as a  symmetry of the QSC under simple Q-system transformations that multiply the ``fermionic'' indices of Q-functions, namely the lower indices of type $i,j = 1,\dots, 4$,
by multiplication with one of the following matrices (each one defining a new solution): 
\begin{align}\label{eq:MWeyls}
&\left(\mathbb{M}_1 \right)_i^{\,\,\,j} =  \left(
\begin{array}{cccc}
 1 & 0 & 0 & 0 \\
 0 & 0 & 1 & 0 \\
 0 & -1 &  0 & 0 \\
 0 & 0 & 0 & 1
 \end{array}\right)\quad\text{or} \quad
\left(\mathbb{M}_2 \right)_i^{\,\,\,j} =  \left(
\begin{array}{cccc}
 0 & 0 & 0 & 1 \\
 0 & 1 & 0 & 0 \\
 0 & 0 & 1 & 0 \\
 -1 & 0 & 0 & 0
\end{array}\right).
\end{align}
In particular, the action on the Q-functions $\bQ_{ij}$ is explicitly described by the following transformation,
\beq \label{eq:mapQM}
\mathbf{Q}\rightarrow \mathbb{M}\cdot\mathbf{Q}\cdot\mathbb{M}^T,
\eeq
while upper indices, e.g. the ones in $\bQ^{ij}$,  would instead be transformed by multiplying by the inverse transposed matrices. 

From the QQ-relations point of view, these transformations preserve the form of the Q-system\footnote{In particular, this follows from the fact that $\mathbb{M} \cdot \kappa \cdot \mathbb{M}^T = \kappa$.}, 
and at the same time preserve the gluing equations (\ref{eq:basicgluing}) upon an appropriate redefinition of the gluing matrix $\mathcal{G} \rightarrow \mathbb{M} \cdot \mathcal{G} \cdot \mathbb{M}^{-1}$. However, the transformations of the form (\ref{eq:mapQM})  also modify the asymptotics of the Q-functions, and thus allow to construct a new solution.

In particular, notice that conjugation by $\mathbb{M}_1$ corresponds to swapping the indices\footnote{A part for the appropriate $\pm 1 $ signs coming from the components of $\mathbb{M}$. These signs are needed to preserve the QQ-relations, but play no role in the present argument on the redefinition of the charges. } 
\beq\label{S:sym}
i=2 \,\leftrightarrow\, i=3 ,
\eeq
and thus the charges of the new solution are obtained from the original ones by the swap (the notation will be clear in a minute):
\beq\label{M1:M2}
\texttt{W}_{S}:\quad \hat{M}_1\,\leftrightarrow\, \hat{M}_2 .
\eeq
The conjugation by $\mathbb{M}_2$, on the other hand, affects the permutation of indices
\beq
 i=1 \,\leftrightarrow\, i=4 ,\eeq
thus corresponding to a swap of the charges
\beq\label{eq:WeylDelta}
\texttt{W}_{\Delta}:\quad\hat{M}_1\,\leftrightarrow\, -\hat{M}_2 . 
\eeq
At the level of quantum numbers, these transformations act in a very simple way: one of them corresponds to the shadow transformation for the scaling dimension of the \emph{bottom} operators in the multiplet
\beq\label{map:Delta}
\texttt{W}_{\Delta}: \quad\left(\Delta_{\text{bottom}}, S_{\text{bottom}}\right)\,\rightarrow\, \left(3-\Delta_{\text{bottom}}  , S_{\text{bottom}}\right) ,
\eeq
while the other corresponds to a spin-shadow transformation for the \emph{top} component of the multiplet:
\beq\label{map:spin}
\texttt{W}_S : \quad\left(\Delta_{\text{top}}, S_{\text{top}}\right)\,\rightarrow\, \left(\Delta_{\text{top}} , -1-S_{\text{top}}\right).
\eeq
These are individually Weyl reflections that leave the conformal algebra invariant.\footnote{In general dimension, the two Weyl reflections would act us $
\texttt{W}_{\Delta}:\,\Delta \,\leftrightarrow\, d-\Delta $, and $\texttt{W}_S:\, S\,\leftrightarrow\, 2-d - S $. } Following the CFT literature, we will call the two transformations the \emph{shadow transform} and the \emph{spin-shadow}. 
 The reason why one acts on the bottom component and the other on the top component is due to the supersymmetric structure. Indeed, these two particular transformations, acting at the levels of the multiplets specified in (\ref{map:Delta}) and (\ref{map:spin}), are Weyl reflections of the whole $OSp(6|4)$ superalgebra, as can be seen from the form of the Casimir eigenvalue, see e.g. \cite{Binder:2021cnk}.  

The twist $\tau$ and co-twist $\tilde{\tau}$ defined in the introduction are interchanged by one of the Weyl reflection, as they can be written as  
\beq
\tau = \left. \Delta - S \right|_{\text{top}}=\Delta_{\text{top}} - S_{\text{top}}, \quad \tilde{\tau} = \left. \Delta - (\texttt{W}_S[S])\right|_{\text{top}} = \Delta_{\text{top}} + 1 + S_{\text{top}} .
\eeq
 
Let us briefly summarize our findings in this section:
we showed that, for any solution of the QSC, one can automatically build two other solutions (simply by applying the symmetry transformations (\ref{eq:MWeyls}) on indices of Q-functions). The quantum numbers of the shadow solutions are transformed as in (\ref{map:Delta}) or (\ref{map:spin}), thus they represent shadow states related by Weyl symmetries. Notice also that, if the original solution satisfies the unitarity bound (\ref{eq:magnitudes}), the shadow solutions will be below the unitarity bounds. 

So far, these look like almost trivial statements, and it is certainly not too surprising that the QSC should encode the existence of Weyl reflections. 
However, this structure will play a central role when we classify possible gluing conditions for non-integer spin. It will imply that the moment we start taking the spin to be non-integer, we will have two distinct ways to draw a trajectory through the same local physical operator: one will be a Regge trajectory, and the second one will be a spin-shadow reflection of a \emph{different} Regge trajectory. 
This will imply that the same operator will appear \emph{twice} in the $\Delta_{\text{bottom}} > d/2$ half of the Chew-Frautschi plot: once on its own Regge trajectory, and another time its spin-shadow will appear on a different, subleading Regge trajectory. 

\subsection{Local vs non-local operators}\label{local:vs}
All the discussion above did not specify whether the spin should take particular values, and applies equally well for local operators as well as for the non-local light-ray operators which constitute their analytic continuation in spin. 
Based on the significant experience accumulated in $\mathcal{N}$$=$$4$ SYM~\cite{Alfimov:2014bwa,Gromov:2014bva,Gromov:2015wca,Gromov:2015vua,Alfimov:2018cms,Klabbers:2023zdz,Ekhammar:2024neh}, we can formulate  a very simple criterion to distinguish the two cases:
\begin{framed}
\beq\label{eq:localvsnonlocal}
\texttt{$u$-independent gluing matrix}\,\leftrightarrow\, \texttt{local operators (or their shadows)}. 
\eeq
\end{framed}
Indeed, also in the ABJM case it was shown in \cite{Bombardelli:2017vhk} that assuming that the gluing matrix is constant immediately forces the spin to be integer or half integer. 
In the following section we will constrain the possible form of non-constant gluing matrices, which will allow to compute general points along Regge trajectories. 

Notice also that there are points of integer spin on Regge trajectories which do not belong to any local operators,  for example, the points with odd spin on the Regge trajectory interpolating between the even-spin operators.
 The condition above (\ref{eq:localvsnonlocal}) is sharp and distinguishes these two cases, i.e. inspecting whether the gluing matrix is constant or not gives a test of locality. 
This is interesting in view of recent studies of ``missing operators'' and how they put constraints on conformal Regge theory~\cite{Homrich:2022mmd,Henriksson:2023cnh}. In a way, the constant or non-constant nature of the gluing matrix is the QSC analogue of the norm $\langle \mathbb{O} \mathbb{O} \rangle$ constructed in \cite{Henriksson:2023cnh}, which discriminates the local nature of a light-ray operator. It would be fascinating to connect this quantity to the parameters in the QSC gluing matrix. 
\section{Classifying the gluing matrices}\label{sec:classify}
To be able to actually solve the QSC equations, for instance numerically, it is very useful to have insight on the form of the gluing matrix. In the following we use the general QSC axioms, plus some minimal assumptions, to find the possible form of the gluing matrix for generic light-ray operators in terms of a small number of parameters. We will then use these equations for the numerical study of section \ref{sec:numerics}.

A notable finding of this section is that, when relaxing in the simplest way the analytic requirements that force the spin to be integer, two possibilities arise, i.e. the two gluing matrices $\mathcal{G}_{\text{Regge}}(u)$ and $\mathcal{G}_{\text{Bridge}}(u)$ discussed in the introduction. They correspond to two distinct ways to take the spin off its integer quantized values, and are related by the spin-shadow Weyl reflection $\texttt{W}_S$. This structure underlies the twist/co-twist symmetry. 

\subsection{A symmetric sector}
We will specialize the discussion to a sector where solutions of the QSC are invariant under two independent $\mathbb{Z}_2$ symmetries. While this is a restriction, it is still a very interesting infinite-dimensional portion of the spectrum which encompasses in particular the quantum numbers of the Pomeron~\cite{Papathanasiou:2009zm}.

\paragraph{LR symmetry. } 
LR-symmetry is related to an isomorphism of the $OSp(6|4)$ superconformal algebra where one applies Hodge duality to the $SU(4)$ R-symmetry indices. The states that are eigenvalues of this transformation correspond to QSC solutions satisfying \beq
\bP_{14}(u) = \bP_{23}(u),\label{eq:foldP}
\eeq
which implies for the quantum numbers
\beq
M_5 = 0\, \leftrightarrow\, r_3 = 0.
\eeq
The identification of Q-functions in (\ref{eq:foldP}) can also be rewritten as
\beq\label{tau:period}
\bP^{ab}(u) =\kappa^{ac} \bP_{cd}(u)\kappa^{db}.
\eeq 
This symmetry then percolates to other elements of the construction. In particular, by studying  \eqref{eq:shiftQ}, we realize that now $Q^{a|k} \kappa_{ki}$ and $\kappa^{ab} Q_{b|i}$ satisfy the same difference equation.\footnote{In this paragraph, we omit $\downarrow/\uparrow$ for simplicity.} Identifying solutions via their asymptotics, one can \emph{choose} an appropriate normalization of the Q functions such that (see,  \cite{Bombardelli:2017vhk}):
\footnote{We are not adopting exactly the same conventions as in \cite{Bombardelli:2017vhk}. The canonical normalization of Q-functions used here for the LR-symmetric case differs from that paper. In the LR symmetric case we see that the constant $e^{i \mathcal{P}}$ defined in (\ref{tau:up}) reduces to two possible values  $e^{\i \mathcal{P}} = \mathbf{s} =\pm 1$. A redefinition related to this sign connects the convention of this paper to the one in  \cite{Bombardelli:2017vhk}, i.e. 
 $Q_{a|i}^{\text{here}} $ is equivalent to $ Q_{a|i} $ in \cite{Bombardelli:2017vhk}
 if $\mathbf{s} =1$, while if $\mathbf{s}=-1$ we multiply the $a$ index by a diagonal matrix $\left\{\i, -\i, -\i, \i\right\}$.
}
\beq\label{subsector}
Q^{a|j}(u) \kappa_{ji}= 
\kappa^{ab}Q_{b|j}(u)\,\mathbb{K}_i^{j} ,
\eeq
where 
$\mathbb{K}$ is a constant matrix given by
\beq
\mathbb{K}_i^{j}=\begin{pmatrix}
1&0&0&0\\
0&-1&0&0\\
0&0&-1&0\\
0&0&0&1
\end{pmatrix}.
\eeq
As explained in \cite{Bombardelli:2017vhk}, another consequence of LR symmetry is that the periodicity \eqref{tau:up} of the functions $\tau_i(u)$ is enhanced to
\beq \label{eq:tauperiod}
\tau_i(u+i)=\tau_k(u)\,\mathbb{K}_i^{k} ,
\eeq
i.e. $\tau_{1,4}$ functions are  $\imath$-periodic, while $\tau_{2,3}$ are $\imath$-anti-periodic. Another consequence of LR symmetry following immediately from the ones above is 
\beq
\kappa \cdot \bQ^{-1} \cdot \kappa=  \mathbb{K} \cdot \bQ \cdot \mathbb{K}^T ,\label{eq:LRsymQ}
\eeq
which is simply saying that $\bQ_{14} = -\bQ_{23}$. 

\paragraph{Spin chain parity symmetry.}
A second $\mathbb{Z}_2$ symmetry of the system is an internal symmetry acting on the spin chain representation of the states  (based on the planar gauge theory diagrams) by a reflection across the midpoint of the spin chain. We will restrict our analysis to states that are eigenvalues of this transformation. 
At the level of Q functions this $\mathbb{Z}_2$ parity is realized very simply as the change of spectral parameter $u \leftrightarrow -u$ of the $\bP$ functions.\footnote{The transformation $u \leftrightarrow -u$ should be performed on a sheet where the branch cut on the real axis is drawn as short, i.e. along $[-2h, 2h]$.  } 

For parity-symmetric states, this sends the QSC solution to itself which means that the set of $\bP$ functions is invariant under $u\leftrightarrow -u$ up to rescalings.\footnote{Notice also that this symmetry implies that for parity-symmetric states the large-$u$ expansions of all quantities will go in steps of $1/u^2$ only.} 
 Concretely, as we review in \ref{sec:methodrecap}, these functions are defined as series expansions in $1/x(u)$, where the Zhukovsky map is defined as follows:
\beq\label{zhukovsky}
x(u) = \frac{u + \sqrt{u - 2 h}\sqrt{u + 2 h}}{2 h}.
\eeq
In particular, $x(-u) = -x(u)$, which implies that under this parity symmetry the $\bP$ functions pick a sign based on the charges in their asymptotics. For example, $\bP_{12}(-u) = (-1)^{M_1} \bP_{12}(u)$ since the asymptotic behavior is $\bP_{12}(u) \simeq u^{-M_1}$ with $M_1 \in \mathbb{Z}$.
 
We can represent these changes of sign conveniently as a Q-system transformation, i.e. 
\beq\label{P:R} 
\bP_{ab}(-u)=\mathbb{R}_a^{c} \bP_{cd}(u)\mathbb{R}_b^{\,d},
\eeq
where $\mathbb{R}_a^{\,b}$  is a constant matrix given by
\beq\label{R:matrix}
\mathbb{R}_a^{\,b}=
\left(
\begin{array}{cccc}
 i^{3 M_1-M_2} & 0 & 0 & 0 \\
     0 & i^{M_2-M_1} & 0 & 0 \\
 0 & 0 & i^{3 M_2-3 M_1 }& 0 \\
 0 & 0 & 0 & i^{M_2-3 M_1 }\\
\end{array}
\right).
\eeq
Notice $\mathbb{R}\cdot\kappa\cdot\mathbb{R}^T=\kappa$, which ensures that this symmetry is compatible with LR symmetry. 

In presence of spin chain parity, it is very simple to construct the Q-functions  satisfying the Q-system holomorphic in the lower half plane, i.e. the objects we already encountered, $Q_{a|i}^{\uparrow}$ and  $\bQ_{ij}^{\uparrow}$. In fact, given the symmetry they can simply be taken as 
\beq
Q_{a|i}^{\uparrow}(u) = (\mathbb{R}^{-1} )_a^{\,c}\, Q_{c|i}^{\downarrow}(-u) ,\label{Q:Rsym}
\eeq
since it is simple to verify that, in virtue of (\ref{P:R}), the combinations on the r.h.s. satisfy the same functional equations as the ones defining the functions on the l.h.s., and are analytic in the correct half plane. 
 Following the definition (\ref{eq:defQij}), one can then establish the following identity:
\beq\label{eq:Qminusu}
\bQ^{\uparrow}_{ij}(u) = \kappa_{ik} \bQ^{km \downarrow}(-u) \kappa_{mj} \,.
\eeq
From now on, in this simplified sector we omit all symbols $\downarrow$/$\uparrow$,  with the understanding that by default we are always referring to $\bQ^{\downarrow}$ and $Q_{a|i}^{\downarrow}$. The functions $\bQ^{\uparrow}$ and $Q_{a|i}^{\uparrow}$ are then simply related to these as above. 

When classifying the admissible gluing matrices compatible with this $u \leftrightarrow -u$ symmetry, we find solutions only with
\beq  
(-1)^{M_1+M_2}=-1 .
\eeq 
For local operators (or constant gluing matrix), this condition on the R-charges is equivalent  to the spin being integer  rather than half-integer (see the general quantization condition for the spins in \cite{Bombardelli:2017vhk}). In this paper we restrict our attention to trajectories interpolating between integer spins, and we leave for future work the study of more general cases. 

\subsubsection{Constraints on the gluing matrix}\label{sec:2}
Let us now see what additional conditions should be satisfied by the gluing matrix in presence of the two symmetries above. We start by rewriting the gluing equation (\ref{eq:basicgluing}), using the $u \leftrightarrow -u$ symmetry (\ref{eq:Qminusu}), as
\beq\label{gluing:umeq}
\tilde{\bf Q}(u) =- \mathcal{G}(u)\cdot\kappa\cdot{\bf Q}^{-1}(-u)\cdot\kappa^T\cdot\mathcal{G}^T (u) ,
\eeq
or, combining the two symmetries with the help of (\ref{eq:LRsymQ}),
\beq\label{gluing:umeq2}
\tilde{\bf Q}(u) = \mathcal{G}(u)\cdot\mathbb{K}\cdot{\bf Q}(-u)\cdot\mathbb{K}^T\cdot\mathcal{G}^T (u) .
\eeq
We already know that the gluing matrix satisfies the Q-system invariance condition (\ref{eq:constrG}) and should be an entire function. As shown in Appendix \ref{proof}, in this symmetric sector it satisfies some extra conditions.  
In summary, the properties of $\mathcal{G}(u)$ are:
\begin{itemize}
\item An extra periodicity condition (due to LR symmetry)
\beq
\mathcal{G}(u+\i)=\mathbb{K}\cdot\mathcal{G}(u)\cdot\mathbb{K}.\label{eq:constrperiodG}
\eeq
\item An extra $u\leftrightarrow -u$ relation (due to parity):
\beq\label{eq:constrK1} 
\mathcal{G}(-u)=-\kappa\cdot\mathcal{G}^T(u)\cdot\kappa^{-1}.
\eeq
\item The usual Q-system invariance  (valid in all sectors)
\begin{equation}
\mathcal{G}(u)  \cdot\kappa  \cdot\mathcal{G}^T(u+\i) = \kappa.\label{eq:repeatQinv}
\end{equation}
\end{itemize}
In particular, the last condition together with (\ref{eq:constrG})  
implies
\begin{equation*}
\mathcal{G}(-u)\cdot\mathcal{G}(u+\i)=-\1 .
\end{equation*}

Further subtle properties come from the decomposition into the quantities $f$ and $\Omega$, see (\ref{eq:defOmega}) and (\ref{eq:deff}). For the quantity $\Omega$, we will change notation to $\Theta(u) \equiv \Omega^+(u)$, namely the definition (\ref{eq:defOmega}) is translated as
\beq\label{period:Theta}
(\mathbb{R}^{-1} )_a^{\, c}  Q_{c|i}(-u) =Q_{a|j}(u)\Theta_i^{j}(u) , 
\eeq
and the second quantity is defined by
 \beq \label{G:def:extra}
\mathcal{G}(u)\equiv f(u)\cdot\Theta^{-1}\(u-\frac{\i}{2}\),
\eeq
which, in our case, can be conveniently rewritten using \eqref{eq:repeatQinv} as
\beq\label{def:K} 
\mathcal{G}(u)\equiv f(u)\cdot\kappa\cdot\Theta^{\,T}\(u+\frac{\i}{2}\)\cdot\kappa^{-1}.
\eeq
The nontrivial new information coming from this decomposition comes in the following forms
\begin{itemize}
\item The fact that the building block $f_i^{\, j}(u)$ should be a function of only 4 independent functions $\tau_i(u)$, as in (\ref{f:form}-\ref{tau:up}).
Moreover, these blocks satisfy the periodicity (\ref{eq:tauperiod}) due to LR symmetry.
In this particular sector, using definitions \eqref{G:def:extra} and \eqref{def:K} together with the elementary property \eqref{eq:repeatQinv},  we also have the following useful relation
\beq \label{eq:extraconstr}
\mathcal{G}(u) \cdot\Theta\(u-\frac{\i}{2}\) +\kappa\cdot\Theta^T\(u+\frac{\i}{2}\)\cdot\mathcal{G}^T(u+\i)\cdot\kappa^{-1} = 2\times\1.
\eeq

\item Two new constraints for the matrix $\Theta(u)$  in this sector (see Appendix \ref{proof}), namely
\beqa
\Theta(u)\cdot\Theta(-u)&=&\1 \times (-1)^{M_1+M_2},\label{theta:prod}\\
\Theta(u+\i)&=&\mathbb{K}\cdot\Theta(u) \cdot\mathbb{K}\,,\label{theta:period}
\eeqa
in addition to $\Theta(u) \cdot \kappa \cdot \Theta^T(u+\i) = \kappa$ which is valid in any sector. 

\item The asymptotics of $\Theta(u)$ at $\pm \infty$ is fixed by the charges.     
\end{itemize}
On the last point, we review in Appendix \ref{proof} how one fixes the asymptotics of this matrix. In particular, inspecting the definition reveals that $\Theta(u)$ has to be asymptotically diagonal at $\pm \infty$, with eigenvalues related to the charges as follows:
\beqa\label{theta:asym}
\lim_{u\rightarrow + \infty}\Theta_i^{\,\,\,j}(u) &=&
\begin{pmatrix}
\e^{ \i\pi\hat{\mathcal{N}}_1}&0&0&0\\
0&\e^{  \i\pi\hat{\mathcal{N}}_2}&0&0\\
0&0&\e^{ \i\pi\hat{\mathcal{N}}_3}&0\\
0&0&0&\e^{  \i\pi\hat{\mathcal{N}}_4}
\end{pmatrix} ,\\\label{theta:asym:2}
\lim_{u\rightarrow-\infty}\Theta_i^{\,\,\,j}(u) &=&(-1)^{M_1+M_2}\times
\begin{pmatrix}
\e^{ -\i\pi\hat{\mathcal{N}}_1}&0&0&0\\
0&\e^{  -\i\pi\hat{\mathcal{N}}_2}&0&0\\
0&0&\e^{ -\i\pi\hat{\mathcal{N}}_3}&0\\
0&0&0&\e^{  -\i\pi\hat{\mathcal{N}}_4}
\end{pmatrix}.
\eeqa
When imposing all constraints with these symmetry assumptions, we will find that only solutions with $(-1)^{M_1+M_2} = -1$ are allowed. 

\subsection{Fixing the gluing matrix}
In this section, we implement the set of constraints outlined in the previous section, which gives an algorithm capable of determining the admissible gluing matrices.

\subsubsection{Strategy of the derivation}\label{strategy}  
Since $\mathcal{G}(u)$ is a periodic matrix of holomorphic functions with no singularities anywhere for finite $u$, it is severely constrained. In particular, if we require that it has non-growing asymptotics at $u \rightarrow \pm \infty$, then it should be a constant by Liouville's theorem: this is the case for QSC solutions corresponding to local operators. To have a nontrivial dependence on $u$, the most general class of functions which do not grow worse than exponentials are polynomials in $e^{\pm \pi u}$.  We will make the following
\beq\label{eq:assumption}
\texttt{Assumption: }\text{ The elements of $\mathcal{G}(u)$ grow at most like $e^{2 \pi |u|}$ at infinity.}
\eeq
Within this hypothesis,  $\mathcal{G}(u)$ has an explicit ansatz in terms of a few parameters. 
Imposing the constraints then fixes some of the parameters and leaves us with the most general admissible form of the gluing matrix.

Before discussing how the classification goes, we notice that through (\ref{eq:assumption}) we are assuming the lowest nontrivial degree of the polynomials in the ansatz. There are currently no examples, either in this theory or in $\mathcal{N}$$=$$4$ SYM, of situations where more than this would be needed to describe some Regge trajectories. It remains, however, not impossible that gluing matrices with more general structure might be needed for more general sectors -- or possibly to describe other types of non-local operators. 

\paragraph{Classification algorithm. }\label{class}

Practically, we found it helpful to implement the constraints in this order:
\begin{itemize}
\item Write down a parametrization of the gluing matrix $\mathcal{G}(u)$ as a polynomial of degree 2 in $e^{\pm \pi u}$, in such a way that it satisfies the linear constraints: (\ref{eq:constrperiodG}) and (\ref{eq:constrK1}). 
\item Impose the compatibility of this ansatz with the properties of $f(u)$ and $\Theta(u)$. 
The most immediate constraints come from the equation (\ref{eq:extraconstr}), which splits into 2 equations once we expand it at $\pm \infty$ and we use the asymptotic values of $\Theta(u)$. 
\item If more constraints are needed, construct 
\begin{equation*}
f(u) \equiv \mathcal{G}(u)\cdot\Theta\(u-\frac{\i}{2}\),
\end{equation*}
and impose that it can be parametrized in terms of 4 $\tau$-functions in the correct way. This equation should be studied asymptotically at both $+\infty$ and $-\infty$
.
\item Once the parameter space is sufficiently tiny, impose the quadratic constraint (\ref{eq:repeatQinv}).
\end{itemize}
These steps will be illustrated below. 

\subsubsection{Review of the result for local operators }\label{local}
Let us start by summarising the case of the gluing matrix for local operators, i.e. if we assume $\mathcal{G}(u)\equiv \mathcal{G}$ to be constant. This was discussed in full generality in \cite{Bombardelli:2018bqz}, and we review how this follows from the method of this paper in Appendix \ref{case:local}. Here we specify to the symmetric sector discussed in this paper. 

Since we assume all $\tau$-functions to have power-like asymptotics, from the condition of periodicity \eqref{tau:period}, it follows that
\begin{align}\la{asy:23}
&\tau_{2,3}\simeq0,\\\la{asy:14}
&\tau_{1,4}\simeq\texttt{const}\,,
\end{align}
as $u\rightarrow\pm\infty$.
 
Using the strategy described above and in particular the algorithm explained in Appendix \ref{case:local}, one gets as a result a matrix depending on the charges of the state (here given in $sl(2)$-like grading) and just one unfixed parameter~\cite{Bombardelli:2017vhk}
\beq\label{K:inter} 
\(\mathcal{G}_\text{local}\)_{i}^{\,\,\,j}=\begin{pmatrix}
 -\i \sec \(\pi\Delta_{sl(2)}\) & 0 & 0 & \i\,\delta_1\\
 0 & \pm\,\i & 0 & 0 \\
 0 & 0 & \mp\,\i & 0 \\
 -\i\,\delta_2 & 0 & 0 & \i \sec \(\pi\Delta_{sl(2)}\)  \\
\end{pmatrix},
\eeq
where $\delta_1\delta_2=\tan{ \(\pi\Delta_{sl(2)}\)}^2$ and 
$S_{sl(2)}$ is explicitly quantized according to 
\beq\label{discr}
\e^{\imath\pi S_{sl(2)}}=\pm1,
\eeq
where this is the same $\pm$ sign appearing in (\ref{K:inter}).  We see that the cases of even/odd-spin operators are explicitly disjoint. Moreover, the following relation is required for consistency,
\beq\label{M1M2} 
(-1)^{M_1+M_2}=-1 ,
\eeq
in accordance with the spin being integer rather than half-integer~\cite{Bombardelli:2017vhk}. 
We note that, while here we obtained (\ref{K:inter}) under special symmetry assumptions, this form for  a constant gluing matrix is actually not relegated to the particular sector we considered but is valid more generally~\cite{Bombardelli:2017vhk}.

\subsubsection{Result for non-local operators}\label{sec:classifyresult}
To avoid spin quantization, we need to relax the asymptotic behaviour of $\tau$-functions \eqref{asy:23}-\eqref{asy:14}, which will open up the possibility for the gluing matrix to depend non-trivially on $u$. Following the strategy described above, we focus on the simplest non-trivial assumption compatible with the enhanced periodicity of the sector \eqref{eq:tauperiod}, i.e., as $u\rightarrow\pm\infty$, we allow $\tau_2$, $\tau_3$ to have a nontrivial asymptotically growing behaviour,\footnote{We have included in (\ref{ansatz})-(\ref{eq:ansatz22}) the leading and the first subleading terms. It is a general result that the corrections should have this form, due to the periodicity of the $\tau$ functions. }
\begin{align}\label{ansatz}
&\tau_{2,3}\simeq A_{2,3}^{\pm} \text{e}^{\pi\,|u|}+B_{2,3}^{\pm}  \text{e}^{-\pi\,|u|},\\
&\tau_{1,4}\simeq C_{1,4}^\pm +D_{2,3}^{\pm}  \text{e}^{-2\pi\,|u|} ,\label{eq:ansatz22}
\end{align}
 where $A,B,C,D\in\mathbb{C}$ are some parameters. In this work we analyze in detail two disjoint configurations of the general ansatz \eqref{ansatz}, e.g.
\begin{align}
&\textit{``Regge'' case:}\quad A_2^\pm\ne0,\, A_3^\pm=0,\label{a}\\
&\textit{``Bridge'' case:}\quad A_2^\pm=0,\, A_3^\pm\ne0.\label{b}
\end{align}
Notice that $A_3^\pm=A_2^\pm=0$ reduces to what discussed in Section \ref{local}.

\paragraph{Consistent gluing matrices. }

Following the strategy discussed in \ref{strategy}, let us examine in detail the steps we used to fix the form of the $\mathcal{G}(u)$. We describe step by step the simplifications in the case of the first assumption, the ``Regge'' case \eqref{a}. 
\begin{itemize}
\item A general solution to the linear constraints (\ref{eq:constrperiodG}) and (\ref{eq:constrK1}) looks like
\begin{align}
&\left(
\begin{array}{cccc}
 g^{(0)}_{11} & 0 & 0 & g^{(0)}_{14} \\
 0 & g^{(0)}_{22} & g^{(0)}_{23} & 0 \\
 0 & g^{(0)}_{32} & -g^{(0)}_{22} & 0 \\
 g^{(0)}_{41} & 0 & 0 & -g^{(0)}_{11} \\
\end{array}
\right)+e^{\pi u}\times \left(
\begin{array}{cccc}
 0 & g_{34} & -g_{24} & 0 \\
 g_{43} & 0 & 0 & -g_{13} \\
 -g_{42} & 0 & 0 & g_{12} \\
 0 & -g_{31} & g_{21} & 0 \\
\end{array}
\right)\notag\\
&\quad+e^{-\pi u}\times \left(
\begin{array}{cccc}
 0 & g_{12} & g_{13} & 0 \\
 g_{21} & 0 & 0 & g_{24} \\
 g_{31} & 0 & 0 & g_{34} \\
 0 & g_{42} & g_{43} & 0 \\
\end{array}
\right)+e^{2\pi u}\times\left(
\begin{array}{cccc}
 -g_{44} & 0 & 0 & g_{14} \\
 0 & -g_{33} & g_{23} & 0 \\
 0 & g_{32} & -g_{22} & 0 \\
 g_{41} & 0 & 0 & -g_{11} \\
\end{array}
\right)\notag\\
&\quad\quad+e^{-2\pi u}\times\left(
\begin{array}{cccc}
 g_{11} & 0 & 0 & g_{14} \\
 0 & g_{22} & g_{23} & 0 \\
 0 & g_{32} & g_{33} & 0 \\
 g_{41} & 0 & 0 & -g_{44} \\
\end{array}
\right).
\end{align}
\item Studying equation (\ref{eq:extraconstr}) at large $u\rightarrow \pm \infty$,
and exploiting the knowledge of the limit values $\Theta(\pm \infty)$ in terms of charges, we get  
\beq
\begin{aligned}
g_{33}=-e^{-2\pi\i \hat{\mathcal{N}}_2}g_{22},\\
g_{44}=-e^{-2\pi\i \hat{\mathcal{N}}_1}g_{11}.
\end{aligned}
\eeq
Up to this point, we have not yet used the specific assumptions differentiating the ``Regge'' and ``Bridge'' case.

\item We use the parametrization in terms of the 4 $\tau$-functions\textcolor{black}{, \eqref{ansatz}, \eqref{eq:ansatz22} and the ``Regge'' case \eqref{a},} to force further relations. 
In particular, we  impose the relations (\ref{G:def:extra}) and (\ref{def:K}) at $|u|\gg1$, which give
\beqa\label{expansion} 
&& f(u) \simeq \mathcal{G}(u)\cdot\Theta\(u-\i/2\), \quad u\rightarrow \pm \infty,\\
&&\lim_{u\rightarrow\pm\infty}\Theta(u+\i/2)\equiv\lim_{u\rightarrow\pm\infty}\left(\kappa ^{-1}\cdot f^{-1}(u) \cdot\mathcal{G}(u) \kappa \right)^T .\label{prob}
\eeqa
These equations allow us to establish relations between the parameters, $A$, $B$, $C$ entering the asymptotic behaviour of $f(u)$ at $\pm \infty$ through the $\tau$ functions,  $g_{ij}$ parametrising the gluing matrix, and the charges $\hat{\mathcal{N}}$.  

In particular, we find
\begin{equation}
g_{11}=g_{14}=g_{22}=g_{32}=g_{41}=g_{12}=g_{31}=g_{34}=g_{42}=0,
\end{equation}
along with functional dependencies of the form $g_{ij} = g_{ij}(A,B,C,\hat{\mathcal{N}})$ for certain pairs $(i,j)$, as well as specific relations among the parameters $A$, $B$, $C$, and $\hat{\mathcal{N}}$.

\item Although the previous constraint \eqref{prob} incorporates the quadratic condition \eqref{eq:repeatQinv}, the latter is only used in the limit $u\rightarrow\pm\infty$. As a final step, we impose the constraint \eqref{eq:repeatQinv} on the exact form of $\mathcal{G}(u)$. The use of condition \eqref{prob} for $u\rightarrow\pm\infty$ before imposing \eqref{eq:repeatQinv} on the exact gluing matrix is merely a convenience in the computation of the constraints. Alternative orderings of the algorithmic steps are possible, but they prove to be less advantageous for implementation.  Finally we get
\begin{equation}
g_{22}^{(0)}=\i,\quad
g_{32}^{(0)}=0
\end{equation}
together with some supplemental relations $g_{ij}=g_{ij}(A,B,C,\hat{\mathcal{N}})$ and between the coefficients $A,B,C$ and $\hat{\mathcal{N}}$. E.g. we find the constraint
\begin{equation}\label{quant:1}
\e^{\i\pi \hat{\mathcal{N}}_2}=\mp\i\( 1+A_2^+ B_3^+\),
\end{equation}
which replaces the condition (\ref{discr}): in the case of non-constant gluing matrix with $A_2^+$, $A_3^+ \neq 0$, we see that the spin is no longer quantized. Furthermore, the following condition still holds,
\beq
(-1)^{M_1+M_2}=-1.
\eeq
\end{itemize}
Ultimately, we can parametrize the gluing matrix using only three parameters, $\beta,\gamma,\sigma\in\mathbb{C}$,\footnote{The $\pm$ sign in this equation is correlated to the trajectory passing through even or odd integer spin local operators, in accordance with (\ref{K:inter}). Notice that, while an overall sign in $\mathcal{G}$ is invisible at the level of the gluing equations, definitions such as (\ref{def:K}) pick a definite choice of sign. }
\begin{align}\label{K:final}
&\pm\(\mathcal{G}_{\text{Regge}}\)_i^{\,\,\,j}=\left(
\begin{array}{cccc}
 -\i \alpha & 0 & 0 & -\frac{\i \left(\alpha^2-1\right)}{\beta} \\
 0 & \i & \sigma & 0 \\
 0 & 0 & -\i & 0 \\
 \i \beta & 0 & 0 & \i \alpha \\
\end{array}
\right)+e^{\pi u}\times
\left(
\begin{array}{cccc}
 0 & 0 & \gamma & 0 \\
 -\frac{\beta \gamma \rho}{\alpha^2-1} & 0 & 0 & \gamma\frac{1- \rho}{\alpha} \\
 0 & 0 & 0 & 0 \\
 0 & 0 & -\frac{\beta \gamma \left(\alpha^2+\rho-1\right)}{\alpha \left(\alpha^2-1\right)} & 0 \\
\end{array}
\right)+e^{-\pi u}\notag\\
&\times\left(
\begin{array}{cccc}
 0 & 0 & \gamma \frac{\rho-1}{\alpha} & 0 \\
 -\frac{\beta \gamma \left(\alpha^2+\rho-1\right)}{\alpha \left(\alpha^2-1\right)} & 0 & 0 & -\gamma \\
 0 & 0 & 0 & 0 \\
 0 & 0 & -\frac{\beta \gamma \rho}{\alpha^2-1} & 0 \\
\end{array}
\right)+\left(e^{2\pi u}+e^{-2\pi u}\right)\times\left(
\begin{array}{cccc}
 0 & 0 & 0 & 0 \\
 0 & 0 & -\frac{\i \beta \gamma^2 (\rho-1)}{\alpha \left(\alpha^2-1\right)} & 0 \\
 0 & 0 & 0 & 0 \\
 0 & 0 & 0 & 0 \\
\end{array}
\right),
\end{align}
where $\rho^2=1 - \alpha^2\equiv1-\text{csc}^2 \(\pi  \hat{\mathcal{N}_1} \)$.

If we instead use the other assumption, the ``Bridge'' case \eqref{b}, the gluing matrix emerging from the algorithm can be formally taken to be the transpose of the above one, with its own set of parameters, $\beta', \gamma', \sigma' \in \mathbb{C}$,
\begin{align}
&\pm\(\mathcal{G}_{\text{Bridge}}\)_i^{\,\,\,j}=\left(
\begin{array}{cccc}
 -\i \alpha' & 0 & 0 & \i \beta' \\
 0 & \i & 0 & 0 \\
 0 & \sigma' & -\i & 0 \\
 -\frac{\i \left(\alpha'^2-1\right)}{\beta'} & 0 & 0 & \i \alpha' \\
\end{array}
\right)+e^{\pi u}\times
\left(
\begin{array}{cccc}
 0 & -\frac{\beta' \gamma' \rho'}{\alpha'^2-1} & 0 & 0 \\
 0 & 0 & 0 & 0 \\
 \gamma' & 0 & 0 & -\frac{\beta' \gamma' \left(\alpha'^2+\rho'-1\right)}{\alpha' \left(\alpha'^2-1\right)} \\
 0 & \gamma'\frac{1- \rho'}{\alpha'} & 0 & 0 \\
\end{array}
\right)+e^{-\pi u}\notag\\
&\times\left(
\begin{array}{cccc}
 0 &  -\frac{\beta' \gamma' \left(\alpha'^2+\rho'-1\right)}{\alpha' \left(\alpha'^2-1\right)} & 0 & 0 \\
0 & 0 & 0 & 0 \\
 \gamma' \frac{\rho'-1}{\alpha'} & 0 & 0 & -\frac{\beta' \gamma' \rho'}{\alpha'^2-1} \\
 0 & -\gamma' & 0 & 0 \\
\end{array}
\right)+\left(e^{2\pi u}+e^{-2\pi u}\right)\times\left(
\begin{array}{cccc}
 0 & 0 & 0 & 0 \\
 0 & 0 & 0 & 0 \\
 0 & -\frac{\i \beta' \gamma'^2 (\rho'-1)}{\alpha' \left(\alpha'^2-1\right)} & 0 & 0 \\
 0 & 0 & 0 & 0 \\
\end{array}
\right),
\end{align}
where $\(\rho'\)^2=1 - \(\alpha'\)^2\equiv1-\text{csc}^2 \(\pi  \hat{\mathcal{N}_1}\)$.
The parameters $\beta,\gamma$ and $\sigma$, which are implicitly $S$ and $\Delta$-dependent, will be part of the numerical output (see following  Section \ref{sec:numerics}). 

\paragraph{Reduction to the case of local operators. }
Notice also that both these matrices reduce explicitly to the case for local operators if we demand that they lose the $u$-dependence, setting to zero the terms proportional to $e^{\pm \pi u}$, $e^{\pm 2 \pi u}$. This can be achieved by setting, in the case of $\mathcal{G}_{\text{Regge}}$, $\gamma=\sigma=0$, and in the case of $\mathcal{G}_{\text{Bridge}}$, $\gamma'=\sigma'=0$. With these choices of parameters they both collapse to the form $\mathcal{G}_{\text{local}}$ \eqref{K:inter}. Thus, the case of local operators sits precisely at the intersection of the cases described by the two more general gluing matrices. 

\paragraph{Interpretation and implications. }\label{par}

We notice that the formal structure of the two gluing matrices are related precisely by  conjugation with the Q-system symmetry $\mathbb{M}_1$ introduced in Section \ref{sec:Weyl}, i.e.,
\beq\label{eq:gluingM1map}
\mathcal{G}_{\text{Bridge}} \cong -\mathbb{M}_1 \cdot \mathcal{G}_{\text{Regge}} \cdot \mathbb{M}_1^{-1} ,
\eeq
upon an appropriate identification of the parameters $\alpha$, $\beta$, $\gamma$, $\delta$ and $\alpha'$, $\beta'$, $\gamma'$, $\delta'$ entering the gluing matrices on the two sides. Above,  $\mathbb{M}_1$ is the matrix defined in  (\ref{eq:MWeyls}), which realizes the Weyl symmetry $\texttt{W}_S$ at the level of the QSC. Notice that the minus sign on the r.h.s. of (\ref{eq:gluingM1map}) has no observable effect, since the gluing matrix enters quadratically into the basic relation (\ref{eq:basicgluing}). 

As was anticipated in the Introduction, the existence of two distinct gluing matrices $\mathcal{G}_{\text{Regge}}$ and $\mathcal{G}_{\text{Bridge}}$ corresponds to two separate directions for performing the analytical continuation in spin starting from physical local operators. Since the two are related by the transformation (\ref{eq:gluingM1map}), any solution obtained with $\mathcal{G}_{\text{Bridge}}$ can be interpreted as a solution obtained with $\mathcal{G}_{\text{Regge}}$ subject to the Weyl transformation $\texttt{W}_S$. This will lead us straight to the existence of the twist/co-twist symmetry. The evidence for this conclusion is discussed in detail in the Section \ref{sec:numerics}. 

On the other hand, the two gluing matrices are self-invariant in form under the transformation with $\mathbb{M}_2$ (with parameters redefined appropriately), i.e., 
\beq\label{eq:gluingM2map}
\mathcal{G}_{\text{Regge}} \cong 
\mathbb{M}_2 \cdot \mathcal{G}_{\text{Regge}} \cdot \mathbb{M}_2^{-1},\quad\mathcal{G}_{\text{Bridge}} \cong 
\mathbb{M}_2 \cdot \mathcal{G}_{\text{Bridge}} \cdot \mathbb{M}_2^{-1}.
\eeq
Since $\mathbb{M}_2$ is a Q-system symmetry associated to the Weyl reflection in $\Delta$, this indicates that a solution with $\Delta$ and one with $\Delta \rightarrow d - \Delta$ are obtained by imposing the same form of gluing equations, just with redefined parameters. 

\subsection{Two gluing matrices also in $\mathcal{N}$$=$$4$ SYM}\label{sec:gluingSYM}

Finally, we would like to show that also in the case of $\mathcal{N}$$=$$4$ SYM the QSC admits two alternative forms of the gluing equations, related by the Weyl symmetries. Indeed it does: let us present a simple illustration considering the $sl(2)$ sector.  

In $\mathcal{N}$$=$$4$ SYM there are gluing equations very much analogous to (\ref{eq:basicgluing}). The Q-functions in this theory have a different index structure as compared to ABJM. The gluing equations for the simplest case of LR and parity-symmetric states read\footnote{Here we are discussing a slightly different (but related) formulation of gluing where we glue Q-functions related by complex conjugation rather than by $u\rightarrow -u$. }
\beq
\tilde{\mathbf{Q} }_i(u) = (\mathcal{G}^{\mathcal{N}=4} ) _i^{j} \cdot\bar{\bQ}_j(u) \quad\text{ (SYM case) }.
\eeq
Above, the basis of Q-functions is defined by their asymptotics, which again encodes the conformal charges:\beq
\bQ_i(u) \simeq \#\,u^{\hat{M}_i - 1} , \quad\text{ (SYM case) },
\eeq
where for this theory we have,
\beq\label{eq:MisSYM}
\hat{M}_i^{\mathcal{N}=4}=\left\{\frac{\Delta -S_1-S_2+2}{2},\frac{\Delta +S_1+S_2}{2},\frac{-\Delta -S_1+S_2+2}{2},\frac{S_1-S_2-\Delta }{2}\right\},
\eeq
in terms of the scaling dimension and spins in the $sl(2)$ grading. In particular, we will specify the discussion to LR symmetric states for which we have a single spin:
\beq 
S_1\big|_{sl(2)}\equiv S_{sl(2)}\,,\quad S_2\big|_{sl(2)}\equiv0 .
\eeq
The form of the gluing matrix  is explicitly known, in particular for Regge trajectories in the $sl(2)$ sector of the theory it takes the form (see, \cite{Gromov:2017blm})
\beq\label{eq:GSYM}
\(\mathcal{G}^{\mathcal{N}=4}_\text{Regge}\)_i^{\,\,\,j}=\left(\begin{array}{cccc}0 & 0 & \chi & 0 \\ \xi(u) & 0 & \nu & -\bar{\chi} \\ 1/\bar{\chi} & 0 & 0 & 0 \\ \nu/(\chi \bar{\chi}) & -1/\chi & \bar{\xi}(u) & 0\end{array}\right),
\eeq
where $\xi=\xi_1+\xi_2 \cosh (2 \pi u)+\xi_3 \sinh (2 \pi u)$, with parameters $\chi,\xi_1,\xi_2\in{\mathbb{C}}$ and $\nu\in\mathbb{R}$. This form was used to compute the Regge trajectories  with the same quantum numbers as the Pomeron in \cite{Gromov:2015wca,Gromov:2017blm}. Notice that, just like in the ABJM case, when this matrix collapses to the constant case, i.e. when $\xi_2 = \xi_3 = 0$, we find the form of the gluing conditions relevant for local operators (in this case one also has automatically $\gamma = \nu = 0)$. 

What was previously unnoticed is that there is a second possible consistent form of the gluing matrix, formally related to (\ref{eq:GSYM}) by transposition:
\beq\label{eq:GBridgeSYM}
\mathcal{G}^{\mathcal{N}=4}_{\text{Bridge} }\cong \(\mathcal{G}^{\mathcal{N}=4}_\text{Regge}\)^T ,
\eeq
but with possibly independent parameters. This second gluing matrix satisfies all constraints that a healthy gluing matrix in the theory should obey (see \cite{Alfimov:2018cms,Gromov:2017blm}), just like in the ABJM case. Indeed, also in this case we can flip from one form to the other by acting with a constant Q-system transformation, namely the one where the indices of Q-functions are multiplied by
\beq
\mathbb{M}_i^{\,\,\,j} =\left(\begin{array}{cccc}0 & 1 & 0 & 0 \\ 1 & 0 & 0 & 0 \\ 0 & 0 & 0 & -1 \\ 0 & 0 & -1 & 0\end{array}\right).
\eeq
As evident from this form, this matrix swaps the components of the Q-functions $1 \leftrightarrow 2$ and $3 \leftrightarrow 4$, and from (\ref{eq:MisSYM}) correspondingly the charges are redefined (for $S_2 = 0$ as appropriate for LR symmetric states) as 
\beq
S_1\,\rightarrow\,2-S_1 ,\,\quad \Delta \rightarrow \Delta ,
\eeq
where $S_1 = S_{sl(2)}$ is the spin in $sl(2)$ grading. This agrees with the Weyl reflection flipping the spin of the top component (\ref{eq:spinflip}) when we consider that $S_{sl(2)} = S_{\text{top}} + 2$. 
This is in complete analogy with the ABJM case, and we can verify that this transformation exchanges twist and co-twist defined as in the Introduction. 

Beyond the LR symmetric sector, in $\mathcal{N}$$=$$4$ SYM  there exists a second Weyl reflection corresponding to $S_2\rightarrow -S_2$. This too should correspond to a further transformation for the gluing matrix. Alternative gluing matrices related to $S_2$ and $-S_2$ were indeed found by the authors of \cite{Alfimov:2018cms}, where they were used to move in a neighbourhood of the point $S_2 = 0$ in two directions. The analogue of what we are discussing in this paper would be to use the alternative gluing matrix around a generic physical operator with $S_2 \neq 0$, to generate a new $S_2$-flipped trajectory intersecting the first one, thus highlighting a new symmetry.\footnote{We thank Nikolay Gromov for communications.}

\section{Results}\label{sec:numerics}
In this section we present our numerical studies. After a brief review of the numerical method which was first developed in  \cite{Gromov:2015wca} we present studies of the minimum of the leading trajectories interpolating odd and even spins. In Section \ref{sym:sec} we present numerical evidence on the twist/co-twist symmetry. We describe in detail how it operates on trajectories and how it is related to the two gluing matrices found in the previous section. 

\subsection{Setup}\label{setup} We solved numerically the QSC equations for several Regge trajectories focusing on an interesting sector of the theory corresponding to the operators exchanged in the OPE decomposition of the correlator of four stress tensors. This is related by supersymmetry to the correlator of four half-BPS scalar operators in the $[0,1,1]_{SU(4)}$ representation of R-symmetry.\footnote{This is the $\mathbf{15}$ irrep of $SU(4)$. } As shown by the analysis of \cite{Binder:2020ckj}, the non-protected multiplets exchanged in this OPE  are those with top components neutral under R-symmetry, i.e.,
\beq
\left[ p_1, q, p_2 \right]_{\text{top}} = \left[ 0,0,0\right]. 
\eeq
At the level of the QSC, as discussed in Section \ref{sec:review} we should then look for solutions where the powers of $\bP$ functions are specified by
\beq\label{sector}
M_5 = 0,\,M_1 = 1,\,M_2 = 2,
\eeq
while the charges $\hat{M}_1$ and $\hat{M}_2$ are related to the conformal charges as
\beq
\Delta_{\text{top}} = \frac{\hat{M}_1 + \hat{M}_2 - 3}{2}, \quad  S_{\text{top}} = \frac{\hat{M}_2  - \hat{M}_1 - 1}{2} .
\eeq
From this, the conformal charges (and R-symmetry) for any other representative of the multiplet can be reconstructed by acting with the appropriate supercharges. We will often report results in the familiar  $sl(2)$-like grading, which is related to the charges of the top component by the shifts (\ref{eq:sl2shifts}). 

We will also restrict our attention to states with LR parity symmetry as well as $u \leftrightarrow -u$ symmetry. 

\subsection{Recap on the numerical method}\label{sec:methodrecap}
To solve the equations numerically we use the method introduced in \cite{Gromov:2015wca} which was already used for a numerical study of local operators in ABJM theory (including operators without any discrete symmetry, more general than the ones studied here) in \cite{Bombardelli:2018bqz}. A recent fully automated and optimized version of the method was developed for $\mathcal{N}$$=$$4$ SYM~\cite{Gromov:2023hzc}, allowing in that case to compute automatically the dimensions of hundreds of operators.\footnote{While here we focus on the method for finite coupling, progress was also made recently on solving the equations at strong coupling~\cite{Ekhammar:2024rfj}. }
We give a schematic review of how this works (safely skipped by the experts), and highlight some subtleties needed to compute the Pomeron eigenvalue. 

Our numerical algorithm is implemented in \texttt{Mathematica} and is an update of the code already published with \cite{Bombardelli:2018bqz}.

\paragraph{Standard method for non-coincident charges. }

Let us describe the method which works in all situations where the asymptotics of Q-functions are distinct, i.e. in particular $\hat{\mathcal{N}}_i \neq \hat{\mathcal{N}}_j $ for $i\neq j$.

The idea of the method follows the logical flow of equations (\ref{eq:shiftQ}), (\ref{eq:defQij}): the $\bP$ functions are used as in input, parametrised in terms of some numerical constants-- one then constructs $\bQ$ functions using QQ-relations and imposes that they should satisfy the correct gluing equation. Imposing the gluing equations gives a nonlinear condition which fixes the original parameters on a discrete set of solutions of the QSC. 

To parametrise the $\bP$ functions, we use the fact that they should have a single cut on the first Riemann sheet, with prescribed asymptotics. This implies that they should have the form:
\beq\label{eq:Pseries}
\bP_{ab}(u) =  x^{\mathcal{N}_a + \mathcal{N}_b} \left( 
 \mathcal{A}_{ab} + \sum_{n\ge1} \frac{c_{ab, n} }{x^n} \right) , 
\eeq
where $\left\{\mathcal{A}_{ab}, c_{ab, n} \right\}$ are parameters and $x \equiv x(u)$ is the Zhukovsky map \eqref{zhukovsky}, which we repeat here for clarity:
\beq
x(u) = \frac{u + \sqrt{u - 2 h}\sqrt{u + 2 h}}{2 h}.
\eeq
This map is a two-sheeted function with a single cut and asymptotics $x \simeq u/h$ on the first sheet. The series (\ref{eq:Pseries}) converges for $u$ on the first Riemann sheet and in a finite region of the second sheet. This depends on the fact that branch points of $\bP$ functions are quadratic and thus are resolved by the Zhukovsky map.\footnote{The lack of this property is what it makes it more difficult to solve the QSC for the AdS$_3$ case: however see \cite{Cavaglia:2022xld} for a way around this problem which works in some regimes. } 

The parameters $\left\{\mathcal{A}_{ab}, c_{ab, n} \right\}$ satisfy some constraints, among them the ones needed to implement the quadratic relation (\ref{eq:quadraticP}). In addition one needs to fix some residual freedom in the redefinition of the basis of $\bP$ functions, see e.g. \cite{Bombardelli:2017vhk,Bombardelli:2018bqz}. We omit the details here. The main point is that, fixing this residual freedom, one can pick these constants to be in one-to-one correspondence to solutions. The conformal charges $\Delta$ and $S$, in particular can be read off from the leading order coefficients $\mathcal{A}_{ab}$, since the latter satisfy the relation (\ref{eq:AArels}). 

In the numerical method we truncate the series expansion (\ref{eq:Pseries}), thus we use as parameters
\beq
\texttt{params} = \left\{\Delta, S, p_1, q, p_2 \right\} \cup \left\{ c_{ab, n} \right\}_{n=1}^{N_{\text{trunc}}}\cup \left\{ 
 \texttt{gluing constants} \right\},
\eeq
where in anticipation of their role below we added to the list of parameters the \texttt{gluing constants}, i.e. the constant parameters $\alpha$, $\beta$, etc appearing in the gluing matrix, see (\ref{K:final}). 

There is then a well-developed numerical  procedure (see \cite{Gromov:2015wca,Bombardelli:2018bqz} for details) to construct the basis of functions $\bQ^{\downarrow}_{ij}$ and $\bQ^{\uparrow}_{ij}$, defined as solutions of the QQ-relations (\ref{eq:shiftQ}) and (\ref{eq:defQij}). The numerical method proceeds by first computing the coefficients of the asymptotic expansion (\ref{eq:pure1}) or (\ref{eq:pure2}), which will depend on the input $\texttt{parameters}$; these asymptotic expansions then allow to evaluate the functions very far from the real axis in their respective half plane of holomorphicity; using the QQ-relations, one can move towards the real axis in shifts of $-\i$ or $\i$, and finally get to reconstructing the functions on the real axis. In particular, thanks to the Zhukovsky parametrisation we have easy access to these functions both above and below the branch cut $u \in [-2h, 2h]$, see \cite{Gromov:2015wca,Bombardelli:2018bqz} for details.

In summary, one can build a numerical routine that takes the $\texttt{parameters}$ as input and evaluates on the real axis the Q-functions we are interested to glue:
\beq
\bQ^{\downarrow}_{ij}(u\, ; \, \texttt{params}), \quad \bQ^{\uparrow}_{ij}(u\, ; \, \texttt{params}), \quad \tilde{\bQ}^{\downarrow}_{ij}(u\, ; \, \texttt{params} ), \quad \tilde{\bQ}^{\uparrow}_{ij}(u\, ; \, \texttt{params} ).
\eeq

One then builds a cost function expressing the validity of the gluing conditions, e.g. the simplest possibility is to discretise the branch cut by sampling points $\left\{u_i \right\}_{i=1}^{N_{\text{points}}}$ and evaluate
\beq
\texttt{COST}
\left[ \texttt{params} \right] \equiv \sum_{k = 1}^{N_{\text{points}}}  \Big| \tilde{\bQ}^\downarrow(u_k ; \texttt{params}) - \left( \mathcal{G} \cdot \bQ^{\uparrow} \cdot \mathcal{G}^T \right)(u_k; \texttt{params} )\Big|^2 . 
\eeq
A routine evaluating this function is the core of the numerical method. 
One then has to use a root-finding method to find solutions for the parameters making the cost function as close to zero as possible compatibly with the truncations used. A Newton-type algorithm is typically used and works very well, thanks to its quadratic rate of convergence,  when one has a seed value of the parameters close enough to a solution.\footnote{Thus, when we build a continuous plot by variying the coupling or moving along a Regge trajectory, one has to vary the parameter along the curve very slowly, in order to be able to remain inside the basin of convergence of Newton's method. }

In the recent work \cite{Gromov:2023hzc}, a different form of the $\texttt{COST}$ function was advocated, and shown to make the algorithm more stable. The idea is not to evaluate the gluing equations at some sampling points but instead to reconstruct $\bP$ using the glued values for $\bQ$. I.e., we can start from the following equation, which is a consequence of (\ref{eq:shiftQ})-(\ref{eq:defQij}):
\beq
\bP_{ab} = -Q_{a|k}^+ \kappa^{ki} \bQ_{ij} \kappa^{jm} Q_{b|m}^+ ,
\eeq
and using the gluing equations to replace $\bQ_{ij}$ we get
\beq
\left. \bP_{ab}(u; \texttt{params})\right|_{\text{updated}} \equiv  -Q_{a|k}^{\downarrow}\left(u +\frac{i}{2}\right)\, \kappa^{ki} (\mathcal{G} \cdot \tilde{\bQ}^{\uparrow}(u) \cdot \mathcal{G}^T )_{ij}  \,\kappa^{jm} \,Q_{b|m}^{\downarrow}\left(u +\frac{i}{2} \right)
\eeq
One then can compare this reconstructed version of $\bP$ with the original $\bP$ function, coefficient by coefficient in the $1/x$ expansion. This implementation of the method shows more stability and we used it also in our case in the more delicate regions, e.g. close to the minimum of the trajectories where convergence is more fragile. 

Notice that the algorithm uses the form of the gluing matrix -- thus, we need to decide in advance if we want to use the form $\mathcal{G}_{\text{Regge}}$ or $\mathcal{G}_{\text{Bridge}}$ introduced before. After convergence, the algorithm will also fix the constants in these matrices, giving us the parameters $\alpha$, $\beta$, $\gamma$, $\sigma$ in (\ref{K:final}) as an output.  This means that we will be able to track whether this matrix remains a nontrivial function of $u$ or whether it collapses to a form $\mathcal{G}_{\text{local}}$ independent of $u$,  signalling that we have encountered a point corresponding to a local operator or its shadows. 

For our numerical study, we typically used a truncation parameter  $N_\text{trunc} \simeq 20$, which allows us to reach a precision of  e.g. $\simeq 20-30$ digits for the leading trajectory at coupling $h = 0.27$ (some precision is lost around the pomeron point). {The primary source of error in the solution obtained through this approach arises from the systematic bias introduced by truncations. We estimate the error by discarding the digits affected by increasing the truncation parameters.
See Appendix E for a representative sample of our results.}

\subsubsection{How to ride the Pomeron}\label{sec:riding}
It is particularly interesting, for their role in Regge theory, to study the intercepts where Regge trajectories meet the point $\Delta_{\text{bottom}} = d/2$. The leading trajectory in particular is symmetric under $\Delta_{\text{bottom}}\rightarrow d - \Delta_{\text{bottom}}$. Its intercept at the symmetric point is called the Pomeron and is the parameter controlling the Regge limit of the correlator. 

If we want to directly set $\Delta_{\text{bottom}} = d/2$ and compute as an output $S$, we encounter some challenges with the method explained above since the asymptotics of the QSC become degenerate at this point:  in fact this is the fixed point of the Weyl reflection (\ref{eq:WeylDelta}) and thus there we have
\beq\label{eq:pomeronasy}
\hat{M}_1 = - \hat{M}_2,\quad \text{  i.e.  } \quad\hat{\mathcal{N}}_i = \frac{1}{2} \left( 0 , 2 \hat{M}_1, -2 \hat{M}_1 , 0 \right) ,
\eeq
where we note that the remaining parameter, $\hat{M}_1$, is related to the intercept as
\beq
\hat{M}_1 + \frac{5}{2} = \mathcal{S}_0 \equiv \left. S_{\text{bottom}} \right|_{\Delta_{\text{bottom}} = \frac{3}{2}}.
\eeq
Given the degeneracy of the asymptotics (\ref{eq:pomeronasy}), where $\hat{\mathcal{N}}_1 = \hat{\mathcal{N}}_4$, the definition of 
 the Q-functions $Q_{a|i}$ becomes ambiguous.  
 To proceed\footnote{In previous studies, this problem was circumvented by taking a tiny imaginary part for $\Delta$, which lifts the degeneracy. This is also a viable procedure, provided some care is taken to estimate the error introduced by this imaginary shift.}, we need to redefine the Q-functions. 

It is convenient to think of what happens if we take a small deviation from the intercept, i.e. $\Delta = d/2 + \epsilon$ with small $\epsilon$. Because of the degenerate asymptotics, the two columns $i=1$ and $i=4$ of $Q_{a|i}$ become linearly dependent and simultaneously explode (since the determinant is normalised) as $\epsilon \rightarrow 0$. However, we can perform a change of basis by considering the sum and difference of the two columns, normalising them appropriately by factors of $\epsilon$ such that the resulting quantities are finite as $\epsilon \rightarrow 0$. This procedure leads to a new column (coming from the sum) where the asymptotics is powerlike, and another one (coming from the difference\footnote{Before taking the difference, we exploit the freedom to rescale the columns to choose  $B_{a|1} = B_{a|4}$.}) where the asymptotics contains also logarithmic terms coming from the expansion of the asymptotic powers $u^{\mathcal{N}_a + \epsilon }$. It is important to remark that these logarithmic terms only affect asymptotic expansions within specific sectors at infinity, but they do not change anything in the nature of the branch points of the functions. The only branch points remain at $\pm 2 h + \i \mathbb{Z}$ and they are all quadratic.

Thus, to build a working method at $\Delta = d/2$ we should simply modify the asymptotics requirements. Let us make a conventional choice for labelling the columns in the new basis. Then, we will make the usual demand of power-like asymptotics for the first three columns
\beq
Q_{a|i}(u) \simeq B_{a|i} u^{\mathcal{N}_a + \hat{\mathcal{N}}_i }  , \quad i = 1,2,3 ,
\eeq
and these functions will have a regular asymptotic expansion of the form (\ref{eq:pure1}) or (\ref{eq:pure2}) in the appropriate half of the complex plane,\footnote{ depending on whether we are studying $Q_{a|i}^{\downarrow}$ or $Q_{a|i}^{\uparrow}$.} while for the fourth column  instead we require an asymptotic expansion of the form 
\beq
Q_{a|4}(u) \simeq u^0  \sum_{j\ge0} \frac{F_{a, j} }{u^j} + \log(u) \sum_{k\ge0} \frac{G_{a, k} }{u^k} ,\quad u\rightarrow +\infty ,
\eeq
again valid in the appropriate half of the complex plane. In order to remove residual ambiguities in mixing the first and fourth columns, we demand that $F_{1,0}= 0$. 

Except for this modification, the method proceeds just as before. In particular, the coefficients in the modified asymptotic expansion are again fixed (once we fix the leading ones compatibly with their consistency conditions) in one-to-one correspondence with the coefficients in the expansion of $\bP$ functions, which is not modified. One then builds the Q-functions $\bQ_{ij}$, which are also affected by the change of basis, in particular $\bQ_{12}$, $\bQ_{13}$ and $\bQ_{14}$ will now have log-type asymptotics. 

One then proceeds imposing the gluing equations as before. The structure of the gluing matrix is formally very similar to the one of generic points on a Regge trajectory, i.e. it has a form similar to  (\ref{K:final}) (with zeros in the same places). 
However, we can simply exploit the fact that one of the gluing equations is very simple, i.e., at the intercept we found that, in the conventions of this section
\beq
\tilde{\bQ}_{13}^{\downarrow}(u) +\hat{\alpha}\, \bQ_{13}^{\uparrow}(u) +\frac{1-\hat{\alpha}^2}{\hat{\beta}}\,\bQ_{34}^{\uparrow}(u)=0 ,
\eeq
and this subset of the gluing equations is sufficient for the convergence of the algorithm.

\subsection{Leading trajectory and Pomeron eigenvalue}

\paragraph{Leading trajectory. }
As a first application of the method, we traced parts of the \emph{leading} Regge trajectory exchanged in the setup considered, i.e. the trajectory with highest spin for given $\Delta$. 

\begin{figure}[t!]
\centering
\includegraphics[width=9cm,height=9cm]{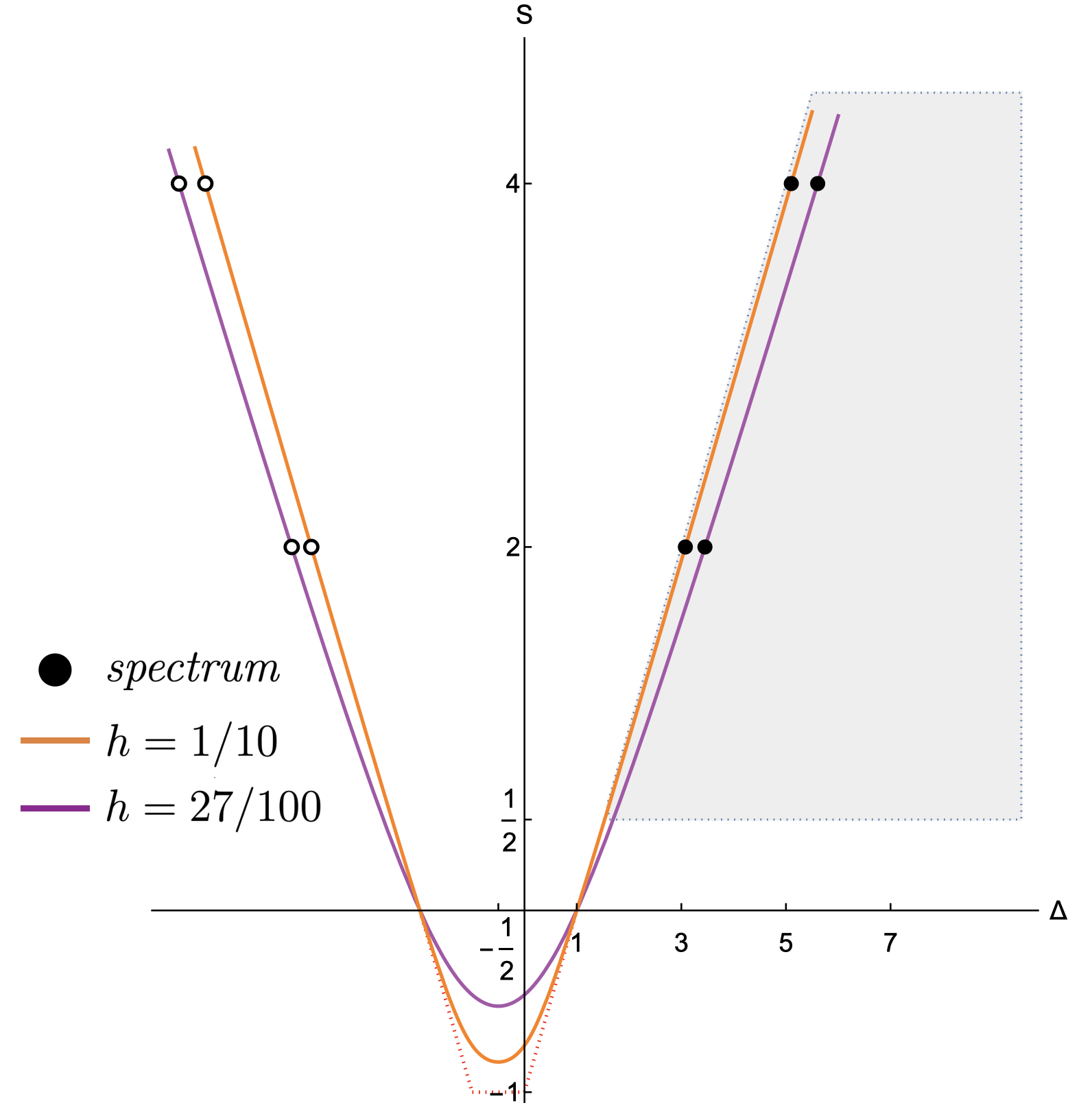}
\caption{Leading Regge trajectory (or rather its image in the $sl(2)$-like grading) at two values of the coupling constant: $h = 0.1$ and $0.27$. The unitarity-allowed region is depicted in gray. Physical states belonging to the spectrum are identified by black bullets, their shadow by circles. The odd-$S$ points crossed by the trajectory correspond to non-local light-ray operators. 
The dotted red piecewise line denotes the BFKL-type singular limit reached at zero coupling.
}
\label{leading:Regge}
\end{figure}

\begin{figure}[t!]
\centering
\includegraphics[width=8cm,height=9cm]{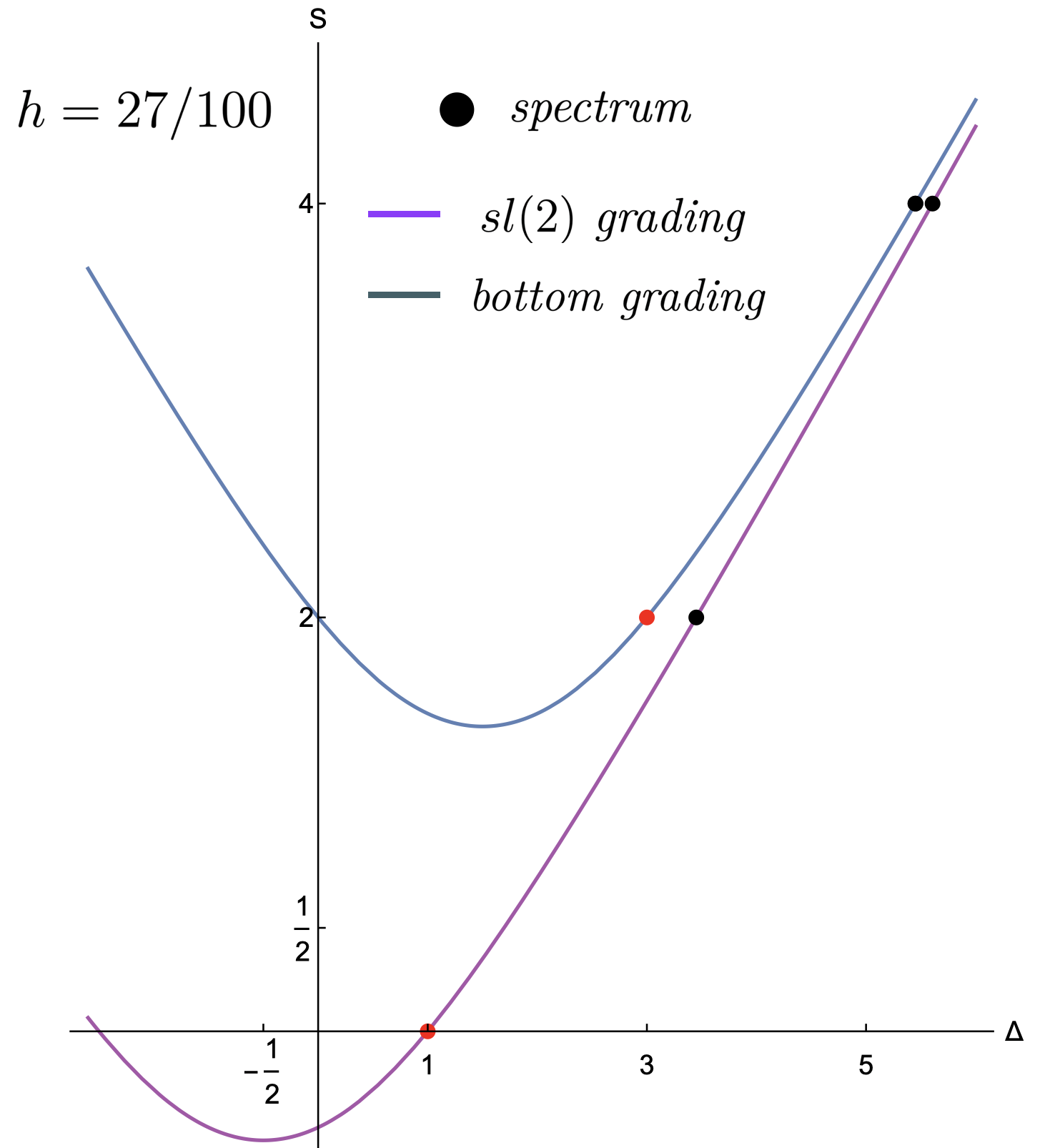}
\caption{The figure illustrates some aspects of the supermultiplet structure, by which the same conformal Regge trajectory is replicated a finite number of times,  each shifted by half-integer steps in the Chew-Frautschi plot generated by the supercharges. In this case, we show together the trajectory interpolating the operators at the \emph{bottom} of the multiplets, together with the trajectory interpolating the operators in $sl(2)$-like grading, which is shifted by an amount $(-2,-2)$. Notice that the \emph{bottom} trajectory is the leading one and, as predicted by conformal Regge theory, it indeed is symmetric under $\Delta \rightarrow 3 - \Delta$. The shifted trajectories do not have individually this symmetry. Their images under $\Delta \rightarrow 3 - \Delta$ are also part of the full Chew-Frautschi plot. In the following, we will be content with depicting one single representative of this multiplet structure. Usually we will plot the $sl(2)$-like grading representative, unless stated otherwise. The red point marks the position of the stress energy tensor (on the bottom trajectory), or correspondingly a BPS scalar operator in the same supermultiplet (on the $sl(2)$ trajectory), i.e. (\ref{eq:opsl2}) with $S=0$.}
\label{shift}
\end{figure}

The results, for two values of the coupling constant, is depicted in Figure \ref{leading:Regge}, where we plot $S$ vs $\Delta$ for a particular level in the supermultiplet, corresponding to the $sl(2)$-like grading. The corresponding physical operators have a simple and well known description, corresponding to 
\beq\label{eq:opsl2}
\mathcal{O}_S = \text{tr}\left( Y^1 D_+^{S_{sl(2)}}Y_{4}^{ \dagger} \right),
\eeq
for $S_{sl(2)}$ an even integer. Notice that the actual leading trajectory is a translation of these data by a constant vector $(+2, +2)$ in the Chew-Frautschi plot, and it corresponds to the interpolation of the spectrum operators at the bottom of the corresponding supermultiplets, see Figure \ref{shift}.

Weak coupling data for the physical operators on this trajectory were obtained in \cite{Anselmetti:2015mda}  up to 12 loops and provide convenient starting points for our algorithm. The continuous trajectory is obtained by solving the QSC equations using the form of the gluing matrix $\mathcal{G}_{\text{Regge}}$. 
 Along the solution, we can track the form of this gluing matrix and we find precisely what anticipated in Section \ref{local:vs}, i.e. the gluing matrix reduces to a constant only on the points of even $S_{sl(2)}$, which correspond to local operators, and remains a nontrivial function of $u$ for the points with odd $S_{sl(2)}$ which are nonlocal light-ray operators. 
The separate trajectory interpolating the operators (\ref{eq:opsl2}) with odd spin will be studied later. 

Notice that there is an interesting BPS point on this trajectory, for $S_{sl(2)} = 0$\footnote{This actually corresponds to $S_{\text{bottom}} = 2$, $\Delta_{\text{bottom}} = 3$ on the translation of this curve which constitutes the actual leading trajectory. There, the protected operator is simply the stress-energy tensor, as expected in general conformal Regge theory.}, where we have exactly $\Delta_{sl(2)} = 1$ (see, red dot in Figure \ref{shift}). The slope of the trajectory at this point is known analytically for any coupling~\cite{Gromov:2014caa}. Making a fit of our data to estimate the slope function, we find perfect agreement with this analytic formula, which is a further consistency test of our method. 

As in $\mathcal{N}$$=$$4$ SYM, it should be possible to compute analytically also the next order in an expansion of the leading trajectory around the BPS point~\cite{Gromov:2014bva}. This result is not yet available in ABJM at finite coupling, but it was recently obtained up to 6 loops~
\cite{Lee:2019oml, Velizhanin:2022faj}:
\beq
\begin{aligned}
\frac{d^2 \gamma(S)}{dS^2}\Big|_{\text{BPS point}}&=2\times4\times \(-\frac{7 h^2 \zeta (3)}{4}+h^4 \left(\frac{7 \pi ^2 \zeta (3)}{12}+\frac{155 \zeta (5)}{8}-\frac{1}{4} \pi ^4 \log (2)\right)\right.\\
&\quad\left.+h^6 \left(\frac{277 \pi ^4 \zeta (3)}{160}-\frac{341 \pi ^2 \zeta (5)}{24}-\frac{13335 \zeta (7)}{64}+\frac{1}{48} \pi ^6 \log (2)\)+\mathcal{O}(h^8)\right),
\end{aligned}
\eeq
where $\gamma(S)$ is the anomalous dimension on the leading trajectory. Our numerics shows very good agreement with this prediction as well. 

\begin{figure}[t!]
\centering
\includegraphics[width=10.2cm,height=9cm]{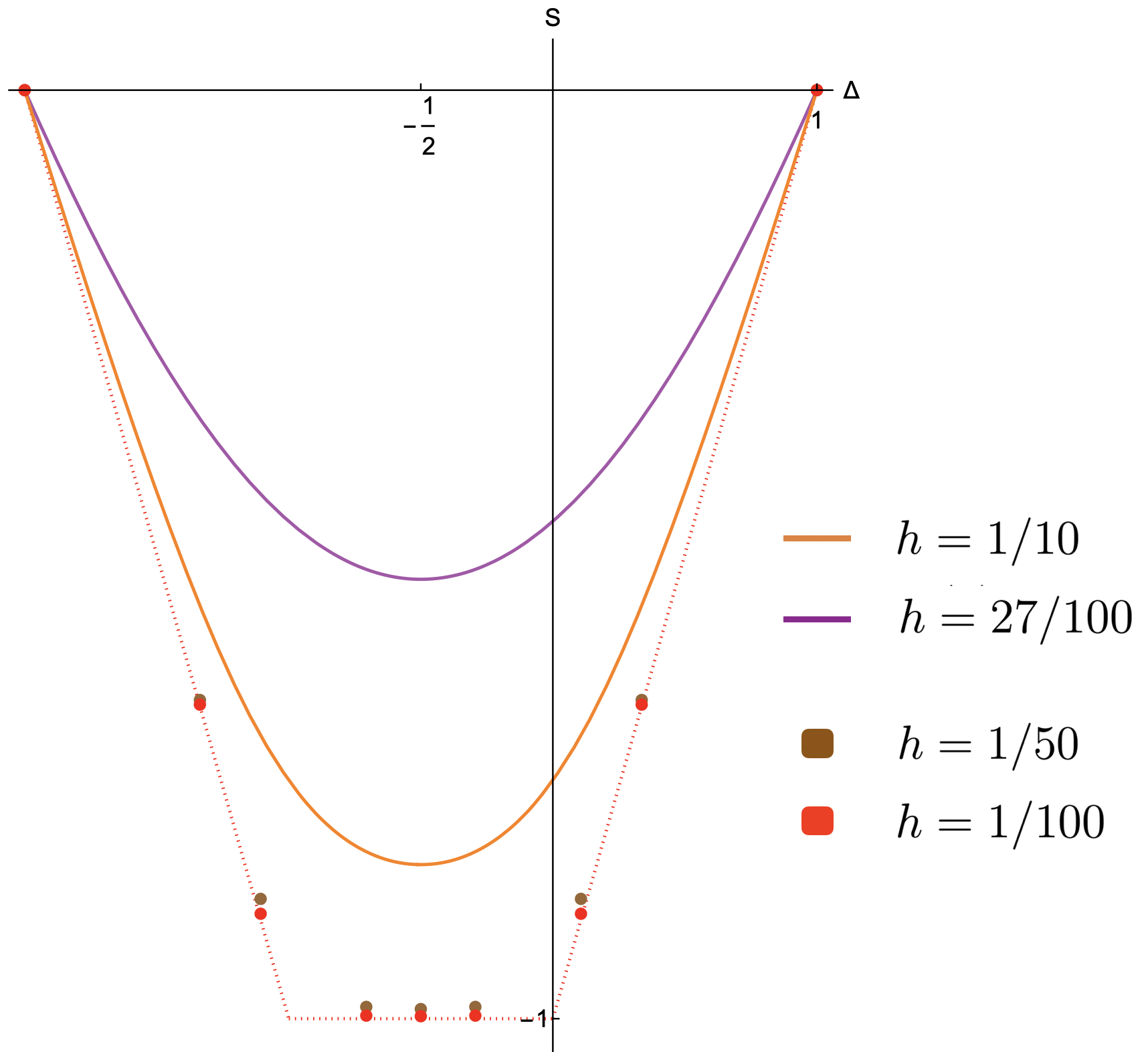}
\caption{The behaviour of the leading trajectory (or rather its image in the $sl(2)$-like grading) at small coupling shows a singular limit similar to the BFKL limit in $\mathcal{N}$$=$$4$ SYM, which shows a broken curve mixing between the linear and the horizontal trajectory. Some sparse $\Delta$-values are driven to coupling $h=0.01$ in order to highlight the BFKL-like flattening of the curve. 
}
\label{plateau}
\end{figure}

\paragraph{BFKL shape at the minimum.}

The shape of the trajectory around the minimum shows a typical BFKL-type behaviour at weak coupling: i.e. going towards zero coupling it tends to a piecewise linear trajectory which mixes the straight-line trajectory interpolating between the tree-level values of local operators and a horizontal trajectory $S_{sl(2)} = -1$, which controls the behaviour for $-1 < \Delta_{sl(2)}<0$. 
 
 As we slightly increase the coupling from zero, the behaviour of the curve in this range of $\Delta$ should take the form $S = -1 + h^2 \chi(\Delta) + O(h^4)$, where $\chi(\Delta)$ is called BFKL eigenvalue. In $\mathcal{N}$$=$$4$ SYM this value is known and equal to a quantity in planar QCD \cite{Alfimov:2020obh}. 

The existence of this type of behaviour was predicted studying the analytic interpolation in spin of perturbative data in \cite{Velizhanin:2022faj}, but, contrary to the case of $\mathcal{N}$$=$$4$ SYM, there is no direct diagrammatic derivation of this behaviour in this theory. However, it is known from the analysis of \cite{Velizhanin:2022faj} that the leading eigenvalue is not simply related to the one of $\mathcal{N}$$=$$4$ SYM.
It would be interesting to study analytically the BFKL-type regime, where we take a double scaling limit of small $\omega \equiv S_{sl(2)} +1 $ and small $h$ with $h^2/\omega$ finite. The shape of the curve in this limit should be describable analytically with the QSC, in particular it would be interesting to generalize to ABJM the recent method of \cite{Ekhammar:2024neh} to obtain several terms in the expansion. We leave this for future work.

\begin{figure}[t!]
\centering
\includegraphics[width=8cm,height=8cm]{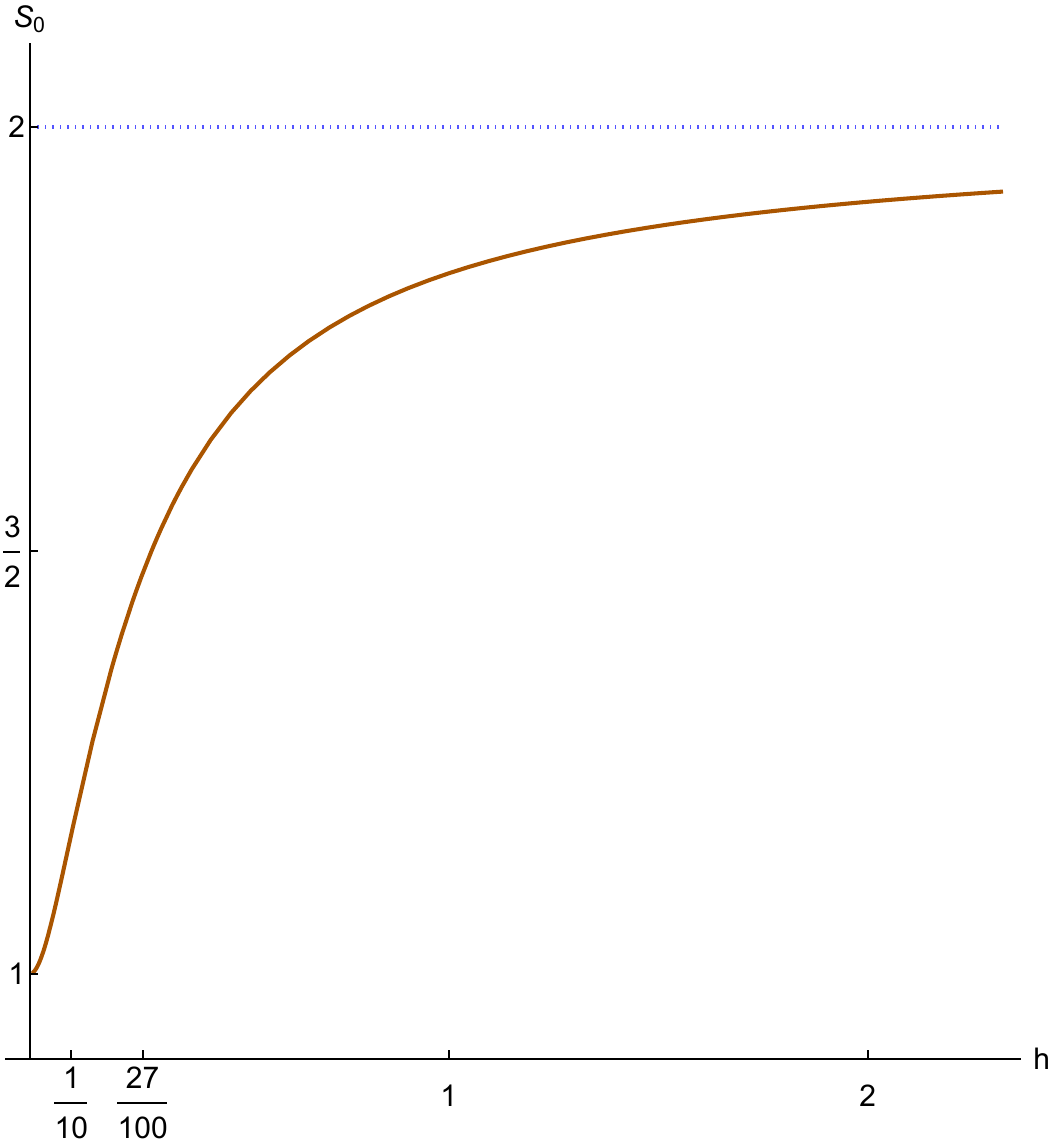}
\caption{Value of the minimum of the leading trajectory (here reported in ``bottom'' grading).}
\label{pomeron}
\end{figure}

\paragraph{Pomeron as function of the coupling. }
Using the method discussed in Section \ref{sec:riding}, we computed the values of the pomeron, i.e. the lowest point where the Regge trajectory intersects its symmetry axis, as functions of the coupling, which is reported in Figure \ref{pomeron}. 

This value has exactly the expected behaviour: $\mathcal{S}_0$ for the actual leading trajectory interpolates between $1$ (which is the lowest limit set by the horizontal trajectory with which it collides at $h\rightarrow 0^+$, and $2$, which is the universal chaos bound for a planar theory~\cite{Maldacena:2015waa, Mezei:2019dfv}.

 We also estimated some coefficients in the expansion of the Pomeron eigenvalue at weak coupling, 
 \beq 
\mathcal{S}_0 = 1+\sum_{n>0}\mathcal{S}_0^{(2n)}h^{2n} ,
\eeq
where fitting our numerical data we estimate 
$\mathcal{S}_0^{(2)}\simeq26.6609(6)$ and $\mathcal{S}_0^{(4)}\simeq-2.11(5)\times10^3$.

\subsection{Trajectory interpolating odd spins}

It is well-known that the analytic continuations interpolating even and odd values of the spin are distinct. We studied the trajectory interpolating the dimensions of the operators (\ref{eq:opsl2}) but with $S_{sl(2)} \in 2 \mathbb{N} + 1$.

\begin{figure}[t!]
\centering
\includegraphics[width=8cm,height=9cm]{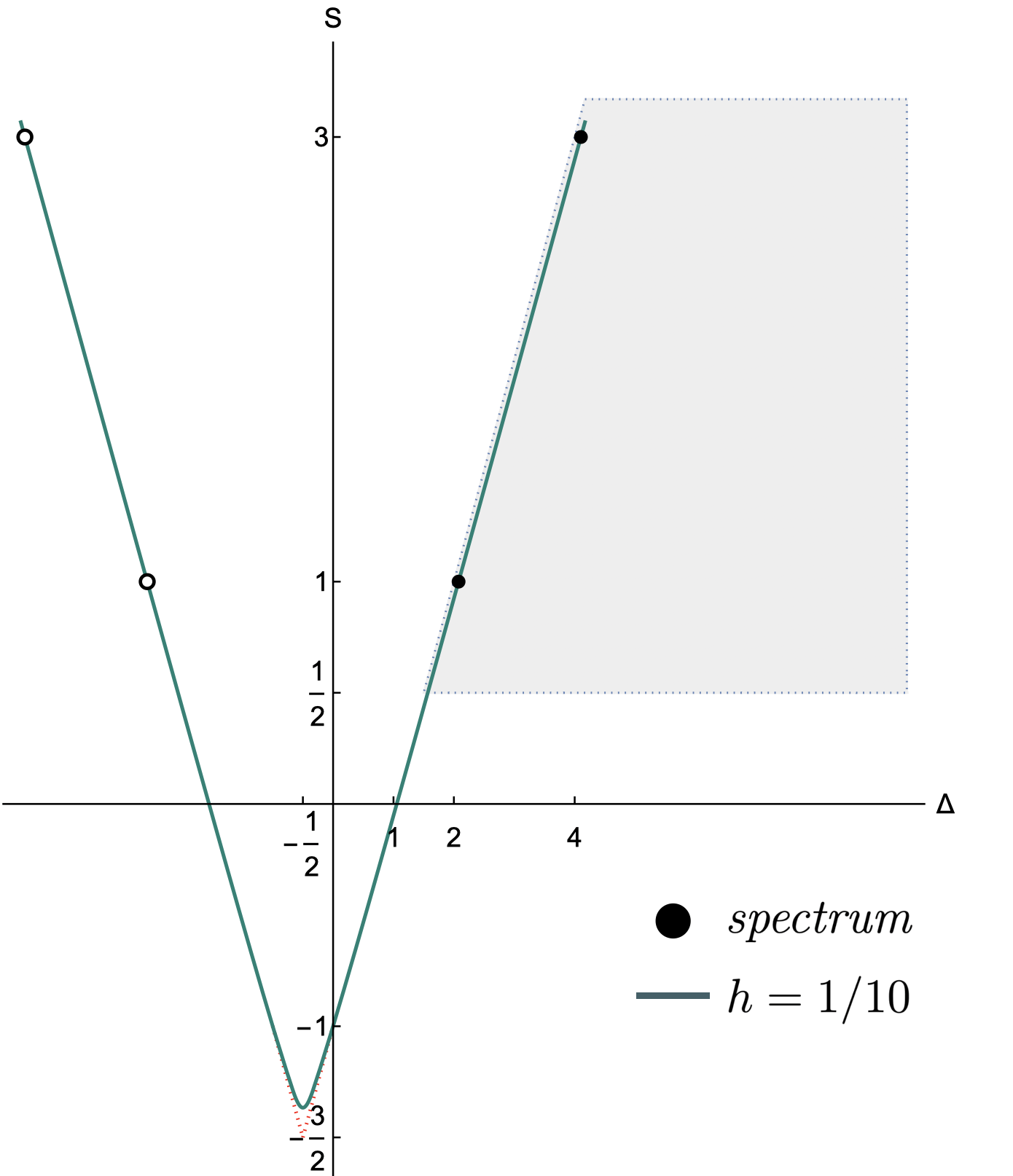}
\caption{Trajectory interpolating  operators (\ref{eq:opsl2}) with odd spins ($sl(2)$-like grading). The unitarity-allowed region is depicted in gray. Physical states belonging to the spectrum are identified by black bullets, their shadow by circles. The weak coupling limit (indicated by the dotted red piecewise line) is in this case a sharp intersection of two linear trajectories.}
\label{leading:odd}
\end{figure}

\paragraph{Trajectory. }
The shape of the trajectory is shown in Figure \ref{leading:odd} for coupling $h = 0.1$. A notable feature is that the trajectory has clearly a very different shape from the even-spins one, in particular it does not have a typical BFKL-type behaviour, but its zero coupling singular limit is simply the intersection of two linear trajectories. 

\begin{figure}[ht!]
\centering
\includegraphics[width=8cm,height=8cm]{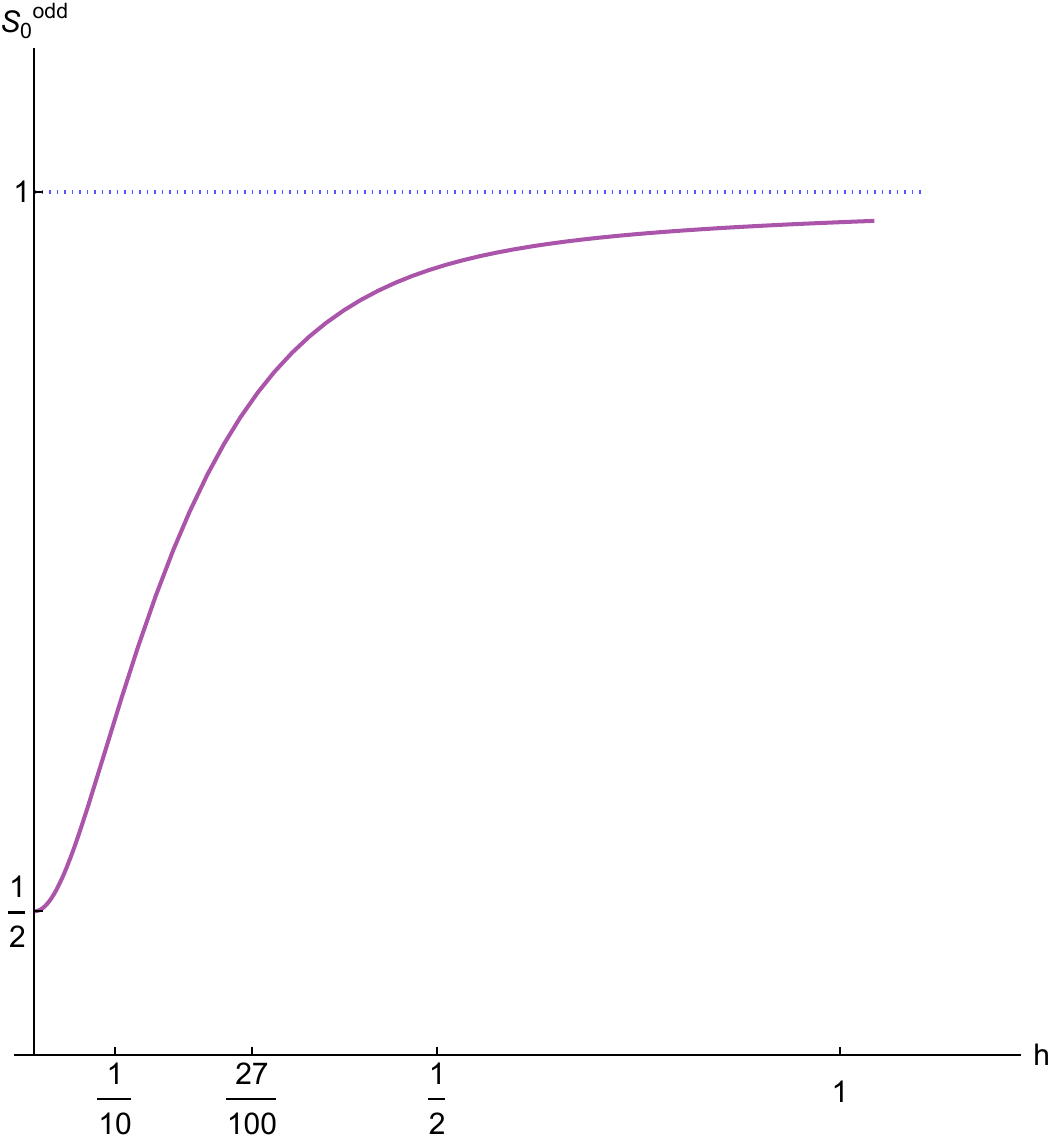}
\caption{Minimum of the leading odd trajectory as a function of the coupling (here reported in ``bottom'' grading).
}
\label{minimum:odd}
\end{figure}

\paragraph{The intercept of the odd spins. } With the same method discussed before, we studied the intercept of the trajectory. We report it more conveniently translated to the ``bottom'' grading, where it is simply the value of the trajectory at its point of symmetry $\Delta_{\text{bottom}} = 3/2$. The result is shown in Figure \ref{minimum:odd}. The value of this Regge pole is consistently subdominant as compared to the one coming from the even$-S_{sl(2)}$ sector, and asymptotes to $\mathcal{S}_0^{\text{odd}}\rightarrow 1$ at strong coupling.
 Extracting some numerical data and comparing them with the weak coupling expansion,
\beq 
\mathcal{S}_0^{\text{odd}} =\frac{1}{2}+\sum_{n>0}\mathcal{S}_0^{\text{odd}\,(2n)}h^{2n} ,
\eeq
 we estimate 
$\mathcal{S}_0^{\text{odd}\,(2)}\simeq21.09289(9)$ and $\mathcal{S}_0^{\text{odd}\,(4)}\simeq-0.1496(1)\times10^4$.

\paragraph{Protected  point. }
Let us observe that also on the odd-spin trajectory there is a second BPS protected point, where the anomalous dimension vanishes, namely the point $\Delta_{sl(2)} = 0$, $S_{sl(2)} = -1$. From the point of view of the QSC, the special nature of this point is signalled by the vanishing of some of the coefficients in front of $\bP$ functions, indeed, due to (\ref{eq:AArels}),  we have
\begin{align}\label{eq:AAzero}
\begin{aligned}
\mathcal{A}_{12} \mathcal{A}_{34}(\Delta_{sl(2)}=0,S_{sl(2)}=-1) &=& 0,\\
\mathcal{A}_{13} \mathcal{A}_{24}(\Delta_{sl(2)}=0,S_{sl(2)}=-1) &=&0,\\
\mathcal{A}_{14} \mathcal{A}_{23}(\Delta_{sl(2)}=0,S_{sl(2)}=-1) &=&1,
\end{aligned}\quad\leftrightarrow\quad
\begin{aligned}
\mathcal{A}_{12} \mathcal{A}_{34}\Big|_{\text{BPS point}} &=& 0,\\
\mathcal{A}_{13} \mathcal{A}_{24}\Big|_{\text{BPS point}} &=&0,\\
\mathcal{A}_{14} \mathcal{A}_{23}\Big|_{\text{BPS point}} &=&1.
\end{aligned}
\end{align}
This is the same behaviour occurring for the BPS point living on the even-spin trajectories, i.e. the one corresponding to the stress-energy tensor, and it is the way the QSC signals a shortening condition.  The vanishing of the anomalous dimension at this point can be verified at 6 loops using the analytic continuation of the perturbative results in \cite{Velizhanin:2022faj}, and is confirmed by our numerics. 

The conditions (\ref{eq:AAzero}) strongly suggest that it should be possible to solve analytically the QSC equations around this point, finding a second slope function in addition to the one in \cite{Gromov:2014eha}. We leave this interesting challenge for the future. 

\subsection{Trajectories and Bridges connecting them: discovering the symmetry}\label{sym:sec}
Let us now come to discuss one of the main findings of this paper.

All the results presented so far were obtained using the gluing matrix (\ref{K:final}), namely the form we have been denoting as  $\mathcal{G}_{\text{Regge}}$. As we discussed in Section \ref{sec:classifyresult}, however, a second consistent possibility, from the point of view of the QSC axioms, is given by
\beq
\mathcal{G}_{\text{Bridge}} \cong ( \mathcal{G}_{\text{Regge}} )^T,
\eeq
which has the same form as the transpose of (\ref{K:final}).
As we discussed, on the one hand this second solution is quickly demystified: it simply corresponds to the gluing matrix obtained for a spin-shadow reflection of a Regge trajectory, i.e. it corresponds to a solution which is transformed with the Weyl reflection (\ref{map:spin}) (given here in various gradings we use commonly):
\beq\label{eq:spinflip}
\texttt{W}_S :\quad S_{\text{top}} \rightarrow -1 - S_{\text{top}} , \quad\text{ i.e. }\quad S_{\text{bottom}} \rightarrow 5 - S_{\text{bottom}} , \quad\text{ i.e. }\quad S_{sl(2)} \rightarrow 1 - S_{sl(2)} ,
\eeq
with $\Delta$ remaining unchanged, which 
implies the following transformation for the twist $\tau \equiv \Delta - S$, given here in the main gradings:
\beq\label{eq:twistmap}
\tau_{\text{top}}\rightarrow \tau_{\text{top}} + 1 + 2 S_{\text{top}} , \quad \tau_{sl(2)}\rightarrow \tau_{sl(2)} -1 + 2 S_{sl(2)}, \quad \tau_{\text{bottom}}\rightarrow \tau_{\text{bottom}} - 5 + 2 S_{\text{bottom}} .
\eeq

\begin{figure}[h!]
\begin{center}
\includegraphics[width=7.5cm,height=8cm]{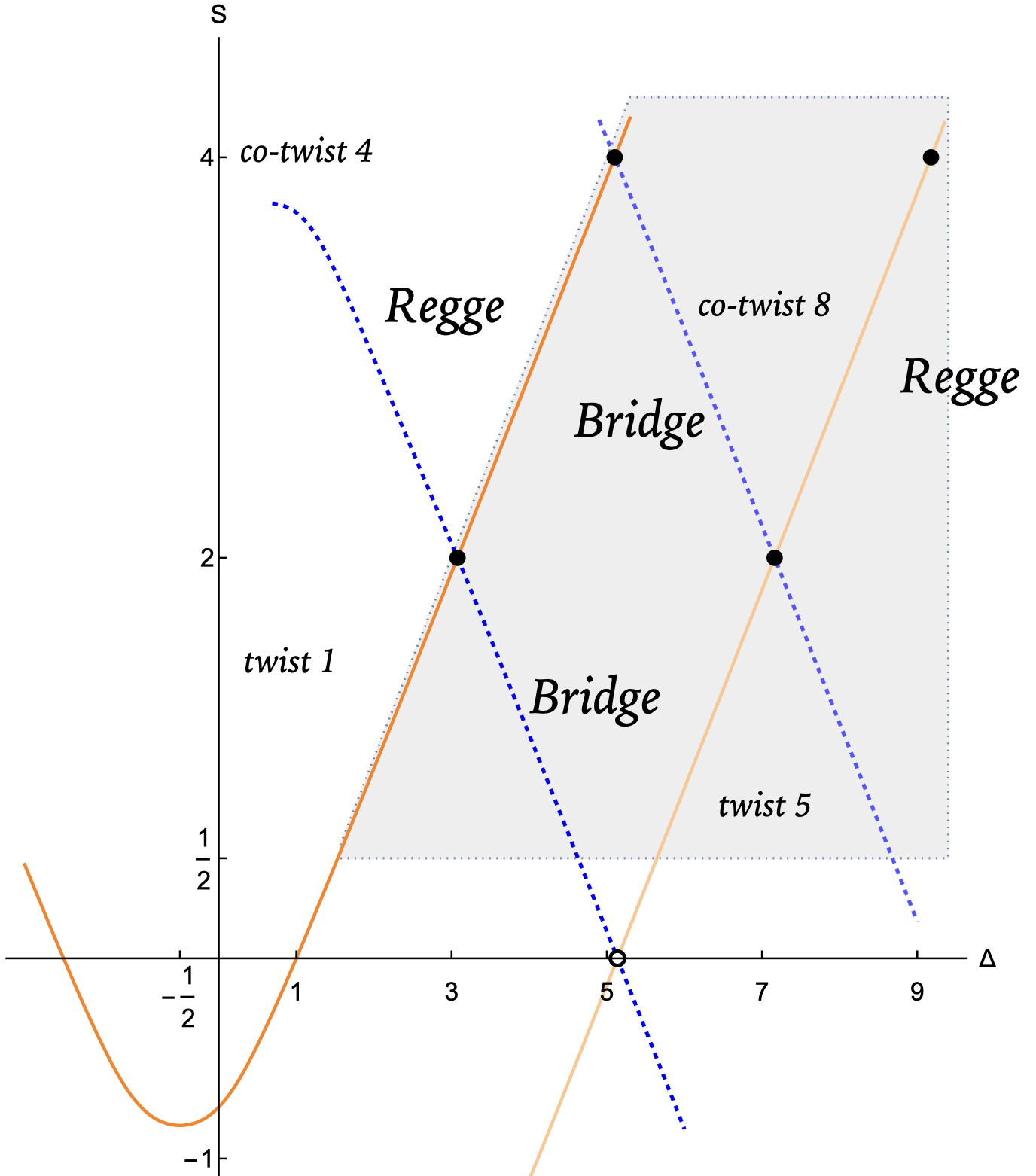}
\end{center}
\caption{
This figure reproduces figure \ref{fig:bridgingintro}, but giving the $sl(2)$-like grading representatives of the trajectories in accordance with the present discussion. 
The coupling is $h=0.1$. The ($sl(2)$-grading copy of) the leading twist-$1$ trajectory is shown, together with two Bridging trajectories connecting it to a subleading sheet with twist $5$. The unitarity-allowed region is depicted in gray. Physical states belonging to the spectrum are identified by black bullets,  shadows of local operators  below unitarity by circles.
}\label{bridging0}
\end{figure}

\begin{figure}[t!]
\centering\includegraphics[width=7.5cm,height=8.5cm]{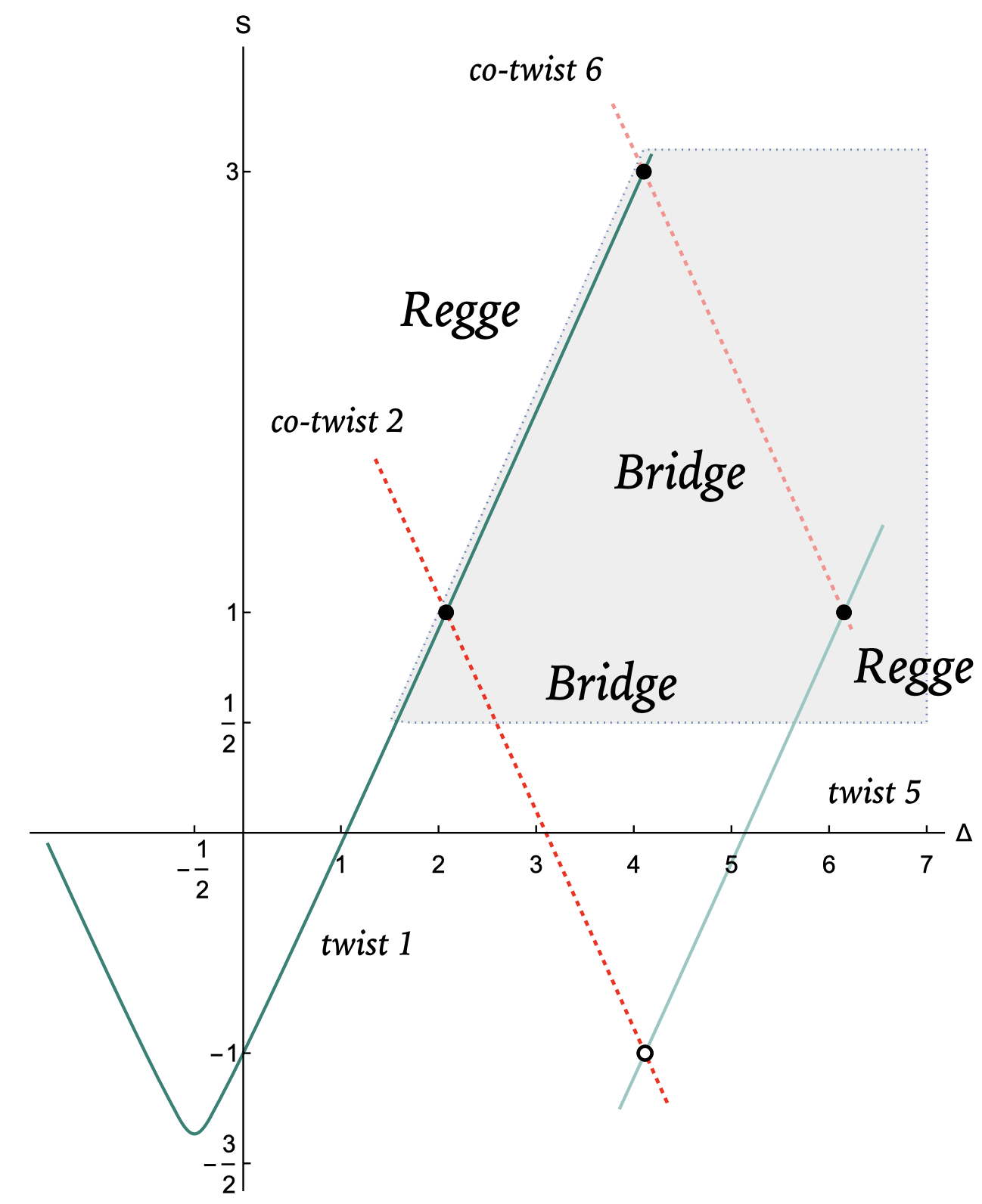}
\caption{The panel shows, at coupling $h=0.1$ ($sl(2)$-like grading), the  twist-$1$ trajectory interpolating between $\emph{odd}$ $S_{sl(2)}$ states, together with two Bridging trajectories connecting it to a subleading sheet with twist $5$. The unitarity-allowed region is depicted in gray. Physical states belonging to the spectrum are identified by black bullets, their shadow by circles.
}
\label{fig:bridgesodd}
\end{figure}

We now want to show the nontrivial phenomenon linked to this formal observation: namely, how the spin-flipped trajectories interlock with the standard Regge trajectories, in this way manifesting the nontrivial twist/co-twist symmetry described in the Introduction. 

\paragraph{Shooting two trajectories out of a local operator. }
To see this, consider starting from a point corresponding to a local operator on the Chew-Frautschi plot. 
As we said many times, solutions of the QSC for local operators obey a gluing condition with a constant gluing matrix, which should take the form (\ref{K:inter}). The latter form is a special case simultaneously of both $\mathcal{G}_{\text{Regge}}$ and $\mathcal{G}_{\text{Bridge}}$, obtained by tuning appropriately their parameters to cancel the $u$-dependent part. This suggests that, starting from any local operator, we may make the spin continuous in two distinct ways, by taking either $\mathcal{G}_{\text{Regge}}$ or $\mathcal{G}_{\text{Bridge}}$ as gluing matrix. 

This is indeed what happens! In Figure \ref{bridging0}, we show in orange the leading trajectory with twist $\tau = 1$. Starting from any (non-BPS) local operator on this trajectory, we can also switch on a second trajectory with negative slope (in blue), obtained by solving the equations with the alternative gluing matrix. We call these negative-slope curves \emph{Bridging trajectories} or simply \emph{Bridges}. 

\paragraph{Bridges between trajectories. }
The name is explained if we follow the Bridging trajectories in the $(S, \Delta)$ plane. As we move $\Delta$, they will cross various integer values of the spin, and for each of these points we can determine whether they correspond to local operators by monitoring the gluing matrix, according to the locality criterion of Section \ref{local:vs}. 
We find that, as the spin decreases in steps of two, they invariably reach local operators (or their spin-shadows) on subleading Regge trajectories, with twist increased in steps of four. At the points of intersection, we can reconstruct the full subleading trajectory by switching to the standard form of gluing matrix, $\mathcal{G}_{\text{Regge}}$. In this way, we can reach subleading sheets of the spectral surface while keeping our equations totally real. 

For example, in Figure \ref{bridging0} is shown the Bridge shooting off from the $S_{sl(2)}=2$ operator on the leading Regge trajectory. We find that at $S_{sl(2)} = 0$ (which is below the unitarity bound), it intersects a \emph{subleading} Regge trajectory with twist $5$. The Figure also shows a second Bridge between the same two trajectories, the one which intersects the local operator with $S_{sl(2)} = 4$ on the leading trajectory and which reaches the same subleading trajectory at a physical local operator with $S_{sl(2)} = 2$. It is obvious to guess that there are in fact infinitely many Bridges between these two Trajectories, one for each non-protected local operator on the first trajectory. We have verified this conjecture in several experiments, finding Bridges passing through every non-BPS local operator we have inspected.\footnote{Notice that there is instead no Bridge passing through the BPS point. } If we continue the Bridges further increasing $\Delta$ past the twist-$5$ subleading trajectory, we find that they allow us to reach further subleading sheets, with twist always increasing in steps of four. 

We tested these numerical findings for various values of the coupling, in particular $h=0.1$ and $h = 0.27$. We believe that the structure is general, and that in this way it is possible to reach infinitely many subleading Regge trajectories by using alternatively the two forms of the gluing matrix. 

In Figure \ref{fig:bridgesodd}, we show that the same phenomenon occurs starting from the trajectory interpolating odd $sl(2)$-like spins. In the figure we show two Bridges which link unto a subleading twist-$5$ trajectory, different from the one found before as it now has local operators for $S_{sl(2)}$ odd. The first two Bridges in this case have co-twists $2$ and $6$.

\paragraph{Bridges = Spin-shadows of Regge trajectories }

 By what we discussed on the symmetries of the QSC and the two gluing matrices, we can expect that each of the Bridges are nothing else but reflections of regular Regge trajectories under the spin-flip transformation (\ref{eq:spinflip}). This, again, is true. 
 
 For instance,  in Figure \ref{bridging0} we see a Bridge passing through the $S_{sl(2)} = 2$ operator on the leading twist-$1$ Regge trajectory. This can also be seen as the spin-flip of a trajectory with twist 4. The Bridge intersecting the next operator with $S_{sl(2)} = 4$ is a spin-flip of a trajectory with twist $8$, and so on: subsequent Bridges, once flipped, will be seen as trajectories of increasing twists in steps of four, as per (\ref{eq:twistmap}).

\paragraph{Twist $\leftrightarrow$ co-twist symmetry. }
The consequence of this observation is that in the Chew-Frautschi plot (which contains \emph{only} Regge trajectories, not their spin-flips), we will have an unexpected symmetry relating points on different Regge trajectories. For each physical local operator sitting on a Regge trajectory of twist $\tau$, we will have an exact image, with the same anomalous dimension, which will sit below the unitarity bound on a subleading trajectory. The image is obtained by applying the Weyl reflection $\texttt{W}_S$, and thus where the original operator sits on a trajectory of twist $\tau$ and has co-twist $\tilde{\tau}$, its image will live on a subleading trajectory of twist $\tilde{\tau}$ and have co-twist $\tau$. On this second trajectory, the original operator (or rather its image) will appear \emph{below the unitarity bound}. Thus we are learning something unexpected on the behaviour of trajectories below the unitarity bound, which is a less accessible region where the Lorentzian inversion formula does not apply. 
To make a very concrete example, the operator $S_{sl(2)} = 2$ on the leading twist-$1$ trajectory will have an image below unitarity on a subleading trajectory with twist $4$. 

Let us state again that the most nontrivial part of this finding comes from the fact that each operator is crossed by both a Regge trajectory and a Bridge. 
It would have been conceivable that this did not happen -- if this had been the case, the Regge trajectories and their spin-flips would have simply been separate curves, and their intersection would not have been at fixed integer spins for all couplings. Then, we would not have had any symmetry. 
 Instead, the symmetry seems to be there, although we do not know its origin. There seems to be something special playing out in this theory, which is neatly encoded in the QSC and in its solutions. In the next section, we present some evidence that exactly the same phenomenon happens also in $\mathcal{N}$$=$$4$ SYM.

\subsection{Bridges between trajectories in $\mathcal{N}$$=$$4$ SYM}\label{sec:numericsSYM}
As discussed in Section \ref{sec:gluingSYM}, also in the case of $\mathcal{N}$$=$$4$ 
 SYM there are two gluing matrices satisfying all the QSC consistency axioms, and they are again simply related to the symmetry interchanging twist $\leftrightarrow$ co-twist. 

\begin{figure}[t!]
\centering
\includegraphics[width=8cm,height=7.2cm]{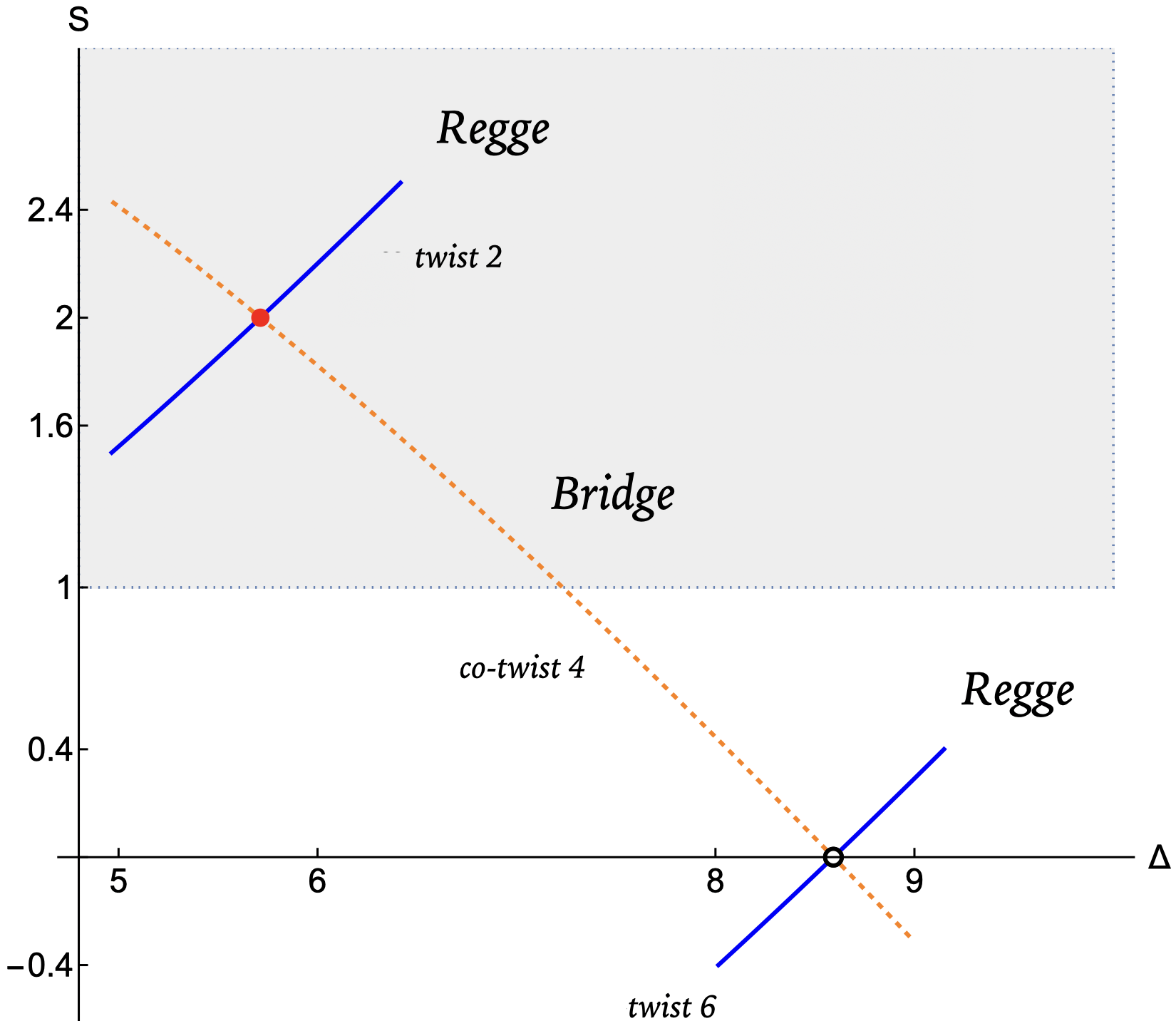}
\caption{ 
Here we show the existence of two continuous curves (i.e. a Regge trajectory and a Bridge) going through the Konishi operator, which is marked by the solid red bullet. 
They are obtained with the two gluing matrices discussed in \ref{sec:gluingSYM}.
A portion of the unitarity-allowed region is depicted in gray: $1<S_{sl(2)}<\Delta_{sl(2)}+1$ \cite{Gromov:2014caa}. 
The value of the coupling $h=0.5$ is considered.
}
\label{Konishi}
\end{figure}

As a preliminary demonstration, we show in Figure \ref{Konishi} a piece of the leading, twist-$2$ Regge trajectory in the $sl(2)$ sector of $\mathcal{N}$$=$$4$ SYM, containing the well known Konishi operator $\mathcal{O}_{\text{Konishi}} = \text{tr}( Z D_+^{2} Z )$ at spin $S_{sl(2)} = 2$. We show that through the Konishi operator pass both its regular Regge trajectory and a Bridge, where the first is obtained by continuing the QSC equations for non-integer spin with the gluing matrix (\ref{eq:GSYM}), while the Bridge is obtained with the alternative, transpose form.  

In the Figure, we see that following the Bridge leads to a subleading Regge trajectory with twist $6$, i.e. shifted by 4 just like in ABJM theory. 

Finally, just like in ABJM theory the Bridge itself can be seen as spin-shadow of a further trajectory with twist $4$. 
Thus, we have found a very nontrivial manifestation of the symmetry: namely the statement that the image of the Konishi operator can be found, at $S_{sl(2)} = 0$ (i.e. below the unitarity bound), on a trajectory of twist-$4$. As already mentioned in the Introduction, this observation was  independently made recently in \cite{Julius:2024ewf}.
The findings of the present paper indicate that the statement extends to all loops.

\section{Discussion and Outlook}\label{sec:conclusions}

In this paper, we have closed a gap in the integrability description of ABJM theory, by showing how to use the QSC in this theory to describe Regge trajectories at continuous spin. The missing ingredient was the form of the gluing matrix, that we have fixed (in the sector we are considering) using a constructive approach based on general principles. 

This has allowed us to obtain new data, in particular we present high-precision numerical data for the intercepts of the leading Regge trajectories in the sectors of even and odd spins, reported in Section \ref{sample}. The method is general and we expect it can be used to systematically generate 
 data for subleading trajectories. 

A surprise of our analysis was revealing a non-perturbative symmetry between non-BPS  \emph{local operators multiplets} and their \emph{spin-flips} with $S_{\text{top}}\rightarrow - d + 2 - S_{\text{top}}$ on the Chew-Frautschi plot\footnote{This definition of the symmetry applies to the spin of the top component of the supermultiplet. The whole supermultiplet gets then carried along.}. 
 The symmetry operates formally through an interchange of twist and co-twist defined in the Introduction, and implies that, for every non-BPS local operator which living on a certain Regge trajectory, its spin shadow will be present on \emph{another} Regge trajectory \emph{below the unitarity bound}, i.e. in the region that is not described by the Lorentzian inversion formula. 
 We stress that this symmetry is not explained trivially by supersymmetry, since it relates Regge trajectories that are not connected by the action of supercharges. 
 Rather, it emerges through a nontrivial interplay of the form of the QSC equations (and the way they encode the Weyl symmetries), and the way their solutions behave. Since it is not possible to track rigorously the solutions of the nonlinear QSC equations, we do not have a complete mathematical proof of the symmetry, but we highlighted that it seems to work by a very simple and elegant mechanism: in particular, the heart of the matter is that \emph{every} non-BPS local operator seems to lie at the intersection of two distinct continuous trajectories, one its own Regge trajectory and the other the spin flip of a separate subleading trajectory. The numerical study of several states has confirmed this property, which then implies the twist/co-twist symmetry. 

We also pointed out that exactly the same symmetry seems to be present in another example of integrable AdS/CFT, i.e. $\mathcal{N}$$=$$4$ SYM, and demonstrated it by a preliminary numerical exploration, where we highlighted that the spin shadow of the Konishi operator (which has twist $2$) lives on a subleading trajectory of twist-$4$. This is precisely what was recently observed at strong coupling in \cite{Julius:2024ewf}.  What seems to be the same symmetry was also observed at weak coupling in \cite{Henriksson:2017eej}. Moreover, in these works the symmetry appears to hold for the full CFT data including OPE coefficients!  We also notice that a slightly different symmetry, also related to Weyl reflections, was observed in non-planar, double trace data in \cite{Aprile:2017qoy}.
 It would be very interesting to understand where the twist/co-twist symmetry comes from in ABJM and $\mathcal{N}$$=$$4$ SYM and possibly other holographic theories (it is natural to conjecture that a version of this symmetry would be present also in integrable AdS$_3$/CFT$_2$). One could also try to explore if something related happens in the Fishnet theory~\cite{Gurdogan:2015csr, Grabner:2017pgm, Gromov:2017cja, Levkovich-Maslyuk:2020rlp}, where the $\mathcal{N}$$=$$4$ supersymmetry but also unitarity are broken. 
 
 Further, it would be nice to find an interpretation of this duality in the theory of light-ray operators, and to understand if there might be versions of the same symmetry in  general  CFTs. Is there a way to construct an object corresponding to the spin-shadow of a local operator\footnote{or, more precisely, in the framework of \cite{Kravchuk:2018htv} one should investigate the shadow transform applied to the light transform of a local operator.} proving that it should sit on the Chew-Frautschi plot? 
 
 An interesting problem is also raised by the increase of degeneracies of trajectories with twist. What does this imply together with the symmetry?
 Two possible mechanisms likely both play a role:  first, some higher-twist trajectories could have points below the unitarity bound that, despite having negative integer spin (in steps of 2), are \emph{not} spin-flipped images of local operators, and, from our point of view, this would be highlighted by the fact that they satisfy QSC equations with non-constant gluing matrix;\footnote{We can expect that such points would also have exact zeros for the analytic continuation of the OPE coefficients. This comes from extending below unitarity the principle that a non-constant gluing matrix is equivalent to the divergence of the norm $\langle \mathbb{O} \mathbb{O} \rangle$ of \cite{Henriksson:2023cnh}, which then implies the vanishing of the OPE coefficient. In Fig. 2 of \cite{Julius:2024ewf} we see already some examples of this mechanism: in the case of the four subleading trajectories studied there, we see that points where the OPE coefficients vanish are precisely the ones that are neither local operators nor their spin-shadows below unitarity, i.e. they are not involved in the twist/co-twist symmetry. 
 } in addition, some  ``extra'' trajectories might bend before reaching these negative values of $S$. In general, it would be important to decode the full pattern in which trajectories are threaded together by the twist/co-twist symmetry. 

We also discussed how this symmetry gives us a nice way to zigzag through the spectral surface: indeed, at each local operator we can take a turn, and start walking along the spin-shadow of another Regge sheet, which then acts as a Bridge allowing us to reach infinitely many subleading sheets without ever needing to leave the real $S$ axis!
This is in contrast to previous studies where reaching subleading sheets always required finding branch points in the complex spin plane. 
An interesting question is whether the connection of these sheets through the twist/co-twist zigzag is the same as the connection of the Riemann surface through branch points. 
On the purely practical level, we notice that this finding gives a useful additional tool to shortcut the numerical exploration of the space of trajectories. 

There are further interesting problems for the future. 

In the spirit of this paper, it would be nice to understand the classification of gluing matrices in general, moving outside the LR-symmetric and $u\leftrightarrow -u$ symmetric sector we considered. One may also wonder if non-constant gluing matrices with asymptotics stronger than $e^{2 \pi |u|}$ should ever be required for some operators, although in $\mathcal{N}$$=$$4$ SYM so far there is no evidence that they should be needed.

Another question concerns the relation between the gluing matrix and the local or non-local nature of the operator represented by the solution to the QSC. As we discussed, the coefficients of the gluing matrix which multiply nontrivial functions of $u$ should vanish when the solution is a local operator --- just like the 2-point function $\langle \mathbb{O} \mathbb{O} \rangle$ of light-ray operators recently constructed in \cite{Henriksson:2023cnh}. This might suggest that these coefficients in the gluing matrix might be proportional to this 2-point function. It would be interesting to understand if there is such a connection, in view of the recent SoV program to connect the QSC and correlation functions in these theories~\cite{Gromov:2016itr,Cavaglia:2018lxi,Giombi:2018hsx,Maillet:2018bim,Ryan:2018fyo,Gromov:2022waj,Basso:2022nny,Bercini:2022jxo,Ekhammar:2023iph}.

On ABJM theory itself, for which we have obtained new data, it would be interesting to understand the analytical description around the flattening behaviour observed around the intercept of the leading trajectory at weak coupling, which is so similar to the BFKL limit in $\mathcal{N}$$=$$4$ SYM. 
It would be very nice to generalize the method of \cite{Ekhammar:2024neh} and determine this behaviour analytically from the QSC. 
 In parallel, is there a description of this limit from field theory methods alone? This would likely be quite different from the one in $\mathcal{N}$$=$$4$ SYM, since the gauge fields have a very different form. 
Computing at least the leading order from field theory alone would give us a quantity known both as function of $h$ (from the QSC) and of $\lambda$ (from field theory): thus, it would provide a path to a proof of the form of the interpolating function $h(\lambda)$. As we pointed out, it looks that another interesting and exactly computable quantity would be a \emph{second slope} function, expanding around the protected point on the trajectory interpolating through odd spins.

Finally, another motivation for developing tools for studying continuous spin is to gather a new type of data to use in the Bootstrability approach~\cite{Cavaglia:2021bnz,Cavaglia:2022qpg,Cavaglia:2023mmu,Caron-Huot:2022sdy}. So far, this program has used the spectrum of low-lying local operators coming from QSC methods as an input for the conformal bootstrap in order to derive constraints on correlation functions. Since continuous spin comes with a tight structure in CFT~\cite{Caron-Huot:2017vep}, it would be very nice to explore if the input on Regge trajectories at any continuous $S$ coming from the QSC could be used to put even more constraints on OPE coefficients. This direction might be important especially to study Bootstrability for the full 3D or 4D theory (see \cite{Caron-Huot:2022sdy}), which is significantly more complex than the most studied defect Wilson line setup. 
{In a parallel development, in the last few years there was significant progress in the study of 4-point correlators in $\mathcal{N}$$=$$4$ SYM at strong coupling~\cite{Alday:2022uxp,Alday:2022xwz,Alday:2023mvu,Alday:2023jdk,Alday:2023flc}.
The approach of these works uses the analytic conformal bootstrap, with input from integrability spectral data on the first Regge trajectories. This analysis has also led to new structural insight on the analytic structure of the spectrum itself \cite{Julius:2023hre,Julius:2024ewf}. The symmetry we are discussing might give further constraints and be helpful in this program, as well as for the one recently initiated for ABJM theory in \cite{Chester:2024esn}. 
}
\section*{Acknowledgements}
We are grateful to Mikhail Alfimov, Francesco Aprile, Till Bargheer, Benjamin Basso, Davide Bonomi, Carlos Bercini, Lorenzo Bianchi, Riccardo Conti, Nikolay Gromov, Tobias Hansen, Johan Henriksson, Julius Julius, Rob Klabbers, Shota Komatsu, Petr Kravchuk, Madalena Lemos, Marco Meineri, Michelangelo Preti, Nicol\`o Primi,  Nika Sergeevna Sokolova, Istvan Sze\'cs\'enyi for discussions related to this work. 

This project received partial support from the INFN, and the PRIN project No. 2022ABPBEY, with the title ``Understanding quantum field theory through its deformations'', funded by the Italian Ministry of University and Research.

\appendix

\section{Weak coupling checks of the symmetry in ABJM theory}\label{app:weakc}

In ABJM theory, we can see clearly the twist/co-twist symmetry in the explicitly known 2 loops perturbative data.\footnote{Two loops are the lowest nontrivial perturbative order.} Using the weak coupling Bethe ansatz equations, \cite{Papathanasiou:2009zm} obtained the explicit 2-loop anomalous dimension for the following sequences of supermultiplets:
{\small
\begin{align}\label{eq:explicittraj}
\gamma_2 &= 4\( S_1(j+1) + S_{-1}(j+1)\)+\frac{2\(1-(-1)^{1+j}\)}{j+2}+8,&&\text{for}\,\, \mathcal{V}^4_{3 + 2 j, 3},\quad ( \tau = 2 , \,j\in \mathbb{N} ) ,\nonumber \\ 
\gamma_2 &= 4\( S_1(j+2) - S_{-1}(j+2)\)+\frac{\(1+(-1)^{j}\)}{j+3}-\frac{\(1-(-1)^{j}\)}{j+2}+4,&&\text{for}\,\,  \mathcal{V}^4_{4 + 2 j, 4},\quad( \tau = 3 , \,j\in \mathbb{N} ),\nonumber \\
\gamma_2 &= 4\( S_1(j+2) + S_{-1}(j+2)\)+\frac{2\(1-(-1)^{j}\)}{j+3}+\frac{32}{3},&& \mathcal{V}^4_{5 + 2 j, 5},\quad ( \tau = 4 , \,j\in\mathbb{N}_\text{odd}),
\end{align}
}
where we used the super Young tableaux notation of \cite{Papathanasiou:2009en,Papathanasiou:2009zm} for the supermultiplets, and $j\equiv S_{\text{top}}$. These are all supermultiplets with top components neutral under R-symmetry. These multiplets in general have the form
\beq
\mathcal{V}^{2 L}_{3 + \delta_1, \dots , 3+\delta_L},
\eeq
where $\delta_i\geq \delta_{i+1}$, $\delta_i\geq 3$. Here, $L$ is the length, i.e. at leading order the state is built with $2 L$ fundamental fields on the alternating spin chain of ABJM. 

The top spin and top dimension at tree level are
\beq
j = \frac{\text{max}\left(\delta_1,0\right) - \text{max}\left(\delta_2,0\right)}{2}, \quad \Delta = L + \frac{\text{max}\left(\delta_1,0\right) + \text{max}\left(\delta_2,0\right)}{2} .
\eeq
The labels in the SYT notation allow us to count trajectories. Thus, the trajectories with $\tau = 1$ and $\tau=2$ in this sector are non-degenerate and are built with $L=1$ and $L=2$ multiplets respectively. The trajectory with $\tau=3$ is doubly degenerate, and the degeneracy then increases with twist. 

In addition, in this sector, the leading trajectories for even/odd spin correspond to the supermultiplets $\mathcal{V}^2_{1+ 2 j}$ with dimension: 
{\small
\begin{align}
\gamma_2 = 4\( S_1(j+1) - S_{-1}(j+1)\), \quad \text{for\,\,} \mathcal{V}^2_{3 + 2 j},\quad ( \tau = 1 , \,j\in \mathbb{N} ) .\nonumber
\end{align}
}
With these data we can test some instances of the twist/co-twist symmetry. For instance, the first physical state on the leading even-$j$ trajectory is the one with $j=0$. This has $(\tau = 1, \tilde{\tau} = 2 )$, and has the same anomalous dimension $\gamma_2 = 8$ as is found on the twist-$2$ trajectory $\mathcal{V}^4_{3+2j, 3}$ continued to unphysical spin $-1$:
\begin{equation}
\left. \mathcal{V}^2_{3+2 j} \right|_{j\rightarrow 0}, \quad(\tau = 1, \tilde{\tau} = 2 ) \quad\leftrightarrow \quad\left. \mathcal{V}^4_{3+2 j, 3} \right|_{j\rightarrow -1} ,\quad (\tau = 2, \tilde{\tau} = 1 ) .
\end{equation}
Similarly, the symmetry connects the spin $j=0$ state on the  $\tau = 2$ trajectory to a point on the twist-$3$ trajectory $\mathcal{V}^4_{4+2j, 4}$ continued to unphysical spin $-1$:
\begin{equation}
\left. \mathcal{V}^4_{3+2 j,3} \right|_{j\rightarrow 0}, \quad(\tau = 2, \tilde{\tau} = 3 ) \quad\leftrightarrow \quad\left. \mathcal{V}^4_{4+2 j, 4} \right|_{j\rightarrow -1} ,\quad (\tau = 2 , \tilde{\tau} = 2 ) ,
\end{equation}
as in this case they both have 2-loops anomalous dimension $\gamma_2 = 10$. 

As a third example, the  spin $j=0$ state on the  $\tau = 3$ trajectory $\mathcal{V}^4_{4+2j, 4}$ is connected to a point on the twist-$4$ trajectory $\mathcal{V}^4_{5+2j, 5}$  with $j\rightarrow -1$:
\begin{equation}
\left. \mathcal{V}^4_{4+2 j,4} \right|_{j\rightarrow 0}, \quad(\tau = 3, \tilde{\tau} = 4 ) \quad\leftrightarrow \quad\left. \mathcal{V}^4_{5+2 j, 5} \right|_{j\rightarrow -1} ,\quad (\tau = 4, \tilde{\tau} = 3 ) ,
\end{equation}
and in this case their 2-loops anomalous dimension is $\gamma_2 = 38/3$. 

We also extracted numerically the two-loop anomalous dimension of the point  $(\tau=5, \tilde{\tau}=2)$ living, below the unitarity bound, on the subleading trajectory shown in Figure \ref{fig:bridgesodd} at $S_{sl(2)} = -1$. This gives an excellent agreement with $\gamma_2 = 12$, which is the two-loop anomalous dimension of the physical operator with $S_{sl(2)} = 2$ ($\tau=2, \tilde{\tau}=5$) belonging to the $\mathcal{V}^4_{3 + 2j, 3}\big|_{j \to 1}$ multiplet~\cite{Papathanasiou:2009zm}. 

The twist/co-twist symmetry implies these coincidences of anomalous dimensions extend to all loops. 

Notice that in general the symmetry connects a state of spin $j$ on a trajectory of twist $\tau$ to a point on a trajectory of twist $\tilde{\tau} = d-2 + \tau + 2 j$, continued to unphysical spin $\tilde{j} = -d + 2 - j$.  
The only instances of the symmetry which we can check based on the analytic trajectories (\ref{eq:explicittraj}) are the ones we have presented, as we do not have the exact expressions in order to perform the analytic continuations to negative $j$ for higher twists. However, it would be interesting to use numerical explorations of the QSC solutions to map the precise pairings of multiplets given by the symmetry.

\section{Useful ancillaria}\label{ancil}

\subsection{Leading coefficients}

Here we report some additional internal constraints of the QSC construction. The leading prefactors coefficients of the expansion (\ref{eq:Pseries}) satisfy (see \cite{Bombardelli:2017vhk})
\begin{align}\label{eq:AArels}
\begin{split}
\mathcal{A}_{12} \mathcal{A}_{34} &= \frac{(M_1^2 - \hat{M}_1^2) (M_1^2 - \hat{M}_2^2)}{(M_1^2 - M_2^2) (M_1^2 - M_5^2)} ,\\
\mathcal{A}_{13} \mathcal{A}_{24} &= \frac{(M_2^2 - \hat{M}_1^2) (M_2^2 - \hat{M}_2^2)}{(M_1^2 - M_2^2) (M_2^2 - M_5^2) },  \\
\mathcal{A}_{14} \mathcal{A}_{23} &=\frac{(M_5^2 - \hat{M}_1^2) (M_5^2 - \hat{M}_2^2)}{(M_1^2 - M_5^2) (M_2^2 -M_5^2)}.
\end{split}
\end{align}
Up to Weyl reflections \ref{sec:Weyl}, the only points satisfying the BPS constraint \eqref{eq:AAzero} are 
\begin{equation}
(\Delta_{sl(2)},S_{sl(2)})\in\{(1,0),(0,-1)\}.
\end{equation}
Studying the Q-system equations at large $u$ one can constrain the form of the coefficients $B_{a|i}$ entering in the expansion \eqref{eq:pure1}. The expressions are provided below in \texttt{Mathematica} format, where $B_{a|i}\equiv\texttt{B}\texttt{[}a,i\texttt{]}$, the charges are denoted as $M_{A=1,2,5}\equiv\texttt{M}\texttt{[}A\texttt{]}$ and $\hat{M}_{I=1,2,5}\equiv\texttt{Mh}\texttt{[}I\texttt{]}$, and the coefficients $\mathcal{A}_{ab}\equiv\texttt{A}\texttt{[}a,b\texttt{]}$.
{\scriptsize
\begin{align*}
&\texttt{\big\{}\texttt{B[2, 1] \texttt{->} ((M[1] + M[2]) (M[2] - M[5]) (M[1] + M[5]) (M[2] + M[5]) A[2, 3] A[2, 4] }\\
&\quad\texttt{B[1, 1]) / ((M[2] + Mh[1]) (M[5] + Mh[1]) (M[2] + Mh[2]) (M[5] + Mh[2])), }\\
&\texttt{B[2, 2] \texttt{->} ((M[1] + M[2]) (M[2] - M[5]) (M[1] + M[5]) (M[2] + M[5]) A[2, 3] A[2, 4] }\\
&\quad\texttt{B[1, 2]) / ((M[2] + Mh[1]) (M[5] + Mh[1]) (M[2] - Mh[2]) (M[5] - Mh[2])), }\\
&\texttt{B[2, 3] \texttt{->} ((M[1] + M[2]) (M[2] - M[5]) (M[1] + M[5]) (M[2] + M[5]) A[2, 3] A[2, 4] }\\
&\quad\texttt{B[1, 3]) / ((M[2] - Mh[1]) (-M[5] + Mh[1]) (M[2] + Mh[2]) (M[5] + Mh[2])), }\\
&\texttt{B[2, 4] \texttt{->} ((M[1] + M[2]) (M[2] - M[5]) (M[1] + M[5]) (M[2] + M[5]) A[2, 3] A[2, 4] }\\
&\quad\texttt{B[1, 4]) / ((M[2] - Mh[1]) (-M[5] + Mh[1]) (M[2] - Mh[2]) (-M[5] + Mh[2])), }\\
&\texttt{B[3, 1] \texttt{->} ((M[1] + M[2]) (M[2] - M[5]) (M[1] + M[5]) (M[2] + M[5]) A[2, 3] A[3, 4] }\\
&\quad\texttt{B[1, 1]) / ((M[2] + Mh[1]) (M[5] + Mh[1]) (M[2] + Mh[2]) (M[5] + Mh[2])), }\\
&\texttt{B[3, 2] \texttt{->} ((M[1] + M[2]) (M[2] - M[5]) (M[1] + M[5]) (M[2] + M[5]) A[2, 3] A[3, 4] }\\
&\quad\texttt{B[1, 2]) / ((M[2] + Mh[1]) (M[5] + Mh[1]) (M[2] - Mh[2]) (M[5] - Mh[2])), }\\
&\texttt{B[3, 3] \texttt{->} ((M[1] + M[2]) (M[2] - M[5]) (M[1] + M[5]) (M[2] + M[5]) A[2, 3] A[3, 4] }\\
&\quad\texttt{B[1, 3]) / ((M[2] - Mh[1]) (-M[5] + Mh[1]) (M[2] + Mh[2]) (M[5] + Mh[2])), }\\
&\texttt{B[3, 4] \texttt{->} ((M[1] + M[2]) (M[2] - M[5]) (M[1] + M[5]) (M[2] + M[5]) A[2, 3] A[3, 4] }\\
&\quad\texttt{B[1, 4]) / ((M[2] - Mh[1]) (-M[5] + Mh[1]) (M[2] - Mh[2]) (-M[5] + Mh[2])), }\\
&\texttt{B[4, 1] \texttt{->} ((M[1] + M[2]) (M[2] - M[5]) (M[1] + M[5]) (M[2] + M[5]) A[2, 4] A[3, 4] }\\
&\quad\texttt{B[1, 1]) / ((M[2] + Mh[1]) (M[5] + Mh[1]) (M[2] + Mh[2]) (M[5] + Mh[2])), }\\
&\texttt{B[4, 2] \texttt{->} ((M[1] + M[2]) (M[2] - M[5]) (M[1] + M[5]) (M[2] + M[5]) A[2, 4] A[3, 4] }\\
&\quad\texttt{B[1, 2]) / ((M[2] + Mh[1]) (M[5] + Mh[1]) (M[2] - Mh[2]) (M[5] - Mh[2])), }\\
&\texttt{B[4, 3] \texttt{->} ((M[1] + M[2]) (M[2] - M[5]) (M[1] + M[5]) (M[2] + M[5]) A[2, 4] A[3, 4] }\\
&\quad\texttt{B[1, 3]) / ((M[2] - Mh[1]) (-M[5] + Mh[1]) (M[2] + Mh[2]) (M[5] + Mh[2])), }\\
&\texttt{B[4, 4] \texttt{->} ((M[1] + M[2]) (M[2] - M[5]) (M[1] + M[5]) (M[2] + M[5]) A[2, 4] A[3, 4] }\\
&\quad\texttt{B[1, 4]) / ((M[2] - Mh[1]) (-M[5] + Mh[1]) (M[2] - Mh[2]) (-M[5] + Mh[2]))}\texttt{\big\}},
\end{align*}
}
together with
{\scriptsize
\begin{align*}
&\texttt{B[1, 4] -> ((M[1] - Mh[1])\textasciicircum 2 (M[2] - Mh[1])\textasciicircum 2 (M[5] - Mh[1])\textasciicircum 2 (M[1] + 
        Mh[1])\textasciicircum 2)}\\
        &\quad\texttt{(M[2] + Mh[1])\textasciicircum 2 (M[5] + Mh[1])\textasciicircum 2 (M[1] - Mh[2])\textasciicircum 2 (M[2] - 
        Mh[2])\textasciicircum 2)}\\
        &\quad\texttt{(M[5] - Mh[2])\textasciicircum 2(M[1] + Mh[2])\textasciicircum 2 (M[2] + Mh[2])\textasciicircum 2 (M[5] + 
        Mh[2])\textasciicircum 2)/}\\
        &\quad\texttt{(16 (M[1] - M[2])\textasciicircum 2(M[1] + M[2])\textasciicircum 4 (M[1] - M[5])\textasciicircum2(M[2] - 
        M[5])\textasciicircum 2)}\\
        &\quad\texttt{(M[1] + M[5])\textasciicircum 4 (M[2] + M[5])\textasciicircum 4 Mh[1]\textasciicircum 2 Mh[2]\textasciicircum 2(Mh[1]\textasciicircum 2 - Mh[2]\textasciicircum 2)}\\
       &\quad\,\texttt{A[
       2, 3]\textasciicircum 2 A[2, 4]\textasciicircum 2 A[3, 4]\textasciicircum 2 B[1, 1] B[1, 2] B[1, 3])}.
\end{align*}
}

\subsection{Generic form of the Weyl reflections}
For completeness, we report a slightly more general form of the Q-system transformations implementing the Weyl reflections, which simply differs from the one given in the main text by a generic unphysical ``gauge transformation'', i.e. a redefinition of the normalisations of various Q-functions. In general, the transformation is implemented as
\begin{equation}
\mathbf{Q} \rightarrow \bQ' \equiv \mathbb{M} \cdot \mathbf{Q} \cdot \mathbb{M}^T ,
\end{equation}
and correspondingly
\begin{equation}
Q_{a|i}\rightarrow Q'_{a|j} = Q_{a|j}\mathbb{M}_i^j\,.
\end{equation}
The most general form of $\mathbb{M}$ implementing the Weyl transformation $\texttt{W}_S$ or $\texttt{W}_{\Delta}$  takes the form, respectively,
\begin{align}\label{eq:newMs}
&\left(\mathbb{M}_1\right)_i^{\,j} =
\begin{pmatrix}
a_1 & 0 & 0 & 0 \\
0 & 0 & a_2 & 0 \\
0 & -\frac{1}{a_2} & 0 & 0 \\
0 & 0 & 0 & \frac{1}{a_1}
\end{pmatrix}\, \leftrightarrow\, \texttt{W}_S,\\
&\left(\mathbb{M}_2\right)_i^{\,j} =
\begin{pmatrix}
0 & 0 & 0 & b_1 \\
0 & b_2 & 0 & 0 \\
0 & 0 & \frac{1}{b_2} & 0 \\
-\frac{1}{b_1} & 0 & 0 & 0
\end{pmatrix}\,\leftrightarrow\, \texttt{W}_\Delta ,
\end{align}
where the matrices given in  (\ref{eq:MWeyls}) are simply a particular gauge choice. In particular, with the transformation $\mathbb{M}_1$ in (\ref{eq:newMs}), on top of the interchange of columns with $i=1$ and $i=4$, the prefactors $B_{a|i}$ in the normalisation of Q-functions are redefined according to 
\begin{align}
a_1 = \frac{\left(B_{a|1}\right)'}{\left(B_{a|1}\right)} = \frac{\left(B_{a|4}\right)}{\left(B_{a|4}\right)'}, \quad
a_2 = \frac{\left(B_{a|2}\right)'}{\left(B_{a|3}\right)} = -\frac{\left(B_{a|2}\right)}{\left(B_{a|3}\right)'}.
\end{align} 
 Similarly, the transformation $\mathbb{M}_2$ interchanges columns $i=2$ and $i=3$ with the normalisations redefined as
\begin{align}
b_1 = \frac{\left(B_{a|1}\right)'}{\left(B_{a|4}\right)} = \frac{\left(B_{a|1}\right)}{\left(B_{a|4}\right)'}, \quad
b_2 = \frac{\left(B_{a|2}\right)'}{\left(B_{a|2}\right)} = -\frac{\left(B_{a|3}\right)}{\left(B_{a|3}\right)'}.
\end{align}
The redefinition of the prefactors of $\bQ_{ij}$ functions also follows easily.

\section{Proof of some QSC constraints in special sector}\label{proof}
In this appendix, we prove the constraints on the gluing matrix in the special LR-symmetric, parity symmetric sector considered in this paper. It is convenient to introduce the standard notation for shifts of the rapidity variable $u$:
\beq
F^{[\pm n]}(u) \equiv F\left( u \pm \frac{\i n}{2} \right),
\eeq
where we will always assume that shifts are performed on the section of the Riemann surface where all cuts are short.

\paragraph{\emph{Proof} $\Theta^{[+2]}(u)=\mathbb{K}\cdot\Theta(u) \cdot\mathbb{K}$.}From the definition of $\mathbb{R}$-matrix, \eqref{period:Theta}, one can write
\beq\label{start} 
Q_{a|i}^{[-2]}(-u)=\mathbb{R}_a^{c}Q_{c|j}^{[+2]}(u)\Theta_i^{j\,[+2]}(u).
\eeq
Using definition \eqref{eq:shiftQ} for $Q_{a|i}$ and the sector property \eqref{subsector} one gets
\beq
\label{rel}
Q_{a|i}^{[-2]}(-u)={\bf P}_{ad}^{[-1]}(-u)\kappa^{de}Q_{e|k}(-u)\,\mathbb{K}_i^{k}
\eeq
that is 
 \beq 
Q_{a|i}^{[-2]}(-u)={\bf P}_{ad}^{[-1]}(-u)\kappa^{de}\mathbb{R}_e^{h}Q_{h|l}(u)\Theta_k^{l}(u)\,\mathbb{K}_i^{k}
\eeq
via definition \eqref{period:Theta}.
Now, using parity condition \eqref{P:R}, one obtain
\beq\label{first} 
Q_{a|i}^{[-2]}(-u)=\mathbb{R}_a^m{\bf P}_{mn}^{[+1]}(u)\mathbb{R}_d^n\kappa^{de}\mathbb{R}_e^hQ_{h|l}(u)\Theta_k^{l}(u)\,\mathbb{K}_i^{k}.
\eeq
From equation \eqref{eq:shiftQ} one could write $Q_{a|i}^{[+1]}={\bf P}_{ab}{\bf P}^{bc\,[-2]}Q_{c|i}^{[-3]}$ (see, \cite{Bombardelli:2017vhk}). In this way we can write
\beq
\mathbb{R}_a^{c}Q_{c|j}^{[+2]}(u)\Theta_i^{j\,[+2]}(u)
=\mathbb{R}_a^{c}{\bf P}_{cr}^{[+1]}(u) {\bf P}^{rt\,[-1]}(u)Q_{t|j}^{[-2]}(u)\Theta_i^{j\,[+2]}(u).
\eeq
Finally, using relation \eqref{rel}, we get
\begin{align} 
\mathbb{R}_a^{c}Q_{c|j}^{[+2]}(u)\Theta_i^{j\,[+2]}(u)&=\mathbb{R}_a^{c}{\bf P}_{cr}^{[+1]}(u){\bf P}^{rt\,[-1]}(u){\bf P}_{tg}^{[-1]}(u)\kappa^{gq}Q_{q|p}(u)\,\mathbb{K}_j^{p}\Theta_i^{j\,[+2]}(u)\\\label{second}
&=\mathbb{R}_a^{c}{\bf P}_{cr}^{[+1]}(u)\kappa^{gq}Q_{q|p}(u)\,\mathbb{K}_j^{p}\Theta_i^{j\,[+2]}(u).
\end{align}
Now, using expression \eqref{start}, we can compare \eqref{first} and \eqref{second}. We get
\beq 
\Theta_i^{j}(u)\,\mathbb{K}_i^j=\mathbb{K}_j^k\,\Theta_i^{j\,[+2]}(u)\quad\Box.
\eeq

\paragraph{\emph{Proof} $\Theta(u)\cdot\Theta(-u)=\1 \times (-1)^{M_1+M_2}$. }
From the definition of the $\Theta$-matrix, \eqref{period:Theta} immediately, 
\beq
Q_{a|i}(u) =\mathbb{R}_a^{c}Q_{c|j}(-u)\Theta_i^{j}(-u)=\mathbb{R}_a^{c}\mathbb{R}_c^{d}Q_{d|k}(u)\Theta_j^{k}(u)\Theta_i^{j}(-u),
\eeq
where
\beq
\mathbb{R} \cdot \mathbb{R} = \1 \times (-1)^{M_1+M_2} .
\eeq
Then,  \eqref{theta:prod} follows simply by multiplying both sides of the equation by the inverse of $Q_{a|i}$ $\Box$.

\paragraph{\emph{Proof} $\Theta(u)$ asymptotics. } 

Due to the analytic properties discussed in \ref{analytic}, 
 the asymptotic expansion of $Q_{a|i}^{\downarrow}(-u)$ as $u\rightarrow+\infty$  can be obtained by analytic continuation in the holomorphic region of the upper half plane, namely it is related to $Q_{a|i}^{\downarrow}(\e^{\i \pi }|u| )$ 
as $|u|\rightarrow+\infty$. Using the  definition \eqref{period:Theta} together with the asymptotics \eqref{eq:pure1}, we can write
\begin{align} 
&\sum_{b=1}^4 (\mathbb{R}^{-1})^b_ae^{\i\pi(\mathcal{N}_b+\hat{\mathcal{N}}_i)} |u|^{\mathcal{N}_b+\hat{\mathcal{N}}_i}B_{b|i}^{\downarrow}=\sum_{j=1}^4 |u|^{\mathcal{N}_a+\hat{\mathcal{N}}_j}\Theta_i^j(u) B_{a|j}^{\downarrow},\\
&(\mathbb{R}^{-1} )^a_ae^{\i\pi(\mathcal{N}_a+\hat{\mathcal{N}}_i)} |u|^{\mathcal{N}_a+\hat{\mathcal{N}}_i}B_{a|i}^{\downarrow}=\sum_{j=1}^4 |u|^{\mathcal{N}_a+\hat{\mathcal{N}}_j}\Theta_i^j(u) B_{a|j}^{\downarrow},\\
&(\mathbb{R}^{-1})^a_a e^{\i\pi(\mathcal{N}_a+\hat{\mathcal{N}}_i)}=\Theta_i^i(u)
\end{align}
as $u\rightarrow+\infty$ (the only sums over indices here are the ones explicitly written). In the derivation we used the fact that $\mathbb{R}$ is diagonal together with \eqref{Nhat}, which imply that $\Theta$ is asymptotically diagonal.
It seems that the LHS depends on index $a$ whereas the RHS does not, but one can note that this dependence disappears, indeed $(\mathbb{R}^{-1})_a^a\e^{\i\pi \mathcal{N}_a}=1$ $\forall a$. The $u\rightarrow+\infty$ asymptotics \eqref{theta:asym} immediately follows. To find the behaviour as $u\rightarrow-\infty$ one can use \eqref{theta:prod}.

\paragraph{\emph{Proof} $\mathcal{G}^{[+2]}(u)=\mathbb{K}\cdot\mathcal{G}(u) \cdot\mathbb{K}$.}
From the definition of the gluing matrix we have $\Theta^{[-1]}(u)=\mathcal{G}^{-1}(u)\cdot f(u)$ and substituting that in \eqref{theta:period} one gets
\beq\label{eq:stepapp}
\mathcal{G}^{[+2]}(u)=f^{[+2]}(u)\cdot\mathbb{K}\cdot f^{-1}(u)\cdot\mathcal{G}(u)\cdot\mathbb{K}.
\eeq
Now the goal is to rewrite $f^{[+2]}(u)\cdot\mathbb{K}\cdot f^{-1}(u)$ in a simpler way using the properties of the $\tau$-functions. Moreover, we notice that $\mathbb{K}=\mathbb{K}^T=\mathbb{K}^{-1}$ and $[\kappa,\mathbb{K}]=0$.
First of all, using (\ref{tau:up}) it holds
\beq
\tau^{j[+2]}=e^{-i \mathcal{P}} \kappa^{ji}\tau^{[+4]}_i= e^{-i \mathcal{P}}  \kappa^{ji}\mathbb{K}_i^{l}\tau^{[+2]}_l= e^{-i \mathcal{P}} \mathbb{K}^j_{i}\kappa^{il}\tau^{[+2]}_l= \mathbb{K}^j_{i}\tau^i.
\eeq
From that, it follows
\begin{align}
\left(f^{[+2]}(u)\cdot\mathbb{K}\cdot f^{-1}(u)\right)_i^{j}&=(\delta_i^{l}-\mathbb{K}_i^{p}\tau_p \mathbb{K}^l_{r} \tau^r)\mathbb{K}_l^{m}(\delta_m^{j}+\tau_m \tau^j)\\
&=(\mathbb{K}_i^{m}-\mathbb{K}_i^{p}\tau_p \tau^m)(\delta_m^{j}+\tau_m \tau^j)\\
&=\mathbb{K}_i^{j}+\mathbb{K}_i^m \tau_m\tau^j-\mathbb{K}_i^p \tau_p\tau^j-\mathbb{K}_i^{p}\tau_p \tau^m\tau_m \tau^j=\mathbb{K}_i^{j}\quad\Box ,
\end{align}
which together with (\ref{eq:stepapp}) proves the result. 

\paragraph{\emph{Proof} $\mathcal{G}(-u)=\pm\kappa\cdot\mathcal{G}^T(u)\cdot\kappa^{-1}$.} It can be proved (using the quadratic nature of branch points) that analytic continuation through the branch cut commutes with the operation $u\rightarrow-u$, see e.g. \cite{Gromov:2014caa}. Starting from the gluing condition \eqref{gluing:umeq}, sending $u \rightarrow-u$, taking the inverse and then the analytic continuation we obtain
\beq
{\bf Q}^{-1}(-u)=\left( \mathcal{G}^T (-u)\right) ^{-1}\cdot\kappa^{-1}\cdot\tilde{{\bf Q}}(u)\cdot \kappa^{-1}\cdot \left( \mathcal{G}(-u)\right)^{-1} 
\eeq  
Substituting again \eqref{gluing:umeq} in the last equation, one get
\begin{equation}
{\bf Q}^{-1}(-u)=\left( \mathcal{G}^T (-u)\right) ^{-1}\cdot\kappa^{-1}\cdot\mathcal{G}(u)\cdot \kappa\cdot {\bf Q}^{-1}(-u)\cdot\kappa \cdot\mathcal{G}^T (u)\cdot \kappa^{-1} \cdot\left( \mathcal{G}(-u)\right)^{-1}.
\end{equation}
It follows
\beq\label{opp:sign} 
\mathcal{G}(-u)=\pm\kappa\cdot\mathcal{G}^T(u)\cdot\kappa^{-1}\quad\Box.
\eeq
We observe here that the choice of the minus sign in \eqref{opp:sign} is fully consistent with the other constraints involved in the algorithm \ref{class} presented in the main text. In contrast, the plus sign is incompatible with these constraints and prevents the algorithm from admitting a solution.

\section{The case of local operators reviewed}\label{case:local}
We review how gluing is fixed in the case of local operators, which was considered in full generality in \cite{Bombardelli:2017vhk} albeit in a slightly different language. Here we show an alternative quick derivation in the symmetric sector with the method of this paper. 

We are considering the gluing matrix for local operators, hence $\mathcal{G}_i^j=\texttt{const}$. Due to LR symmetry and parity, $\mathcal{G}=\mathbb{K}\cdot\mathcal{G}\cdot\mathbb{K}$ and $\mathcal{G}=-\kappa\cdot\mathcal{G}^T\cdot\kappa^{-1}$. This fixes
\begin{equation}\label{form}
\mathcal{G}_{i}^{\,\,\,j}=\begin{pmatrix}
 g_{11} & 0 & 0 & g_{14}\\
 0 & g_{22}& g_{23} & 0 \\
 0 & g_{32} & -g_{22} & 0 \\
 g_{41} & 0 & 0 & -g_{11} \\
\end{pmatrix}\,.
\end{equation}
Firstly, from definition \eqref{G:def:extra}, we can build the matrix $f(u)$, and using the asymptotics of $\Theta(u)$ \eqref{theta:asym}, we deduce this should take the form
\beq
\lim_{u\rightarrow+\infty}f_i^{\,\,\,j}(u)=\mathcal{G}_i^{\,\,\,k}\lim_{u\rightarrow+\infty}\Theta_k^{\,\,\,j}(u)=\begin{pmatrix}
 e^{\i\pi\hat{\mathcal{N}_1}}g_{11} & 0 & 0 & e^{\i\pi\hat{\mathcal{N}_4}}g_{14}\\
 0 & e^{\i\pi\hat{\mathcal{N}_2}}g_{22}& e^{\i\pi\hat{\mathcal{N}_3}}g_{23} & 0 \\
 0 & e^{\i\pi\hat{\mathcal{N}_2}}g_{32} & -e^{\i\pi\hat{\mathcal{N}_3}}g_{22} & 0 \\
e^{\i\pi\hat{\mathcal{N}_1} }g_{41} & 0 & 0 & -e^{\i\pi\hat{\mathcal{N}_4}}g_{11} \\
\end{pmatrix},
\eeq
and similarly at $u\rightarrow-\infty$ using \eqref{theta:prod}. Now, we require the elementary property $f(u)+f^{-1}(u)=2\times\1$, that is \eqref{eq:extraconstr} in the symmetric sector,
\begin{equation}
\mathcal{G}\cdot\lim_{u\rightarrow\pm\infty}\Theta\(u\) +\kappa\cdot\lim_{u\rightarrow\pm\infty}\Theta^T\(u\)\cdot\mathcal{G}^T\cdot\kappa^{-1} = 2\times\1.
\end{equation}
This allows certain coefficients to be expressed in terms of the charges. Now, we can impose that the $f_i^j(u)$ matrix should be built out of only four functions $\tau_i(u)$ (see Section \ref{section}), and at large $u$ we find
\begin{align}\label{use:f}
\lim_{u\rightarrow\pm\infty}f_i^{\,\,\,j}(u)&=\left(
\begin{array}{cccc}
e^{\i\mathcal{P}}t_{1,\pm} t_{4,\pm}+1 & e^{\i\mathcal{P}}t_{1,\pm} t_{3,\pm} & -e^{\i\mathcal{P}}t_{1,\pm} t_{2,\pm} & -e^{\i\mathcal{P}}t_{1,\pm}^2 \\
e^{\i\mathcal{P}}t_{2,\pm}t_{4,\pm} & e^{\imath\mathcal{P}}t_{2,\pm} t_{3,\pm}+1 & -e^{\i\mathcal{P}}t_{2,\pm}^2 & -e^{\i\mathcal{P}}t_{1,\pm} t_{2,\pm} \\
 e^{\i\mathcal{P}}t_{3,\pm} t_{4,\pm} & e^{\i\mathcal{P}}t_{3,\pm}^2 & 1-e^{\i\mathcal{P}}t_{2,\pm}t_{3,\pm} & -e^{\i\mathcal{P}}t_{1,\pm} t_{3,\pm} \\
e^{\i\mathcal{P}}t_{4,\pm}^2 & e^{\i\mathcal{P}}t_{3,\pm} t_{4,\pm} & -e^{\i\mathcal{P}}t_{2,\pm} t_{4,\pm} & 1-e^{\i\mathcal{P}}t_{1,\pm} t_{4,\pm} \\
\end{array}
\right),
\end{align}
where  
\begin{equation} 
t_{i,\pm} \equiv \lim_{u\rightarrow\pm\infty} \tau_i(u),
\end{equation}
and we should assume that $\tau$ has constant asymptotics in order to have a gluing matrix independent of $u$. 
 Above, $e^{\i \mathcal{P}}$ is a constant appearing through (\ref{tau:up}), which in this sector satisfies $e^{2 \i \mathcal{P} }=1$.

The crucial step is now to use the fact that
\beq\label{eq:lasteqapp}
\lim_{u\rightarrow\pm\infty}f(u) = \mathcal{G}\cdot\lim_{u\rightarrow\pm\infty}\Theta(u)
\eeq
 to parametrize $\mathcal{G}$ by the charges and by the asymptotic values of $\tau_i(u)$. The two alternative expressions found using the relation at  $u\rightarrow \pm\infty$ should coincide, since $\mathcal{G}$ is a constant. This poses some additional constraints, and they can have only two possible outcomes, either $\Delta$ or $S$ needs to be quantized to integer values.
In particular, by employing the asymptotic behavior of $\Theta(u)$ along with the prototype form \eqref{form} in \eqref{use:f}, the conditions 
\begin{equation}
t_{1,\pm} = t_{4,\pm} = 0 \,\vee \, t_{2,\pm} = t_{3,\pm} = 0
\end{equation}
are immediately derived. It becomes evident that the case $t_{1,\pm} = t_{4,\pm} = 0$ results in the integer quantization of $\Delta$. In order to avoid this restriction, one must choose $t_{2,\pm} = t_{3,\pm} = 0$. Finally, we impose the condition 
$\text{det}\(\mathcal{G}\)=1$, as discussed in Section \ref{det}. Consequently, the gluing matrix \eqref{K:inter}, as reported in the main text, directly follows. The spin will be quantized to integer values according to
\beq
\e^{\i\pi \hat{\mathcal{N}}_2}=\mp\i ,
\eeq
which is one of the constraints stemming from (\ref{eq:lasteqapp}), together with
\beq 
(-1)^{M_1+M_2}=-1.
\eeq 

\section{Some numerical results}\label{sample}
Some numerical results for $S(\Delta,h)$ are reported below.
We are presenting here a minimal selection of polished data, but are happy to share more data upon request.

\begin{table}[h]
\centering
\caption{
Numerical results for the leading Regge trajectory interpolating the even spin operators in the $\Delta-S$ plane ($sl(2)$-like grading). Coupling is fixed to be $h=0.27$.}
\quad\\
\begin{tabular}{ |c|c| }
\hline
$\Delta_{sl(2)}$ & $S_{sl(2)}$  \\
\hline
\hline
 $-0.5$ & $-0.5267962336589070763252749469202(3)$\\ \hline\hline
 $0.175786$ & $-0.4132683587628778733046406947798962(5)$\\\hline
 $0.890786$ & $-0.0684735330846095852571316551180825(1)
 $\\\hline
 $1.570786$ & $0.40349193595508341712711865204630808(2)$\\\hline
 $2.490786$ & $1.15331291813913362657973321522920687(6)$\\\hline
 $2.930786$ & $1.535895832119957869450052453320828(8)$\\\hline
 $3.65$ & $2.18041725863311622312207884034814(6)$\\\hline
 $4.3$ & $2.77670390118381260576336103962597(5)$\\\hline
 $4.95$ & $3.38189806050410142138805775251816(9)$\\\hline
 $5.7$ & $4.0881362169514661658588998390465(8)$\\
\hline
\end{tabular}
\label{tabv}
\end{table}

\begin{table}[h]
\centering
\caption{Numerical results for the Pomeron as a function of the coupling (``bottom'' grading). The lesser precision in our data for $h \geq 0.75$ is due to a choice of lower truncation parameters.}
\quad\\
\begin{tabular}{ |c|c| }
\hline
$h$ & $\mathcal{S}_0$  \\
\hline
\hline
 $0.005$ & $1.0006652065447467198180(7)
 $ \\
 \hline
 $0.05$ & $1.0565169798036575580020(6)$  \\
\hline
 $0.1$ & $1.16593475741818149563694(3)
 $  \\
\hline
 $0.15$ & $1.2743232289040032420234(0)
 $  \\
\hline
$0.4$ & $1.60653570276846146957507(6)$  \\
\hline
$0.5$ & $ 1.6736828615476622260025(2)$  \\
\hline
$0.75$ & $1.7736017100(7)$  \\
\hline
$1$ & $1.82742249883(1)$  \\
\hline
$1.5$ & $1.8833291288(0)$  \\
\hline
$2$ & $1.91194771316(4)$  \\
\hline
\end{tabular}\\
\end{table}

\clearpage

\begin{table}[h]
\centering
\caption{Numerical results for the minimum of the leading odd trajectory as a function of the coupling (``bottom'' grading).}
\quad\\
\begin{tabular}{ |c|c| }
\hline
$h$ & $\mathcal{S}_0^{\text{odd}}$  \\
\hline
\hline
 $0.003$ & $0.5001897150261351386032301306349(2)$ \\
 \hline
 $0.01$ & $0.5020944885214757093188076580147(7)$ \\
 \hline
 $0.1$ & $0.6336521170620874337755649398308(5)
 $ \\
 \hline
 $0.15$ & $0.7183644121590866032768177141(3)$ \\
  \hline
  $0.2$ & $0.7867554536262786127683358642(8)
  $ \\
  \hline
$0.3$ & $0.87634247722560075689570311(2)$ \\
  \hline
 $0.5$ & $0.94753956996763001(2)$  \\
\hline
$0.8$ & $0.973009902592(3)$ \\
\hline
$0.92$ & $0.97698880504(7)$ \\
\hline
 $1$ & $0.97899057845(7)$ \\
\hline
\end{tabular}
\end{table}

\bibliography{references}

\providecommand{\href}[2]{#2}\begingroup\raggedright\begin{thebibliography}{10}

\bibitem{Regge:1959mz}
T.~Regge, {\it {Introduction to complex orbital momenta}},  {\em Nuovo Cim.} {\bf 14} (1959) 951.

\bibitem{Caron-Huot:2017vep}
S.~Caron-Huot, {\it {Analyticity in Spin in Conformal Theories}},  {\em JHEP} {\bf 09} (2017) 078 [\href{http://arXiv.org/abs/1703.00278}{{\tt 1703.00278}}].

\bibitem{Costa:2012cb}
M.~S. Costa, V.~Goncalves and J.~Penedones, {\it {Conformal Regge theory}},  {\em JHEP} {\bf 12} (2012) 091 [\href{http://arXiv.org/abs/1209.4355}{{\tt 1209.4355}}].

\bibitem{Mezei:2019dfv}
M.~Mezei and G.~S\'arosi, {\it {Chaos in the butterfly cone}},  {\em JHEP} {\bf 01} (2020) 186 [\href{http://arXiv.org/abs/1908.03574}{{\tt 1908.03574}}].

\bibitem{Hartman:2022zik}
T.~Hartman, D.~Mazac, D.~Simmons-Duffin and A.~Zhiboedov, {\it {Snowmass White Paper: The Analytic Conformal Bootstrap}},  in {\em {Snowmass 2021}}, 2, 2022.
\newblock \href{http://arXiv.org/abs/2202.11012}{{\tt 2202.11012}}.

\bibitem{Kravchuk:2018htv}
P.~Kravchuk and D.~Simmons-Duffin, {\it {Light-ray operators in conformal field theory}},  {\em JHEP} {\bf 11} (2018) 102 [\href{http://arXiv.org/abs/1805.00098}{{\tt 1805.00098}}].

\bibitem{Chang:2020qpj}
C.-H. Chang, M.~Kologlu, P.~Kravchuk, D.~Simmons-Duffin and A.~Zhiboedov, {\it {Transverse spin in the light-ray OPE}},  {\em JHEP} {\bf 05} (2022) 059 [\href{http://arXiv.org/abs/2010.04726}{{\tt 2010.04726}}].

\bibitem{Kologlu:2019mfz}
M.~Kologlu, P.~Kravchuk, D.~Simmons-Duffin and A.~Zhiboedov, {\it {The light-ray OPE and conformal colliders}},  {\em JHEP} {\bf 01} (2021) 128 [\href{http://arXiv.org/abs/1905.01311}{{\tt 1905.01311}}].

\bibitem{Henriksson:2023cnh}
J.~Henriksson, P.~Kravchuk and B.~Oertel, {\it {Missing local operators, zeros, and twist-4 trajectories}},  {\em JHEP} {\bf 07} (2024) 248 [\href{http://arXiv.org/abs/2312.09283}{{\tt 2312.09283}}].

\bibitem{Caron-Huot:2022eqs}
S.~Caron-Huot, M.~Kologlu, P.~Kravchuk, D.~Meltzer and D.~Simmons-Duffin, {\it {Detectors in weakly-coupled field theories}},  {\em JHEP} {\bf 04} (2023) 014 [\href{http://arXiv.org/abs/2209.00008}{{\tt 2209.00008}}].

\bibitem{Beisert:2010jr}
N.~Beisert {\em et.~al.}, {\it {Review of AdS/CFT Integrability: An Overview}},  {\em Lett. Math. Phys.} {\bf 99} (2012) 3--32 [\href{http://arXiv.org/abs/1012.3982}{{\tt 1012.3982}}].

\bibitem{10.1093/oso/9780198828150.001.0001}
P.~Dorey, G.~Korchemsky, N.~Nekrasov, V.~Schomerus, D.~Serban and L.~Cugliandolo, {\em {Integrability: From Statistical Systems to Gauge Theory: Lecture Notes of the Les Houches Summer School: Volume 106, June 2016}}.
\newblock Oxford University Press, 07, 2019.

\bibitem{Seibold:2024qkh}
F.~K. Seibold and A.~Sfondrini, {\it {AdS3 Integrability, Tensionless Limits, and Deformations: A Review}},  \href{http://arXiv.org/abs/2408.08414}{{\tt 2408.08414}}.

\bibitem{Gromov:2013pga}
N.~Gromov, V.~Kazakov, S.~Leurent and D.~Volin, {\it {Quantum Spectral Curve for Planar $\mathcal{N} = 4$ Super-Yang-Mills Theory}},  {\em Phys. Rev. Lett.} {\bf 112} (2014), no.~1 011602 [\href{http://arXiv.org/abs/1305.1939}{{\tt 1305.1939}}].

\bibitem{Alfimov:2014bwa}
M.~Alfimov, N.~Gromov and V.~Kazakov, {\it {QCD Pomeron from AdS/CFT Quantum Spectral Curve}},  {\em JHEP} {\bf 07} (2015) 164 [\href{http://arXiv.org/abs/1408.2530}{{\tt 1408.2530}}].

\bibitem{Gromov:2014bva}
N.~Gromov, F.~Levkovich-Maslyuk, G.~Sizov and S.~Valatka, {\it {Quantum spectral curve at work: from small spin to strong coupling in $ \mathcal{N} $ = 4 SYM}},  {\em JHEP} {\bf 07} (2014) 156 [\href{http://arXiv.org/abs/1402.0871}{{\tt 1402.0871}}].

\bibitem{Gromov:2015wca}
N.~Gromov, F.~Levkovich-Maslyuk and G.~Sizov, {\it {Quantum Spectral Curve and the Numerical Solution of the Spectral Problem in AdS5/CFT4}},  {\em JHEP} {\bf 06} (2016) 036 [\href{http://arXiv.org/abs/1504.06640}{{\tt 1504.06640}}].

\bibitem{Gromov:2015vua}
N.~Gromov, F.~Levkovich-Maslyuk and G.~Sizov, {\it {Pomeron Eigenvalue at Three Loops in $\mathcal N=$ 4 Supersymmetric Yang-Mills Theory}},  {\em Phys. Rev. Lett.} {\bf 115} (2015), no.~25 251601 [\href{http://arXiv.org/abs/1507.04010}{{\tt 1507.04010}}].

\bibitem{Alfimov:2018cms}
M.~Alfimov, N.~Gromov and G.~Sizov, {\it {BFKL spectrum of $ \mathcal{N} $ = 4: non-zero conformal spin}},  {\em JHEP} {\bf 07} (2018) 181 [\href{http://arXiv.org/abs/1802.06908}{{\tt 1802.06908}}].

\bibitem{Klabbers:2023zdz}
R.~Klabbers, M.~Preti and I.~M. Sz\'ecs\'enyi, {\it {Regge Spectroscopy of Higher-Twist States in N=4 Supersymmetric Yang-Mills Theory}},  {\em Phys. Rev. Lett.} {\bf 132} (2024), no.~19 191601 [\href{http://arXiv.org/abs/2307.15107}{{\tt 2307.15107}}].

\bibitem{Ekhammar:2024neh}
S.~Ekhammar, N.~Gromov and M.~Preti, {\it {Long Range Asymptotic Baxter-Bethe Ansatz for N=4 BFKL}},  \href{http://arXiv.org/abs/2406.18639}{{\tt 2406.18639}}.

\bibitem{Homrich:2022mmd}
A.~Homrich, D.~Simmons-Duffin and P.~Vieira, {\it {Complex Spin: The Missing Zeroes and Newton's Dark Magic}},  \href{http://arXiv.org/abs/2211.13754}{{\tt 2211.13754}}.

\bibitem{Kuraev:1977fs}
E.~A. Kuraev, L.~N. Lipatov and V.~S. Fadin, {\it {The Pomeranchuk Singularity in Nonabelian Gauge Theories}},  {\em Sov. Phys. JETP} {\bf 45} (1977) 199--204.

\bibitem{Balitsky:1978ic}
I.~I. Balitsky and L.~N. Lipatov, {\it {The Pomeranchuk Singularity in Quantum Chromodynamics}},  {\em Sov. J. Nucl. Phys.} {\bf 28} (1978) 822--829.

\bibitem{Gromov:2014caa}
N.~Gromov, V.~Kazakov, S.~Leurent and D.~Volin, {\it {Quantum spectral curve for arbitrary state/operator in AdS$_{5}$/CFT$_{4}$}},  {\em JHEP} {\bf 09} (2015) 187 [\href{http://arXiv.org/abs/1405.4857}{{\tt 1405.4857}}].

\bibitem{Cavaglia:2014exa}
A.~Cavagli\`a, D.~Fioravanti, N.~Gromov and R.~Tateo, {\it {Quantum Spectral Curve of the $\mathcal N=$ 6 Supersymmetric Chern-Simons Theory}},  {\em Phys. Rev. Lett.} {\bf 113} (2014), no.~2 021601 [\href{http://arXiv.org/abs/1403.1859}{{\tt 1403.1859}}].

\bibitem{Bombardelli:2017vhk}
D.~Bombardelli, A.~Cavagli\`a, D.~Fioravanti, N.~Gromov and R.~Tateo, {\it {The full Quantum Spectral Curve for $AdS_4/CFT_3$}},  {\em JHEP} {\bf 09} (2017) 140 [\href{http://arXiv.org/abs/1701.00473}{{\tt 1701.00473}}].

\bibitem{Gromov:2014eha}
N.~Gromov and G.~Sizov, {\it {Exact Slope and Interpolating Functions in N=6 Supersymmetric Chern-Simons Theory}},  {\em Phys. Rev. Lett.} {\bf 113} (2014), no.~12 121601 [\href{http://arXiv.org/abs/1403.1894}{{\tt 1403.1894}}].

\bibitem{Anselmetti:2015mda}
L.~Anselmetti, D.~Bombardelli, A.~Cavagli\`a and R.~Tateo, {\it {12 loops and triple wrapping in ABJM theory from integrability}},  {\em JHEP} {\bf 10} (2015) 117 [\href{http://arXiv.org/abs/1506.09089}{{\tt 1506.09089}}].

\bibitem{Bombardelli:2018bqz}
D.~Bombardelli, A.~Cavagli\`a, R.~Conti and R.~Tateo, {\it {Exploring the spectrum of planar AdS$_{4}$/CFT$_{3}$ at finite coupling}},  {\em JHEP} {\bf 04} (2018) 117 [\href{http://arXiv.org/abs/1803.04748}{{\tt 1803.04748}}].

\bibitem{Lee:2018jvn}
R.~N. Lee and A.~I. Onishchenko, {\it {Toward an analytic perturbative solution for the ABJM quantum spectral curve}},  {\em Teor. Mat. Fiz.} {\bf 198} (2019), no.~2 292--308 [\href{http://arXiv.org/abs/1807.06267}{{\tt 1807.06267}}].

\bibitem{Lee:2019oml}
R.~N. Lee and A.~I. Onishchenka, {\it {ABJM quantum spectral curve at twist 1: algorithmic perturbative solution}},  {\em JHEP} {\bf 11} (2019) 018 [\href{http://arXiv.org/abs/1905.03116}{{\tt 1905.03116}}].

\bibitem{Ekhammar:2023cuj}
S.~Ekhammar, J.~A. Minahan and C.~Thull, {\it {The ABJM Hagedorn Temperature from Integrability}},  {\em JHEP} {\bf 10} (2023) 066 [\href{http://arXiv.org/abs/2307.02350}{{\tt 2307.02350}}].

\bibitem{Beisert:2005tm}
N.~Beisert, {\it {The SU(2|2) dynamic S-matrix}},  {\em Adv. Theor. Math. Phys.} {\bf 12} (2008) 945--979 [\href{http://arXiv.org/abs/hep-th/0511082}{{\tt hep-th/0511082}}].

\bibitem{Beisert:2005fw}
N.~Beisert and M.~Staudacher, {\it {Long-range psu(2,2|4) Bethe Ansatze for gauge theory and strings}},  {\em Nucl. Phys. B} {\bf 727} (2005) 1--62 [\href{http://arXiv.org/abs/hep-th/0504190}{{\tt hep-th/0504190}}].

\bibitem{Arutyunov:2009ur}
G.~Arutyunov and S.~Frolov, {\it {Thermodynamic Bethe Ansatz for the AdS(5) x S(5) Mirror Model}},  {\em JHEP} {\bf 05} (2009) 068 [\href{http://arXiv.org/abs/0903.0141}{{\tt 0903.0141}}].

\bibitem{Gromov:2009tv}
N.~Gromov, V.~Kazakov and P.~Vieira, {\it {Exact Spectrum of Anomalous Dimensions of Planar N=4 Supersymmetric Yang-Mills Theory}},  {\em Phys. Rev. Lett.} {\bf 103} (2009) 131601 [\href{http://arXiv.org/abs/0901.3753}{{\tt 0901.3753}}].

\bibitem{Bombardelli:2009ns}
D.~Bombardelli, D.~Fioravanti and R.~Tateo, {\it {Thermodynamic Bethe Ansatz for planar AdS/CFT: A Proposal}},  {\em J. Phys. A} {\bf 42} (2009) 375401 [\href{http://arXiv.org/abs/0902.3930}{{\tt 0902.3930}}].

\bibitem{Bombardelli:2009xz}
D.~Bombardelli, D.~Fioravanti and R.~Tateo, {\it {TBA and Y-system for planar AdS(4)/CFT(3)}},  {\em Nucl. Phys. B} {\bf 834} (2010) 543--561 [\href{http://arXiv.org/abs/0912.4715}{{\tt 0912.4715}}].

\bibitem{Gromov:2009at}
N.~Gromov and F.~Levkovich-Maslyuk, {\it {Y-system, TBA and Quasi-Classical strings in AdS(4) x CP3}},  {\em JHEP} {\bf 06} (2010) 088 [\href{http://arXiv.org/abs/0912.4911}{{\tt 0912.4911}}].

\bibitem{Cavaglia:2010nm}
A.~Cavaglià, D.~Fioravanti and R.~Tateo, {\it {Extended Y-system for the $AdS_5/CFT_4$ correspondence}},  {\em Nucl. Phys. B} {\bf 843} (2011) 302--343 [\href{http://arXiv.org/abs/1005.3016}{{\tt 1005.3016}}].

\bibitem{Gromov:2011cx}
N.~Gromov, V.~Kazakov, S.~Leurent and D.~Volin, {\it {Solving the AdS/CFT Y-system}},  {\em JHEP} {\bf 07} (2012) 023 [\href{http://arXiv.org/abs/1110.0562}{{\tt 1110.0562}}].

\bibitem{Ekhammar:2021pys}
S.~Ekhammar and D.~Volin, {\it {Monodromy bootstrap for SU(2|2) quantum spectral curves: from Hubbard model to AdS$_{3}$/CFT$_{2}$}},  {\em JHEP} {\bf 03} (2022) 192 [\href{http://arXiv.org/abs/2109.06164}{{\tt 2109.06164}}].

\bibitem{Cavaglia:2021eqr}
A.~Cavagli\`a, N.~Gromov, B.~Stefa\'nski, Jr., Jr. and A.~Torrielli, {\it {Quantum Spectral Curve for AdS$_{3}$/CFT$_{2}$: a proposal}},  {\em JHEP} {\bf 12} (2021) 048 [\href{http://arXiv.org/abs/2109.05500}{{\tt 2109.05500}}].

\bibitem{Gromov:2017blm}
N.~Gromov, {\it {Introduction to the Spectrum of $N=4$ SYM and the Quantum Spectral Curve}},  \href{http://arXiv.org/abs/1708.03648}{{\tt 1708.03648}}.

\bibitem{BrizioPoster}
N.~Brizio, ``{Regge trajectories in ABJM theory \& Quantum Spectral Curve}.'' Poster presented at Integrability in Gauge and String Theory, 2023.
\newblock Zürich, \url{https://indico.phys.ethz.ch/event/49/}.

\bibitem{JohanH}
J.~Henriksson, ``{High Spin Bootstrap of Conformal Gauge Theories}.'' Oxford transfer of status report Michaelmas term 2017.

\bibitem{Henriksson:2017eej}
J.~Henriksson and T.~Lukowski, {\it {Perturbative Four-Point Functions from the Analytic Conformal Bootstrap}},  {\em JHEP} {\bf 02} (2018) 123 [\href{http://arXiv.org/abs/1710.06242}{{\tt 1710.06242}}].

\bibitem{Julius:2024ewf}
J.~Julius and N.~S. Sokolova, {\it {Unmixing sub-leading Regge trajectories of $\mathcal{N} = 4$ Super-Yang-Mills}},  \href{http://arXiv.org/abs/2409.07529}{{\tt 2409.07529}}.

\bibitem{Lemos:2021azv}
M.~Lemos, B.~C. van Rees and X.~Zhao, {\it {Regge trajectories for the (2, 0) theories}},  {\em JHEP} {\bf 01} (2022) 022 [\href{http://arXiv.org/abs/2105.13361}{{\tt 2105.13361}}].

\bibitem{Papathanasiou:2009en}
G.~Papathanasiou and M.~Spradlin, {\it {The Morphology of N=6 Chern-Simons Theory}},  {\em JHEP} {\bf 07} (2009) 036 [\href{http://arXiv.org/abs/0903.2548}{{\tt 0903.2548}}].

\bibitem{Papathanasiou:2009zm}
G.~Papathanasiou and M.~Spradlin, {\it {Two-Loop Spectroscopy of Short ABJM Operators}},  {\em JHEP} {\bf 02} (2010) 072 [\href{http://arXiv.org/abs/0911.2220}{{\tt 0911.2220}}].

\bibitem{Reshetikhin:1983vw}
N.~Y. Reshetikhin, {\it {A Method Of Functional Equations In The Theory Of Exactly Solvable Quantum Systems}},  {\em Lett. Math. Phys.} {\bf 7} (1983) 205--213.

\bibitem{Dorey:2006an}
P.~Dorey, C.~Dunning, D.~Masoero, J.~Suzuki and R.~Tateo, {\it {Pseudo-differential equations, and the Bethe ansatz for the classical Lie algebras}},  {\em Nucl. Phys. B} {\bf 772} (2007) 249--289 [\href{http://arXiv.org/abs/hep-th/0612298}{{\tt hep-th/0612298}}].

\bibitem{Ferrando:2020vzk}
G.~Ferrando, R.~Frassek and V.~Kazakov, {\it {QQ-system and Weyl-type transfer matrices in integrable SO(2r) spin chains}},  {\em JHEP} {\bf 02} (2021) 193 [\href{http://arXiv.org/abs/2008.04336}{{\tt 2008.04336}}].

\bibitem{Ekhammar:2021myw}
S.~Ekhammar and D.~Volin, {\it {Bethe Algebra using Pure Spinors}},  \href{http://arXiv.org/abs/2104.04539}{{\tt 2104.04539}}.

\bibitem{Cavaglia:2016ide}
A.~Cavagli\`a, N.~Gromov and F.~Levkovich-Maslyuk, {\it {On the Exact Interpolating Function in ABJ Theory}},  {\em JHEP} {\bf 12} (2016) 086 [\href{http://arXiv.org/abs/1605.04888}{{\tt 1605.04888}}].

\bibitem{Binder:2021cnk}
D.~J. Binder, {\em {On $N=6$ Superconformal Field Theories}}.
\newblock PhD thesis, Princeton U., Princeton U., 2021.

\bibitem{Binder:2020ckj}
D.~J. Binder, S.~M. Chester, M.~Jerdee and S.~S. Pufu, {\it {The 3d $ \mathcal{N} $ = 6 bootstrap: from higher spins to strings to membranes}},  {\em JHEP} {\bf 05} (2021) 083 [\href{http://arXiv.org/abs/2011.05728}{{\tt 2011.05728}}].

\bibitem{Gromov:2023hzc}
N.~Gromov, A.~Hegedus, J.~Julius and N.~Sokolova, {\it {Fast QSC Solver: tool for systematic study of N=4 Super-Yang-Mills spectrum}},  \href{http://arXiv.org/abs/2306.12379}{{\tt 2306.12379}}.

\bibitem{Ekhammar:2024rfj}
S.~Ekhammar, N.~Gromov and P.~Ryan, {\it {New Approach to Strongly Coupled N = 4 SYM via Integrability}},  \href{http://arXiv.org/abs/2406.02698}{{\tt 2406.02698}}.

\bibitem{Cavaglia:2022xld}
A.~Cavagli\`a, S.~Ekhammar, N.~Gromov and P.~Ryan, {\it {Exploring the Quantum Spectral Curve for AdS$_{3}$/CFT$_{2}$}},  {\em JHEP} {\bf 12} (2023) 089 [\href{http://arXiv.org/abs/2211.07810}{{\tt 2211.07810}}].

\bibitem{Velizhanin:2022faj}
V.~N. Velizhanin, {\it {Analytic continuation of harmonic sums with purely imaginary indices near the integer values}},  {\em Int. J. Mod. Phys. A} {\bf 38} (2023), no.~06n07 2350036 [\href{http://arXiv.org/abs/2210.14214}{{\tt 2210.14214}}].

\bibitem{Alfimov:2020obh}
M.~Alfimov, N.~Gromov and V.~Kazakov, {\em {Chapter 13: N=4 SYM Quantum Spectral Curve in BFKL Regime}}, pp.~335--367.
\newblock 2021.
\newblock \href{http://arXiv.org/abs/2003.03536}{{\tt 2003.03536}}.

\bibitem{Maldacena:2015waa}
J.~Maldacena, S.~H. Shenker and D.~Stanford, {\it {A bound on chaos}},  {\em JHEP} {\bf 08} (2016) 106 [\href{http://arXiv.org/abs/1503.01409}{{\tt 1503.01409}}].

\bibitem{Aprile:2017qoy}
F.~Aprile, J.~M. Drummond, P.~Heslop and H.~Paul, {\it {Loop corrections for Kaluza-Klein AdS amplitudes}},  {\em JHEP} {\bf 05} (2018) 056 [\href{http://arXiv.org/abs/1711.03903}{{\tt 1711.03903}}].

\bibitem{Gurdogan:2015csr}
O.~G\"urdo\u{g}an and V.~Kazakov, {\it {New Integrable 4D Quantum Field Theories from Strongly Deformed Planar $\mathcal N = $ 4 Supersymmetric Yang-Mills Theory}},  {\em Phys. Rev. Lett.} {\bf 117} (2016), no.~20 201602 [\href{http://arXiv.org/abs/1512.06704}{{\tt 1512.06704}}]. [Addendum: Phys.Rev.Lett. 117, 259903 (2016)].

\bibitem{Grabner:2017pgm}
D.~Grabner, N.~Gromov, V.~Kazakov and G.~Korchemsky, {\it {Strongly $\gamma$-Deformed $\mathcal{N}=4$ Supersymmetric Yang-Mills Theory as an Integrable Conformal Field Theory}},  {\em Phys. Rev. Lett.} {\bf 120} (2018), no.~11 111601 [\href{http://arXiv.org/abs/1711.04786}{{\tt 1711.04786}}].

\bibitem{Gromov:2017cja}
N.~Gromov, V.~Kazakov, G.~Korchemsky, S.~Negro and G.~Sizov, {\it {Integrability of Conformal Fishnet Theory}},  {\em JHEP} {\bf 01} (2018) 095 [\href{http://arXiv.org/abs/1706.04167}{{\tt 1706.04167}}].

\bibitem{Levkovich-Maslyuk:2020rlp}
F.~Levkovich-Maslyuk and M.~Preti, {\it {Exploring the ground state spectrum of \ensuremath{\gamma}-deformed N = 4 SYM}},  {\em JHEP} {\bf 06} (2022) 146 [\href{http://arXiv.org/abs/2003.05811}{{\tt 2003.05811}}].

\bibitem{Gromov:2016itr}
N.~Gromov, F.~Levkovich-Maslyuk and G.~Sizov, {\it {New Construction of Eigenstates and Separation of Variables for SU(N) Quantum Spin Chains}},  {\em JHEP} {\bf 09} (2017) 111 [\href{http://arXiv.org/abs/1610.08032}{{\tt 1610.08032}}].

\bibitem{Cavaglia:2018lxi}
A.~Cavagli\`a, N.~Gromov and F.~Levkovich-Maslyuk, {\it {Quantum spectral curve and structure constants in $ \mathcal{N}=4 $ SYM: cusps in the ladder limit}},  {\em JHEP} {\bf 10} (2018) 060 [\href{http://arXiv.org/abs/1802.04237}{{\tt 1802.04237}}].

\bibitem{Giombi:2018hsx}
S.~Giombi and S.~Komatsu, {\it {More Exact Results in the Wilson Loop Defect CFT: Bulk-Defect OPE, Nonplanar Corrections and Quantum Spectral Curve}},  {\em J. Phys. A} {\bf 52} (2019), no.~12 125401 [\href{http://arXiv.org/abs/1811.02369}{{\tt 1811.02369}}].

\bibitem{Maillet:2018bim}
J.~M. Maillet and G.~Niccoli, {\it {On quantum separation of variables}},  {\em J. Math. Phys.} {\bf 59} (2018), no.~9 091417 [\href{http://arXiv.org/abs/1807.11572}{{\tt 1807.11572}}].

\bibitem{Ryan:2018fyo}
P.~Ryan and D.~Volin, {\it {Separated variables and wave functions for rational gl(N) spin chains in the companion twist frame}},  {\em J. Math. Phys.} {\bf 60} (2019), no.~3 032701 [\href{http://arXiv.org/abs/1810.10996}{{\tt 1810.10996}}].

\bibitem{Gromov:2022waj}
N.~Gromov, N.~Primi and P.~Ryan, {\it {Form-factors and complete basis of observables via separation of variables for higher rank spin chains}},  {\em JHEP} {\bf 11} (2022) 039 [\href{http://arXiv.org/abs/2202.01591}{{\tt 2202.01591}}].

\bibitem{Basso:2022nny}
B.~Basso, A.~Georgoudis and A.~K. Sueiro, {\it {Structure Constants of Short Operators in Planar N=4 Supersymmetric Yang-Mills Theory}},  {\em Phys. Rev. Lett.} {\bf 130} (2023), no.~13 131603 [\href{http://arXiv.org/abs/2207.01315}{{\tt 2207.01315}}].

\bibitem{Bercini:2022jxo}
C.~Bercini, A.~Homrich and P.~Vieira, {\it {Structure Constants in $\mathcal{N} = 4$ SYM and Separation of Variables}},  \href{http://arXiv.org/abs/2210.04923}{{\tt 2210.04923}}.

\bibitem{Ekhammar:2023iph}
S.~Ekhammar, N.~Gromov and P.~Ryan, {\it {Boundary overlaps from Functional Separation of Variables}},  {\em JHEP} {\bf 05} (2024) 268 [\href{http://arXiv.org/abs/2312.11612}{{\tt 2312.11612}}].

\bibitem{Cavaglia:2021bnz}
A.~Cavagli\`a, N.~Gromov, J.~Julius and M.~Preti, {\it {Integrability and conformal bootstrap: One dimensional defect conformal field theory}},  {\em Phys. Rev. D} {\bf 105} (2022), no.~2 L021902 [\href{http://arXiv.org/abs/2107.08510}{{\tt 2107.08510}}].

\bibitem{Cavaglia:2022qpg}
A.~Cavagli\`a, N.~Gromov, J.~Julius and M.~Preti, {\it {Bootstrability in defect CFT: integrated correlators and sharper bounds}},  {\em JHEP} {\bf 05} (2022) 164 [\href{http://arXiv.org/abs/2203.09556}{{\tt 2203.09556}}].

\bibitem{Cavaglia:2023mmu}
A.~Cavagli\`a, N.~Gromov and M.~Preti, {\it {Computing Four-Point Functions with Integrability, Bootstrap and Parity Symmetry}},  \href{http://arXiv.org/abs/2312.11604}{{\tt 2312.11604}}.

\bibitem{Caron-Huot:2022sdy}
S.~Caron-Huot, F.~Coronado, A.-K. Trinh and Z.~Zahraee, {\it {Bootstrapping $ \mathcal{N} $ = 4 sYM correlators using integrability}},  {\em JHEP} {\bf 02} (2023) 083 [\href{http://arXiv.org/abs/2207.01615}{{\tt 2207.01615}}].

\bibitem{Alday:2022uxp}
L.~F. Alday, T.~Hansen and J.~A. Silva, {\it {AdS Virasoro-Shapiro from dispersive sum rules}},  \href{http://arXiv.org/abs/2204.07542}{{\tt 2204.07542}}.

\bibitem{Alday:2022xwz}
L.~F. Alday, T.~Hansen and J.~A. Silva, {\it {AdS Virasoro-Shapiro from single-valued periods}},  {\em JHEP} {\bf 12} (2022) 010 [\href{http://arXiv.org/abs/2209.06223}{{\tt 2209.06223}}].

\bibitem{Alday:2023mvu}
L.~F. Alday and T.~Hansen, {\it {The AdS Virasoro-Shapiro amplitude}},  {\em JHEP} {\bf 10} (2023) 023 [\href{http://arXiv.org/abs/2306.12786}{{\tt 2306.12786}}].

\bibitem{Alday:2023jdk}
L.~F. Alday, T.~Hansen and J.~A. Silva, {\it {Emergent Worldsheet for the AdS Virasoro-Shapiro Amplitude}},  {\em Phys. Rev. Lett.} {\bf 131} (2023), no.~16 161603 [\href{http://arXiv.org/abs/2305.03593}{{\tt 2305.03593}}].

\bibitem{Alday:2023flc}
L.~F. Alday, T.~Hansen and J.~A. Silva, {\it {On the spectrum and structure constants of short operators in N=4 SYM at strong coupling}},  {\em JHEP} {\bf 08} (2023) 214 [\href{http://arXiv.org/abs/2303.08834}{{\tt 2303.08834}}].

\bibitem{Julius:2023hre}
J.~Julius and N.~Sokolova, {\it {Conformal field theory-data analysis for $\mathcal{N}$ = 4 Super-Yang-Mills at strong coupling}},  {\em JHEP} {\bf 03} (2024) 090 [\href{http://arXiv.org/abs/2310.06041}{{\tt 2310.06041}}].

\bibitem{Chester:2024esn}
S.~M. Chester, T.~Hansen and D.-l. Zhong, {\it {The type IIA Virasoro-Shapiro amplitude in AdS$_4$$\times$ CP$^3$ from ABJM theory}},  \href{http://arXiv.org/abs/2412.08689}{{\tt 2412.08689}}.

\end{thebibliography}\endgroup
\end{document}